\documentclass[aps,nofootinbib,onecolumn,groupedaddress,english,eqsecnum,showpacs,11pt]{revtex4-1}
\usepackage{babel,times}
\usepackage{xcolor}
\usepackage{graphicx}
\usepackage{amsmath}
\usepackage{amssymb}
\usepackage{footnote}

\definecolor{darkblue}{rgb}{0.1,0.45,0.6}
\definecolor{darkred}{rgb}{0.65,0,0.2}

\usepackage[colorlinks,citecolor=darkblue,linkcolor=darkred,urlcolor=darkblue]{hyperref} 

\makeatletter
\adddialect\l@English\l@english
\makeatother

\newcommand{\mean}[1]{\langle#1\rangle}

\def\be{\begin{equation}} \def\ee{\end{equation}}
\def\bea{\begin{eqnarray}} \def\eea{\end{eqnarray}}
\newcommand{\dagga}{{\phantom{\dagger}}}

\begin{document} 
\title{\Large QUANTUM ENTANGLEMENT IN\\
CONDENSED MATTER SYSTEMS}
\author{Nicolas Laflorencie}
\address{Laboratoire de Physique Th\'eorique, Universit\'e de Toulouse, CNRS, UPS, France}


\begin{abstract}
\vskip 1cm 
{\large{This review focuses on the field of quantum entanglement applied to condensed matter physics systems with strong correlations, a domain which has rapidly grown over the last decade. By tracing out part of the degrees of freedom of correlated quantum systems, useful and non-trivial informations can be obtained through the study of the reduced density matrix, whose eigenvalue spectrum (the entanglement spectrum) and the associated R\'enyi entropies are now well recognized to contains key features. In particular, the celebrated area law for the entanglement entropy of ground-states will be discussed from the perspective of its subleading corrections which encode universal details of various quantum states of matter, {\it{e.g.}} symmetry breaking states or topological order. Going beyond entropies, the study of the low-lying part of the entanglement spectrum also allows to diagnose topological properties or give a direct access to the excitation spectrum of the edges, and may also raise significant questions about the underlying entanglement Hamiltonian. All these powerful tools can be further applied to shed some light on disordered quantum systems where impurity/disorder can conspire with quantum fluctuations to induce non-trivial effects. Disordered quantum spin systems, the Kondo effect, or the many-body localization problem, which have all been successfully (re)visited through the prism of quantum entanglement, will be discussed in details. Finally, the issue of experimental access to entanglement measurement will be addressed, together with its most recent developments.}}\end{abstract}
\maketitle

\tableofcontents

\newpage \section{Introduction}
The "spooky" nature of quantum entanglement has been a subject of several and intense debates
since the early days of quantum
mechanics~\cite{einstein_can_1935,schrodinger_gegenwartige_1935}. 
At first attached to fundamental questions regarding the formulation
and the foundations of quantum mechanics, in particular after Bell's
theorem~\cite{bell_problem_1966,aspect_experimental_1982}, the concept of entanglement has recently
generated an enormous interest in several communities~\cite{bennett_teleporting_1993,bouwmeester_experimental_1997,shor_polynomial-time_1997,wootters_entanglement_1998,gisin_quantum_2002,ryu_holographic_2006,amico_entanglement_2008,horodecki_quantum_2009,eisert_colloquium:_2010,georgescu_quantum_2014,aolita_open-system_2015}, 
{\it{e.g.}}
atomic physics and quantum optics, condensed matter, mathematical
physics, high energy physics, quantum information and quantum cryptography, cosmology, etc. 

Shortly after the seminal works of Bennett and co-workers~\cite{bennett_teleporting_1993} on quantum teleportation, the study of entanglement in condensed matter model systems has strongly benefited from the rapid development of quantum information~\cite{nielsen_quantum_2000,preskill_quantum_2000}. Built at the interface of quantum information science, condensed matter theory, statistical physics, quantum field theory, the study of many-body entangled states rapidly has become a very active topic, in particular triggered by the discovery that entanglement could serve as a new smoking gun for quantum critical phenomena~\cite{osterloh_scaling_2002,osborne_entanglement_2002,vidal_entanglement_2003,calabrese_entanglement_2004}. In parallel, numerical approaches for quantum many-body systems have also taken advantage of such progresses, with the impressive development of very efficient variational techniques such as matrix-product-state and tensor network methods~\cite{verstraete_2008,cirac_2009,schollwock_density-matrix_2011,orus_2014}.

\begin{figure}[b]
  \centering
  \includegraphics[width=5cm,clip]{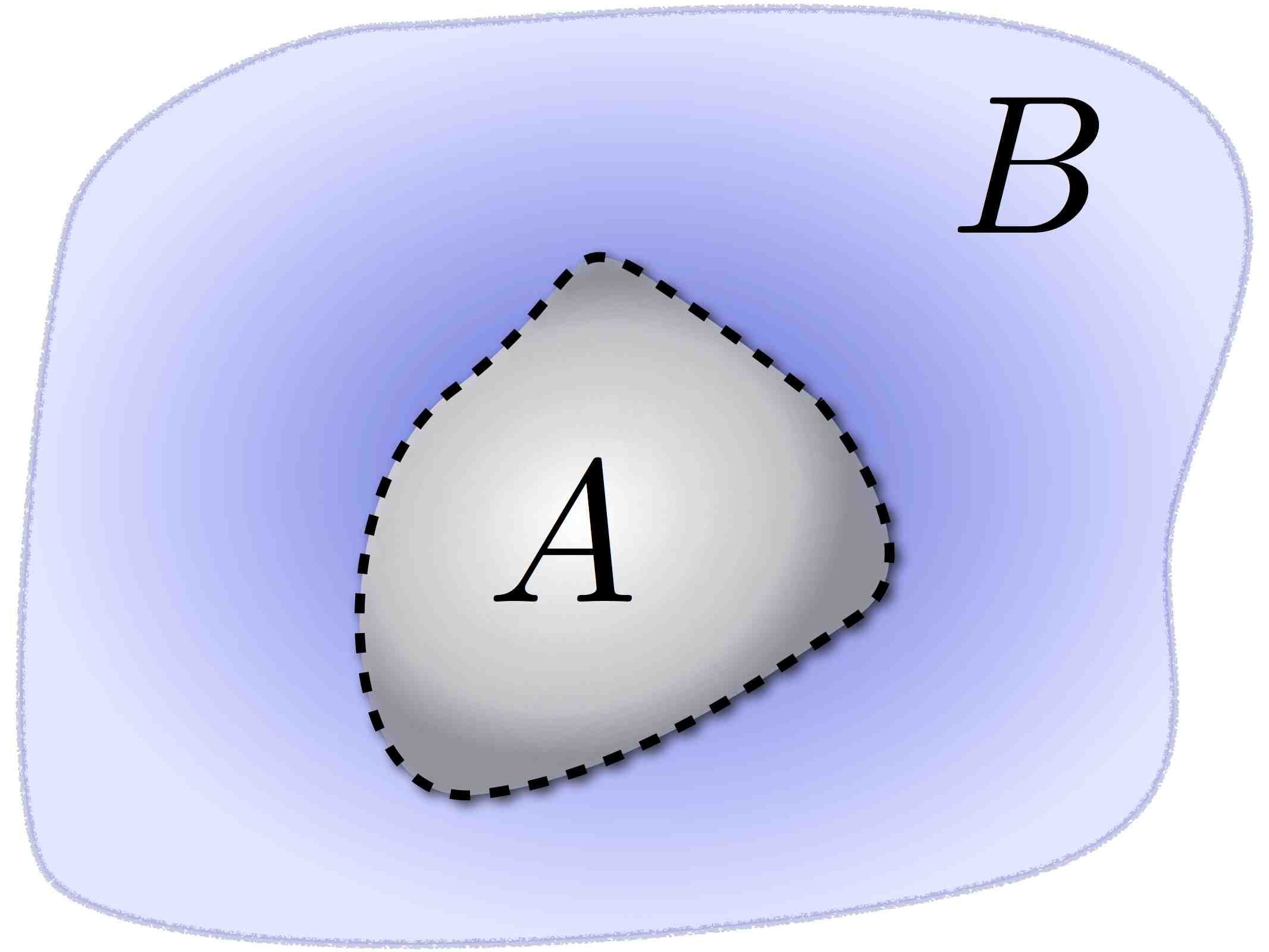}    
  \caption{Real space bipartition of a physical system (here in dimension $d=2$) in two parts.}
  \label{fig:AB}
\end{figure}

The framework of quantum entanglement for condensed matter physics is nowadays very wide, and this topic has certainly acquired now a robust and mature status. It is therefore quite ambitious and perhaps illusory to pretend giving an exhaustive overview. Consequently, some aspects of quantum entanglement in condensed matter systems will not be discussed here. For instance, we will not address multipartite entanglement, focusing on the bipartite case, as exemplified in Fig.~\ref{fig:AB} for a two-dimensional ($d=2$) system. In such a case, assuming $|\Psi\rangle$ is a pure state defined over degrees of freedom of $A \cup B$, the central object of interest in this review article will be the reduced density matrix (RDM) associated with such a real space bipartition\footnote{Other types of bipartitions (momentum or orbital) will also be discussed, see Section~\ref{sec:ES}. Let us also mention the particle-partitioning~\cite{haque_2009}, and also the notion of "field space entanglement" between different fields, studied by several authors~\cite{xu_entanglement_2011,mollabashi_entanglement_2014,mozaffar_entanglement_2016}.}, defined by
\be
\rho_A={\rm{Tr}}_B |\Psi\rangle\langle\Psi|.
\ee
The eigenvalues of $\rho_A$ (the entanglement spectrum) and the corresponding R\'enyi entanglement entropy
\be
S_q(A)=\frac{1}{1-q}\ln {\rm{Tr}}\left(\rho^q_A\right)
\ee
will be the main targets of this overview. 

In a first part (Section~\ref{sec:AL}), after recalling important results in one dimension, the celebrated area law for the ground-state entanglement entropy will be briefly discussed, before focusing on sub-leading correcting terms which carry universality for various quantum states of matter, broken symmetries and topological properties.  In Section~\ref{sec:ES}, entanglement spectroscopy results will be presented for a large variety of low-dimensional correlated ground-states, {\it{e.g.}} gapped and gapless spin chains and ladders, topological ordered phases such as fractional quantum Hall states or spin liquids, broken continuous symmetry states such as superfluids. Disorder quantum systems will then be considered in Section~\ref{sec:dis}, focusing on three important cases. First we address ground-state entanglement for disordered quantum spin models and Kondo impurity problems. Then the many-body localization of excited states will be discussed through entanglement features. Finally, Section~\ref{sec:exp} will address the crucial issue of experimental detection and measure of entanglement. Finally, conclusions and some open questions will be listed in Section~\ref{sec:con}.

\section{The area law, and beyond}
\label{sec:AL}
Originally studied by Bombelli {\it{et al.}}~\cite{bombelli_quantum_1986}, Srednicki~\cite{srednicki_entropy_1993}, Callan and Wilczek~\cite{callan_geometric_1994}, Holzhey {\it{et al.}}~\cite{holzhey_geometric_1994} in the context of black-hole physics~\cite{hawking_desitter_2001}, 
the geometric, or entanglement entropy in the
ground-state of a free scalar bosonic field is well-known to obey the so-called
area law, {\it{i.e.}} in dimension $d$ it scales with the surface of the subsystem $A$:
\be
S_q(A)=a_q L^{d-1}+\ldots\label{eq:AL}
\ee
In this part, we first focus one dimensional systems, and then discuss higher dimensions, in particular the corrections (the ellipses in Eq.~\eqref{eq:AL} above) beyond the area-law.
\subsection{One dimension}
\subsubsection{Entanglement entropy}
\begin{figure}[hb]
  \centering
  \includegraphics[width=.925\columnwidth,clip]{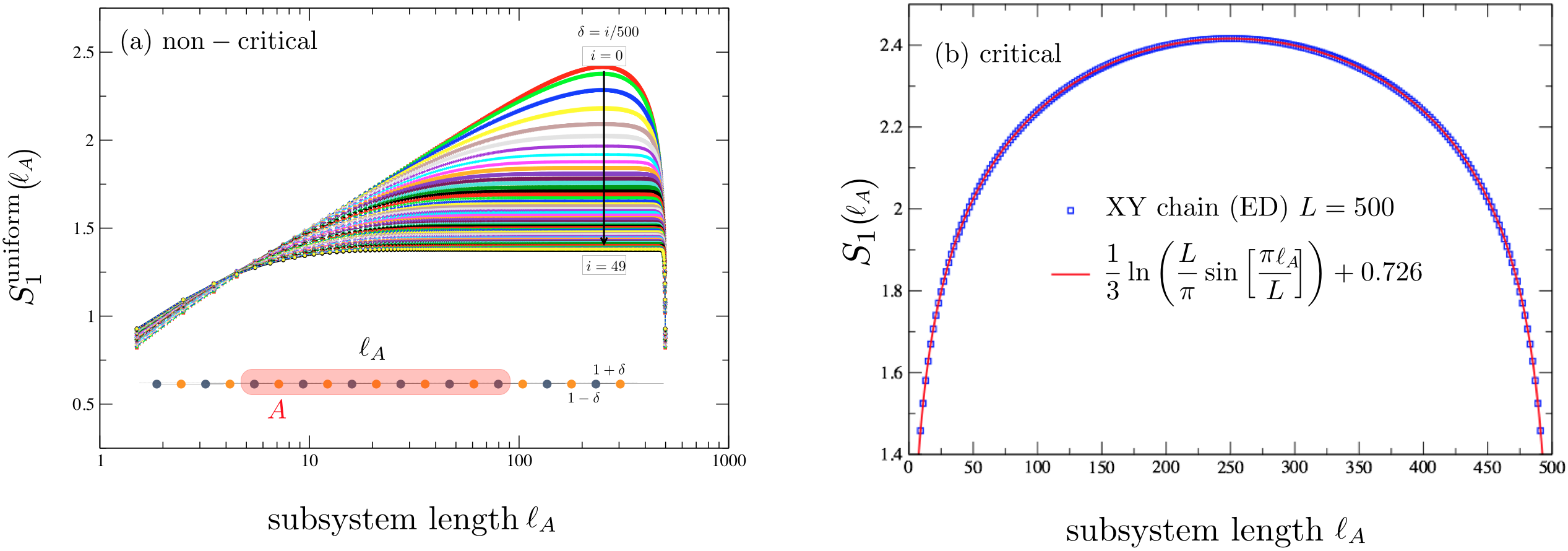}    
  \caption{Left (a): Uniform part of the entanglement entropy $S_1^{\rm uniform}(\ell_A)$ of a non-critical dimerized spin-$\frac{1}{2}$ XY chain [Eq.~\eqref{eq:HXY}] plotted against the subsystem length $\ell_A$ for various dimerization strengths $\delta=i/500$, $i=0,\cdots, 49$. Exact diagonalization results for periodic chains of $L=500$ sites. Right (b): Critical (gapless) spin-$\frac{1}{2}$ XY chain ED results compared to the log scaling from conformal field theory Eq.~\eqref{eq:S1d} with the exact additive constant calculated by Jin and Korepin~\cite{jin_quantum_2004}.}
  \label{fig:xydim}
\end{figure}

For one dimensional quantum systems, one has to make a distinction between critical and non-critical ground-states. Indeed, strictly speaking the area law Eq.~\eqref{eq:AL} for $d=1$ yields a constant entropy, independent of the subsystem size.
As first observed by Calabrese and Cardy~\cite{calabrese_entanglement_2004}, and then proved by Hastings~\cite{hastings_area_2007} this always occurs for non-critical ground-states having a finite correlation length. This is exemplified in Fig.~\ref{fig:xydim} (a) where we show exact diagonalization results obtained for the dimerized quantum spin-$1/2$ chain model 
\be
{\cal{H}}_{{\rm{1dxy}}}(\delta)=\sum_{i=1}^{L} \left[1+\delta(-1)^i\right]\left(S_i^x S_{i+1}^x + S_i^y S_{i+1}^y\right).
\label{eq:HXY}
\ee
This system (studied by several authors, see for instance Ref.~\cite{vidal_entanglement_2003}), equivalent to a free-fermion problem after a Jordan-Wigner transformation~\cite{jordan_uber_1928}, has a gap above its short-range correlated dimerized ground-state for any $\delta\neq 0$~\cite{black_critical_1981}, and is critical with power-law decaying correlations in the absence of modulation $\delta=0$.
The uniform part of the von-Neumann ($q=1$) entanglement entropy is plotted against the subsystem length $\ell_A$ for increasing dimerization strengths in Fig.~\ref{fig:xydim} (a) where we clearly see that $S_1$ gets saturated to a constant $\sim \ln \xi$ (where $\xi$ is the finite correlation length), as expected for $d=1$ with short-range correlations~\cite{hastings_area_2007,brandao_area_2013}. In such a case, corrections beyond the area law are universal~\cite{cardy_form_2007,calabrese_corrections_2010}, which has been also checked numerically~\cite{levi_universal_2013}.

On the other hand, critical chains described by conformal field theory (CFT)~\cite{holzhey_geometric_1994,calabrese_entanglement_2004,calabrese_entanglement_2009-1} display a logarithmic violation of the strict area law, as R\'enyi entropies grow with the subsystem length $\ell_A$ following the universal form
\be
S_q(A)=\frac{c}{6}\left(1+\frac{1}{q}\right)\ln\left(\frac{L}{\pi}\sin\left[\frac{\pi\ell_A}{L}\right]\right)+s_q+\cdots\label{eq:S1d}\ee
for periodic chains of length $L$, where $c$ is the central charge of the CFT. The constant term $s_q$ is non-universal but can be evaluated exactly in some cases~\cite{jin_quantum_2004}, see Fig.~\ref{fig:xydim} (b). The ellipses in Eq.~\eqref{eq:S1d} represent subleading corrections to scaling, vanishing as power-laws of the subsystem size involving the Luttinger liquid exponent~\cite{calabrese_parity_2010,cardy_unusual_2010,calabrese_universal_2010,xavier_renyi_2011,xavier_finite-size_2012}. Open boundary conditions, discussed in more details below (see Section~\ref{sec:OBC}), also leads to finite size corrections to the above logarithmic scaling~\cite{laflorencie_boundary_2006,fagotti_universal_2011}. Interestingly, power-law decaying corrections can be used to estimate the Luttinger exponent~\cite{dalmonte_estimating_2011,dalmonte_critical_2012}. 
For exactly solvable one-dimensional models with multicritical conformal points described by unitary minimal models and ${\mathbb{Z}}_n$ para-fermions, see also~\cite{de_luca_approaching_2013}. Finally, note that finite temperature corrections can also be computed using CFT~\cite{korepin_universality_2004,calabrese_entanglement_2004,cardy_universal_2014}.
\subsubsection{Beyond bipartite entropy}
\begin{figure}[h!]
  \centering
  \includegraphics[width=.35\columnwidth,clip]{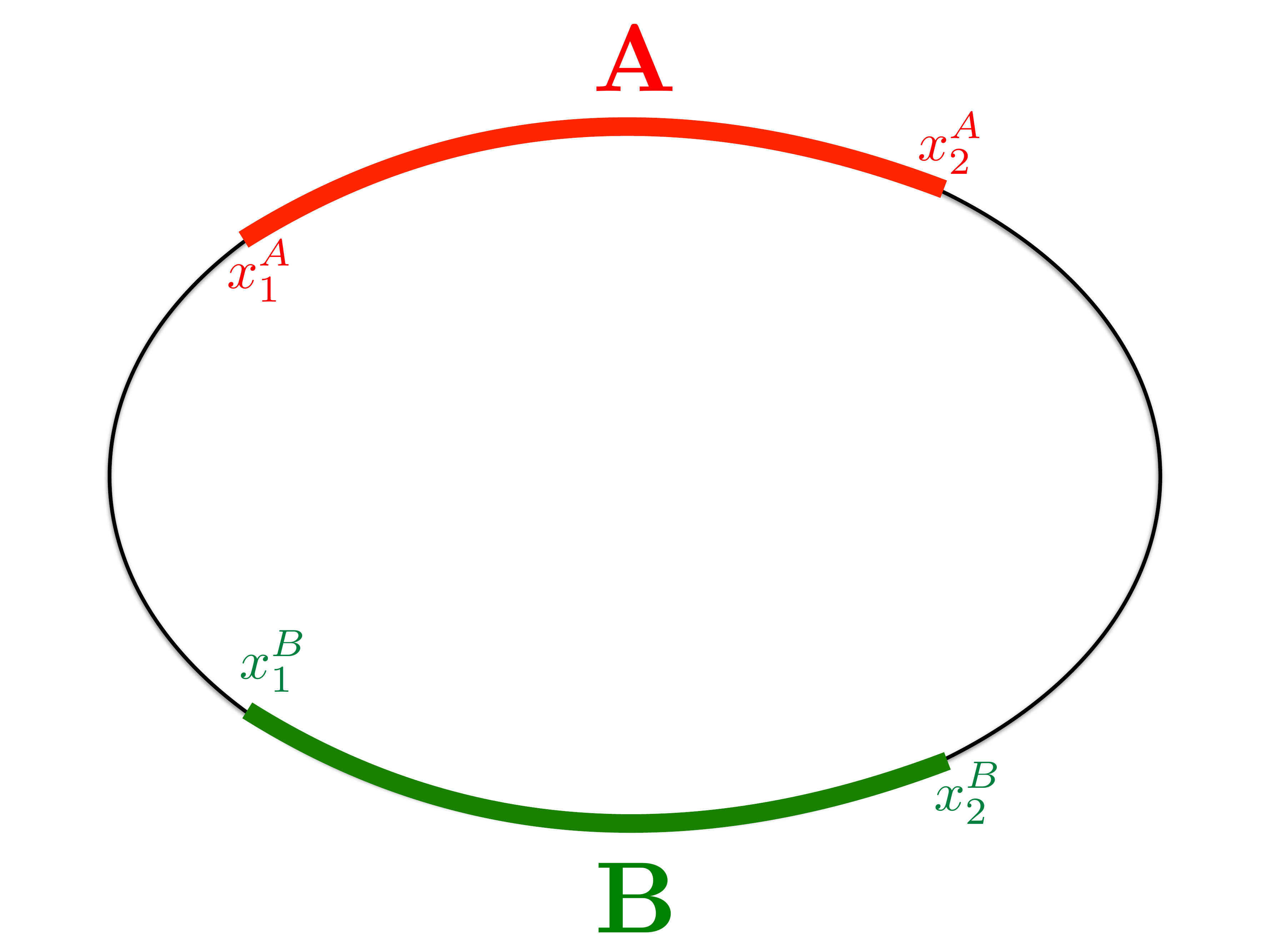}    
  \caption{Schematic picture of a one-dimensional system with A and B disjoint intervals.}
  \label{fig:disjoint}
\end{figure}

Despite the huge interest and the very large amount of works focused on bipartite entanglement entropies in one dimension, let us also mention a few other many-body measures of entanglement.
\paragraph{Entanglement entropy of disjoint intervals---}
Several authors have considered how to estimate the amount of entanglement between disjoint intervals using CFT~\cite{caraglio_entanglement_2008,furukawa_mutual_2009,calabrese_entanglement_2009,alba_entanglement_2010,fagotti_entanglement_2010,calabrese_entanglement_2011,alba_entanglement_2011,rajabpour_entanglement_2012,coser_renyi_2014}. In the case of two intervals $A\cup B = [x_{1A},x_{2A}]
\cup [x_{1B},x_{2B}]$,  as depicted in Fig.~\ref{fig:disjoint}, CFT results lead to 
\be
{\rm{Tr}}\rho_{A\cup B}^q\propto\left(\frac{|x_{1}^{A}-x_{1}^{B}||x_{2}^{A}-x_{2}^{B}|}{|x_{1}^{A}-x_{2}^{A}| |x_{1}^{B}-x_{2}^{B}| |x_{1}^{A}-x_{2}^{B}| |x_{1}^{B}-x_{2}^{A}|}\right)^{\frac{c}{6}\left(q-\frac{1}{q}\right)}{\cal{F}}_q\left(\frac{|x_{1}^{A}-x_{2}^{A}| |x_{1}^{B}-x_{2}^{B}|}{|x_{1}^{A}-x_{1}^{B}||x_{2}^{A}-x_{2}^{B}|}\right),
\ee
where ${\cal{F}}_q$ is a universal function which depends on the compactification radius (related to the Luttinger parameter) of the underlying CFT. Some results have also been obtained in higher dimension~\cite{cardy_results_2013} and for random spin chains~\cite{ruggiero_entanglement_2016}

\paragraph{Entanglement negativity---} 
Another interesting quantity is the entanglement negativity~\cite{zyczkowski_volume_1998,vidal_computable_2002}, defined for any (possibly mixed) state $\rho$ by ${\cal{N}}(\rho)=\frac{||\rho^{T_A}||_1-1}{2}$, where $\rho^{T_A}$ is the partial transpose of $\rho$ with respect to subsystem $A$, and $||\cdot||_1$ is the trace norm. The logarithmic negativity~\cite{plenio_logarithmic_2005} 
$E_{\cal{N}}(\rho)=\log_2||\rho^{T_A}||_1$,
is used more often as a measure of entanglement for thermal states, as well as for disjoint intervals.

CFT approaches for the negativity have been successfully used for ground-states~\cite{calabrese_entanglement_2012-1,calabrese_entanglement_2013}, at finite temperature~\cite{calabrese_finite_2015}, or out-of-equilibrium~\cite{hoogeveen_entanglement_2015}. Numerically, 
Monte Carlo schemes have also been proposed to estimate $E_{\cal{N}}$~\cite{alba_entanglement_2013,chung_entanglement_2014}. Note moreover that the negativity has been proven useful to detect topological order ~\cite{castelnovo_negativity_2013,lee_entanglement_2013} (see also below Sections~\ref{sec:toee} and \ref{sec:esto}), as well as in the context of the
Kondo effect~\cite{bayat_negativity_2010,alkurtass_entanglement_2016} (see also below Section~\ref{sec:kondo}).

\subsection{The area law in higher dimension}
In this section, we simply give a short survey of this topic, which has been recently reviewed in a thorough way by Eisert, Cramer, and Plenio
\cite{eisert_colloquium:_2010}.

\subsubsection{$d>1$ fermions}
Conventional metals, described as free fermions or Fermi liquids are known to weakly violate the area law by a multiplicative logarithmic correction~\cite{wolf_violation_2006,gioev_entanglement_2006,li_scaling_2006,barthel_entanglement_2006,farkas_von_2007} if there is a well-defined Fermi surface, while for non-critical fermions the entropy obeys a strict area law. This logarithmic enhancement can be related to the topology of the Fermi surface~\cite{gioev_entanglement_2006,helling_special_2011,swingle_entanglement_2010,leschke_scaling_2014} through
\be
S_q(A)=\frac{q+1}{24q}\iint {\rm{d}}{\cal A}_x{\rm{d}}{\cal A}_k |n_x\cdot n_k|\left(\frac{\ell_A}{2\pi}\right)^{d-1}\times\ln \ell_A + o(\ell_A^{d-1}\ln\ell_A),
\ee
where $n_x$ and $n_k$ are units normal to the real space boundary ${\cal A}_x$ and the Fermi surface ${\cal A}_k$.
One can also understand such a weak violation of the area law using a low-energy description of the Fermi surface like an ensemble of one dimensional gapless modes, as proposed by Swingle~\cite{swingle_entanglement_2010,swingle_conformal_2012}. A multidimensional bosonization approach gives essentially similar results for Fermi liquids~\cite{ding_entanglement_2012}.

Another very interesting result for free fermions came with the seminal works by Klich and co-workers on quantum noise~\cite{klich_measuring_2006,klich_quantum_2009}. They found a direct link between entanglement entropies and the fluctuations of a globally conserved quantity within a subsystem, for instance the particle number for Fermi gases or the subsystem magnetization for quantum magnets. This has motivated a large number of subsequent studies~\cite{hsu_quantum_2009,song_general_2010,song_entanglement_2011,song_bipartite_2012,calabrese_exact_2012,rachel_detecting_2012,susstrunk_free_2012,swingle_renyi_2012,vicari_entanglement_2012,eisler_universality_2013,klich_note_2014,petrescu_fluctuations_2014}. Remarkably, using the full counting statistics~\cite{blanter_shot_2000} there is an exact expression for all R\'enyi entropies~\cite{song_bipartite_2012}. This is further discussed below in Section~\ref{sec:BF}.

Particle number fluctuations and entanglement entropies have also been explored in the context of ultracold Fermi gases~\cite{calabrese_entanglement_2011-1,vicari_entanglement_2012,calabrese_exact_2012,calabrese_entanglement_2012}. Note also that in a similar context,  connections to random matrix theory have been recently discussed 
\cite{eisler_full_2013,eisler_universality_2013,marino_phase_2014,calabrese_random_2015} to obtain analytic expressions in various regimes of  a one dimensional gas confined by a harmonic trap.

On the numerical side, recent progresses have also been made using quantum Monte Carlo (QMC) techniques for accessing entanglement properties of interacting fermions~\cite{zhang_entanglement_2011,mcminis_renyi_2013,grover_entanglement_2013-1,assaad_entanglement_2014,broecker_renyi_2014,wang_renyi_2014,assaad_stable_2015,drut_hybrid_2015,drut_entanglement_2016}. In particular, emergent fermions (spinons) with a Fermi surface have been detected through the logarithmic enhancement of the area law in a critical quantum spin liquid state using variational Monte Carlo~\cite{zhang_entanglement_2011}.
Nevertheless, a relatively unexplored field concerns non-Fermi liquids for which Swingle and Senthil argued for a similar logarithmically enhanced area law~\cite{swingle_universal_2013}. Using a variational Monte Carlo approach the composite fermion wave function~\cite{halperin_theory_1993} for half-filled Landau level $\nu=1/2$, a non-Fermi liquid state, has been shown to indeed exhibit a logarithmically enhanced area law, but with a  prefactor twice larger~\cite{shao_entanglement_2015}.

Another interesting example is the Bose metal state, the so-called exciton Bose liquid~\cite{paramekanti_ring_2002} which displays a "Bose surface" with gapless excitations, for which a logarithmic enhancement of the area law was found~\cite{lai_violation_2013,lai_probing_2016}.

Let us finally mention that free fermions in a weak random potential, but still in a metallic (diffusive) regime, loose logarithmic enhancement and obey a strict area law for the mean entropy~\cite{pastur_area_2014,potter_boundary-law_2014,pouranvari_entanglement_2015}. This can be understood from the smearing of the sharp Fermi surface by disorder, despite a finite density of gapless states~\cite{potter_boundary-law_2014}.

\subsubsection{Non-interacting bosons}
We now turn to the case of non-interacting bosonic models. In contrast with the free fermions case, free bosons with non-relativistic quadratic dispersion display Bose-Einstein condensation. In this context, the area law term is suppressed, and the dominant scaling of the R\'enyi entropy is logarithmic with the number of particles~\cite{ding_entanglement_2009,alba_entanglement_2013}. This situation is somehow unphysical, and can be understood as a mean-field limit for which we do not expect a well-defined "area". Below we will see in Sec.~\ref{sec:cs} that any interaction will restore the area law term, while additional logarithmic corrections associated to Nambu-Goldstone modes are expected.

For a relativistic free bosonic field theory, which is equivalent to a system of coupled harmonic oscillators, R\'enyi entropies scale with the area in dimension $d>1$~\cite{bombelli_quantum_1986, srednicki_entropy_1993,callan_geometric_1994,casini_entanglement_2005,casini_analytic_2008,casini_entanglement_2010,cardy_logarithmic_2013}. This was clearly confirmed numerically for critical and non-critical regimes where, contrary to free fermions, a strict area law is observed in both cases~\cite{plenio_entropy_2005,barthel_entanglement_2006}. Numerical evaluation for lattice quadratic Hamiltonians is made possible using Wick's theorem, as shown by Chung and Peschel~\cite{chung_density-matrix_2000,peschel_2012}. 
\subsection{Corrections to the area law for various states of matter}
\subsubsection{State of the art}
\begin{figure}[hb]
    \centering
    \includegraphics[width=8cm,clip]{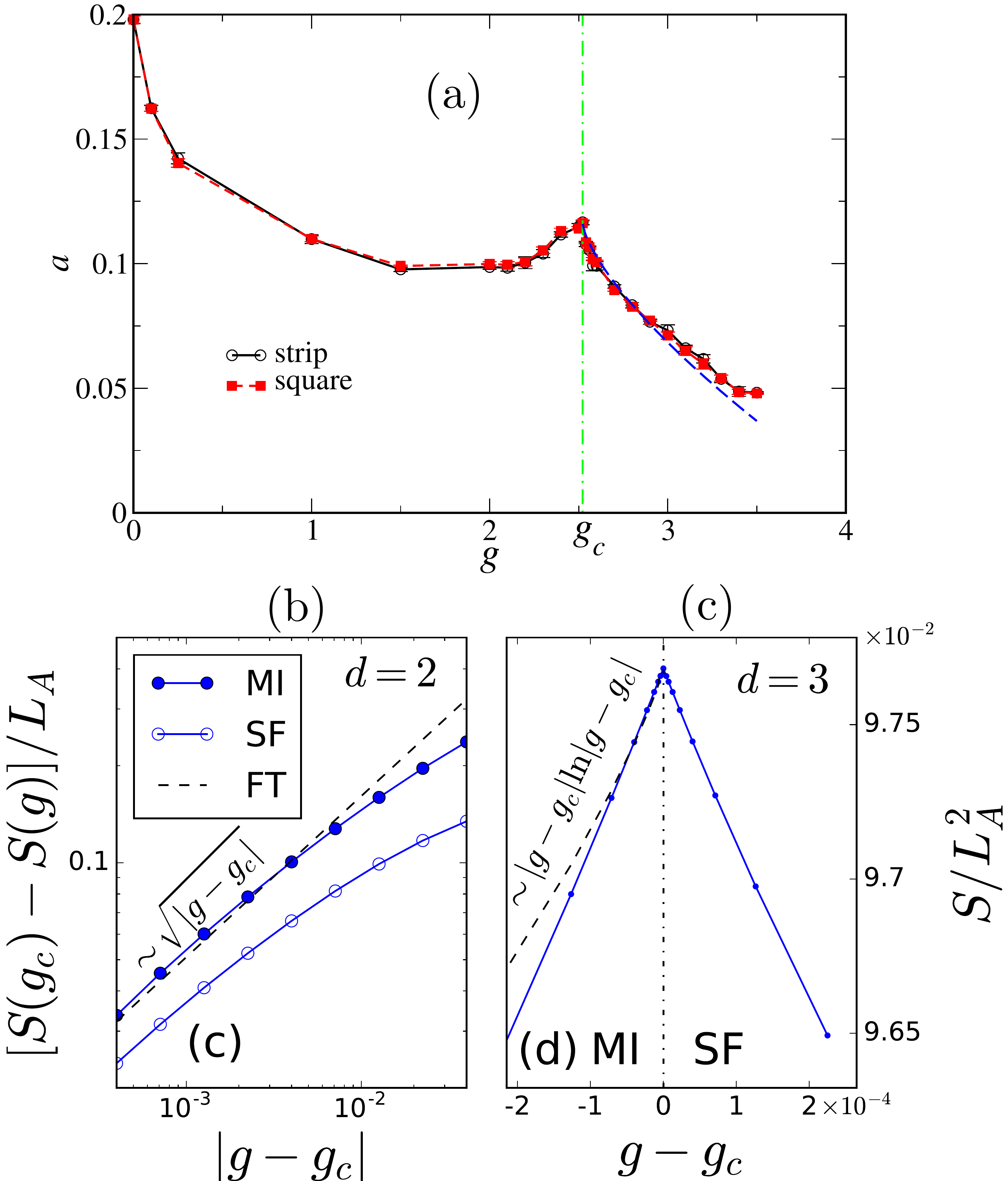}
    \caption{Cusp singularity for the area law prefactor across a quantum phase transition. (a) QMC results for a $s=1/2$ Heisenberg bilayer, from~\cite{helmes_entanglement_2014}. (b) Schwinger-boson results for the superfluid - insulator transition in $d=2$ and (c) $d=3$, from \cite{frerot_entanglement_2016}. }
    \label{fig:area}
\end{figure}

The area law term discussed above does not a priori carry information regarding a given  phase of matter. Nevertheless, it has been shown to display critical behavior when a quantum phase transition is crossed, developing a local extremum at a critical point $g=g_c$~\cite{metlitski_entanglement_2009,casini_renormalization_2012,kallin_entanglement_2013,helmes_entanglement_2014,frerot_entanglement_2016}, with the area-law prefactor $a_q(g_c)-a_q(g)\propto |g-g_c|^{\nu(d-1)}$, where $\nu$ is the correlation length exponent associated to the critical point.  Fig.~\ref{fig:area} shows such cusp singularities for O(3)~\cite{helmes_entanglement_2014} and O(2)~\cite{frerot_entanglement_2016} quantum critical points.
This singular behavior is natural if one interprets the entropy as a surface or boundary free energy~\cite{fradkin_entanglement_2006}.

Going beyond the leading area law term, universal signatures are expected for various quantum phases of matter. The first discussed and celebrated example is for (1+1) CFT which describes critical interacting $d=1$ quantum chains~\cite{vidal_entanglement_2003,calabrese_entanglement_2004,korepin_universality_2004}, with the logarithmic growth Eq.~\eqref{eq:S1d} involving the central charge $c$ of the underlying CFT. In higher dimension, several authors have studied  the problem which turns out to be much richer~\cite{fradkin_entanglement_2006,hsu_universal_2009,metlitski_entanglement_2009,stephan_shannon_2009,casini_entanglement_2009,tagliacozzo_simulation_2009,stephan_phase_2011,metlitski_entanglement_2011,ju_entanglement_2012,stephan_entanglement_2013,inglis_entanglement_2013,
kallin_entanglement_2013,kallin_corner_2014,helmes_entanglement_2014,stoudenmire_corner_2014,chen_scaling_2015,akers_entanglement_2016}.

In Table~\ref{tab:sota}, we give a simplified overview of different quantum states of matter with their associated entanglement entropy scalings, and physical examples of realizations.
Generally speaking, ordered states can be classified in three main families. For discrete symmetry breaking, a trivial additive constant depending only on the degeneracy of the ground-state is expected $\ln({\rm{deg}})$~\cite{stephan_shannon_2009}. In the case of continuous symmetry breaking, {\it{e.g.}} U(1) for superfluids or SU(2) for Heisenberg antiferromagnets, additional logarithmic corrections due to Nambu-Goldstone modes are present~\cite{song_entanglement_2011-1,metlitski_entanglement_2011}, as we discuss in detail below in Section~\ref{sec:cs}. For topological ordered phases, there is a negative constant~\cite{kitaev_topological_2006,levin_detecting_2006}, as will be discussed in Section~\ref{sec:to}. At quantum critical points, for instance O($N$), an additive constant appears~\cite{metlitski_entanglement_2009}. Note also the numerical study on  cubic subsystems for the $d=3$ quantum Ising model~\cite{devakul_entanglement_2014}.

\begin{table}[h]
\begin{tabular}{l|l|l}
Physical state&Entropy&Example\\
\hline
Gapped (brok. disc. sym.)&$aL^{d-1}+\ln({\rm{deg}})$&Gapped XXZ~\cite{stephan_shannon_2009}\\
$d=1$ CFT&$\frac{c}{3}\ln L$&$s=\frac{1}{2}$ Heisenberg chain~\cite{calabrese_entanglement_2004}\\
$d\ge 2$ QCP& $aL^{d-1}+\gamma_{\rm QCP}$&Wilson-Fisher O($N$)~\cite{metlitski_entanglement_2009}\\
Ordered (brok. cont. sym.)&$aL^{d-1}+\frac{n_{\rm G}}{2}\ln L$&Superfluid, N\'eel~order\cite{metlitski_entanglement_2011}\\
Topological order&$aL^{d-1}-\gamma_{\rm top}$&${\mathbb{Z}}_2$ spin liquid~\cite{furukawa_topological_2007}\\
\hline
\hline
\end{tabular}
\caption{\label{tab:sota} Entanglement entropy scaling for various examples of states of matter, either disordered, ordered, or critical, with smooth boundaries (no corners).}
\end{table}

\subsubsection{Corner contributions in dimension $d=2$}
\label{sec:corners}
We first focus on the issue of subsystems having non-smooth boundaries with the particular example of sharp corners in two dimensions
which are responsible for additive logarithmic corrections to the area law, as studied in several works~\cite{fradkin_entanglement_2006,casini_universal_2007,
tagliacozzo_simulation_2009,swingle_mutual_2010,
kallin_entanglement_2013,kallin_corner_2014,
helmes_entanglement_2014,stoudenmire_corner_2014,bueno_universality_2015,laflorencie_spin-wave_2015,bueno_corner_2015,bueno_universal_2015,sahoo_unusual_2016,bueno_bounds_2016,faulkner_shape_2016}. The area law is indeed corrected as follows
\be
\Delta S_{q}^{c}=\left(\sum_{c}l_q^c\right)\ln L,
\label{eq:corner}
\ee
where the sum is taken over sharp corner angles. With smooth boundaries, such a logarithmic divergence occurs only for odd dimensions~\cite{ryu_holographic_2006,nishioka_holographic_2009}.
For even values of $d$, sharp angles lead to logarithmic corrections which are expected to be universal for
all systems with the same type of symmetry breaking/phase transition. In $d=2$ it has been studied in several situations, such as CFT, or Lifshitz with $z\neq 1$, where the prefactor was shown to be directly proportional to the central charge~\cite{fradkin_entanglement_2006}. For free scalar field theory, Casini and Huerta have provided an analytical solution~\cite{casini_universal_2007} for integer $q\ge 2$ while it involves a tricky numerical solution of a set of
non-linear differential equations, valid for $\varphi_c\in [0,\pi]$
($l_q(\varphi)=l_q(2\pi-\varphi)$). Below in Fig.~\ref{fig:corner_sw} we show numerical results obtained in Ref.~\cite{laflorencie_spin-wave_2015} for non-interacting relativistic bosons on a square lattice, reproducing the free scalar field results of Casini and Huerta. 

In order to extract the corner corrections, we study square subsystems (having 4 corners, each with $\varphi=\pi/2$) of perimeter $2L$ (embedded in a $L\times L$ torus) at which one subtracts the entropies from corner-free strips $\ell\times L$ having the same perimeter so that the area law contribution cancels, as well as well as other potential corrections ({\it{e.g.}} Goldstone modes for continuous symmetry breakings, see also below in Section~\ref{sec:cs}).

\begin{figure}
    \centering
    \includegraphics[width=10cm,clip]{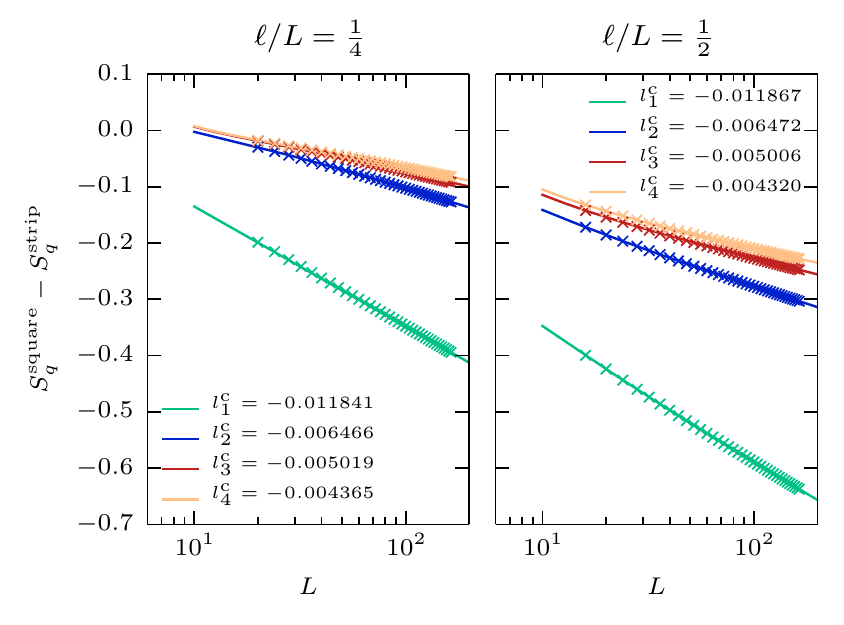}
    \caption{Difference of entanglement entropies for $d=2$ relativistic spin-waves (free bosons) between square and strip geometries, yielding the logarithmic corrections Eq.~\eqref{eq:diff}. Fits (full lines) to the form $8l_q(\pi/2)\ln(L)+b_q + c_q/L +
d_q /L^2$ are shown by full lines. Exact diagonalization results (symbols) are displayed for two different aspect ratios of the strips and various R\'enyi parameters $q=1, 2, 3, 4$. Reprinted from \cite{laflorencie_spin-wave_2015}.}
    \label{fig:corner_sw}
\end{figure}

Working with spin-wave (SW) corrections of an SU(2) model~\cite{laflorencie_spin-wave_2015}, there are two Nambu-Goldstone modes from which we 
expect the leading term of this difference to be given by 
\be
S^{\rm square}_q-S^{\rm strip}_q=8l_q({\pi}/{2})\ln L +\cdots\label{eq:diff}\ee

\noindent Numerical diagonalization results of the non-interacting SWs Hamiltonian~\cite{laflorencie_spin-wave_2015} are plotted in Fig.~\ref{fig:corner_sw} where we clearly see
that the above difference Eq.~\eqref{eq:diff} is clearly dominated by a
logarithmic scaling which allows us to extract $l_q(\pi/2)$. Small variations of the
results for different aspects ratios of the strips (see left and right
panels of Fig.~\ref{fig:corner_sw}) can be used as a measure of the error due to finite
size effects and fitting procedure. 
Our results, in perfect agreement with those of Casini and Huerta~\cite{casini_universal_2007}, are displayed in
Table~\ref{tab:corner} together with other estimates for interacting field theories obtained from numerical simulations using series expansion~\cite{singh_thermodynamic_2012}, numerical linked cluster expansion~\cite{kallin_entanglement_2013,kallin_corner_2014,stoudenmire_corner_2014} or QMC~\cite{helmes_correlations_2015}. Note that extracting such small log corrections is very challenging for interacting fixed points and series or numerical linked cluster expansion turns out to be more controlled than QMC for this task.

\begin{figure}
    \centering
    \includegraphics[width=.55\columnwidth,clip]{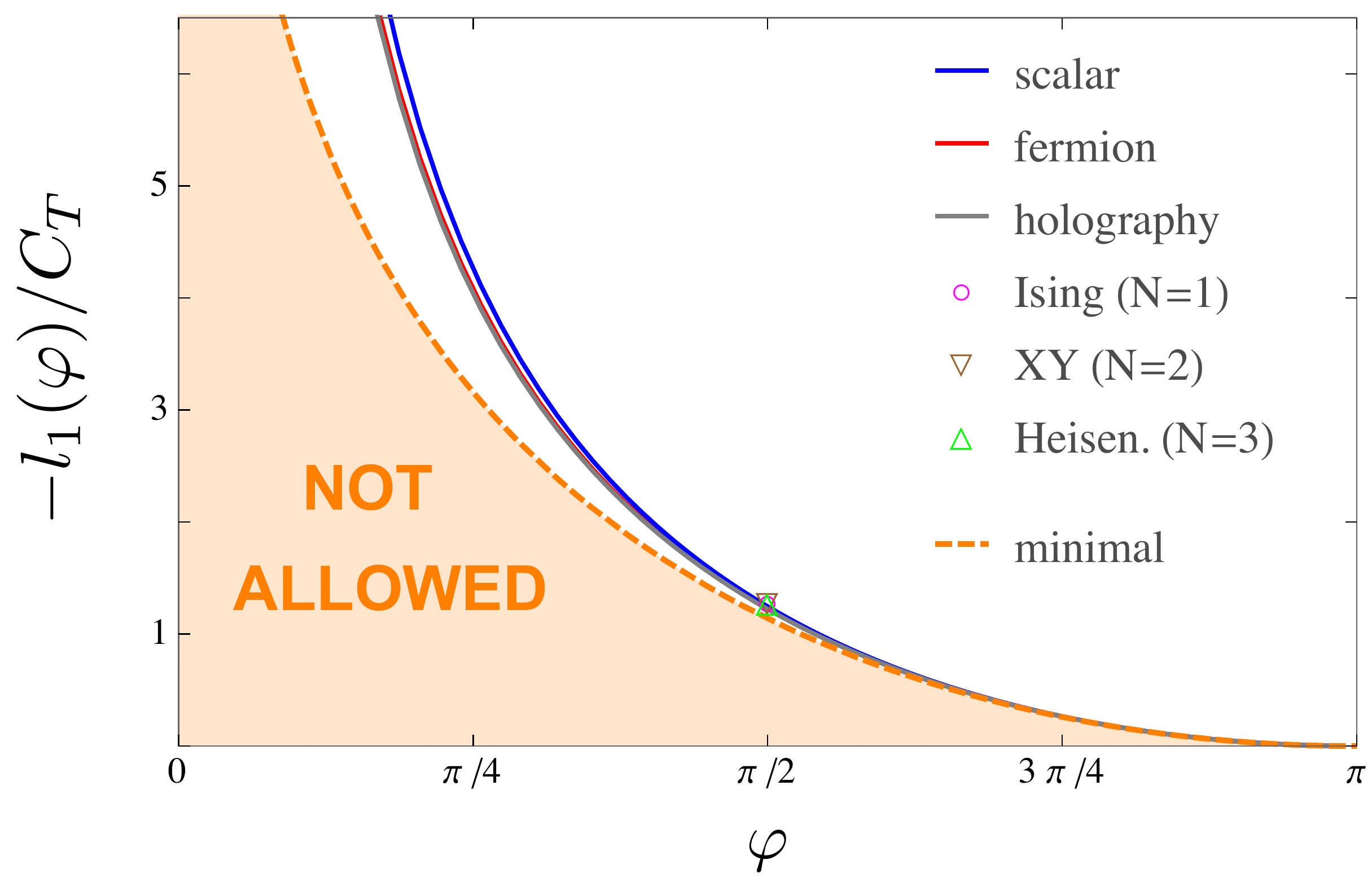}
    \caption{Universal ratio $l_1(\varphi)/C_T$ for various $2+1$ CFT: holography (gray), free Dirac fermion (red), scalar field (blue). Numerical results for Wilson-Fisher O($N$) fixed points: Ising~\cite{kallin_entanglement_2013}, XY~\cite{stoudenmire_corner_2014}, and Heisenberg~\cite{kallin_corner_2014} are shown at $\varphi=\pi/2$. The minimal function (orange) computed by Bueno and Witczak-Krempa is a lower bound. Reprinted from~\cite{bueno_bounds_2016}.}
    \label{fig:bueno}
\end{figure}

The fact that universality emerges in the corner logarithmic terms is remarkable. It was then proposed that the prefactor of the logarithmic correction is an effective measure of the degrees of freedom of the underlying CFT~\cite{casini_universal_2007,nishioka_holographic_2009,kallin_corner_2014}. Using CFT and holographic calculations, Bueno and co-workers have suggested that the ratio $l_1(\varphi)/C_T$ is universal for a broad range of CFT, meaning almost the same for all CFTs, including Wilson-Fisher fixed points of the O($N$) model~\cite{bueno_universality_2015,bueno_corner_2015,bueno_universal_2015} (see Fig.~\ref{fig:bueno}), where $C_T$ is the central charge of the stress tensor correlator. In the smooth limit $\varphi\to \pi$, it was conjectured~\cite{bueno_universality_2015,bueno_corner_2015} and then proved~\cite{faulkner_shape_2016} that for all CFTs $l_1(\varphi\to \pi)=(\pi-\varphi)^2 C_T (\pi^2/24)$.
Beyond the smooth limit, $l_1(\varphi)/C_T$ slightly deviates for different theories (see Fig.~\ref{fig:bueno}). A lower bound for general $l_q(\varphi)$ was also derived recently by Bueno and Witczak-Krempa~\cite{bueno_bounds_2016}, and is shown in Fig.~\ref{fig:bueno}.

Estimates for this ratio are shown  in Table~\ref{tab:corner} for $q=1$ and $\varphi=\pi/2$, using the central charge from conformal bootstrap~\cite{kos_bootstrapping_2014}, which all satisfy the lower bound of  $1.14$~\cite{bueno_bounds_2016}. 
\begin{table}
\begin{center}
\begin{tabular}{r|l|l}
&$-l_1({\pi}/{2})$&$-{l_1({\pi}/{2})}/{C_T}$\\
\hline
O(1)&$0.013(1)$~\cite{kallin_entanglement_2013,stoudenmire_corner_2014} &$1.4(1)$\\
O(2)&$0.024(1)$~\cite{stoudenmire_corner_2014}&$1.34(6)$\\
O(3)&$0.036(2)$~\cite{kallin_corner_2014} &$1.34(8)$\\
Free scalar&$0.0118(1)$~\cite{casini_universal_2007,laflorencie_spin-wave_2015}&$1.24(1)$\\
\hline
\end{tabular}
\caption{\label{tab:corner} Prefactor $l_1(\pi/2)$ and ratio ratio $l_1/C_T$~\cite{bueno_universality_2015,bueno_corner_2015} of the corner logarithmic correction in Eq.~\eqref{eq:corner}. Numerical results for free scalar and O(N) Wilson-Fisher fixed points. }
\end{center}
\end{table}
%
Beyond the von Neumann index at $q>1$, $C_T$ should be replaced by $h_q/(q-1)$~\cite{bueno_universal_2015}, with $h_q$ the scaling dimension of the twist operator.\footnote{This also appears for 1+1 CFT as the prefactor of the logarithmic scaling in Eq.~\eqref{eq:S1d}.}

\begin{figure}[h]
    \centering
    \includegraphics[width=.95\columnwidth,clip]{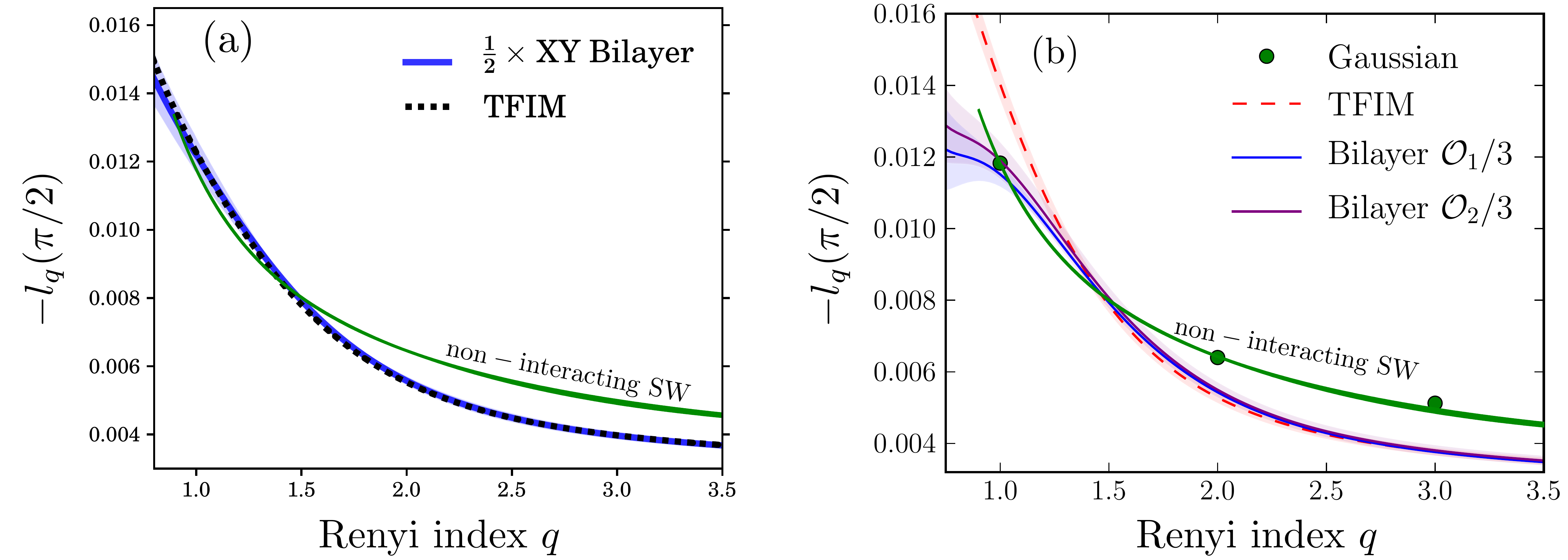}
    \caption{Corner coefficient of the logarithmic correction in Eq.~\eqref{eq:corner} for a $\pi/2$ angle, plotted as a function of the R\'enyi index $q$ for various critical theories in $2+1$ dimensions. (a) Estimates for an O(2) critical point obtained with numerical linked cluster expansion calculations on an XY bilayer are twice the coefficient of the O(1) Ising fixed point. Reprinted from~\cite{stoudenmire_corner_2014}. (b) The Heisenberg O(3) case studied with numerical linked cluster expansion with two types of cluster expansions (${\cal{O}}_1$ and ${\cal{O}}_2$), is seemingly thrice the O(1) result, at least for $q\ge 1.3$. Reprinted form~\cite{kallin_corner_2014}. For both cases, the free scalar result from non-interacting SW~\cite{laflorencie_spin-wave_2015} is shown (green line) for comparison. Note the difference in the estimates for Ising between the two panels, as commented in Ref.~\cite{stoudenmire_corner_2014} where larger clusters were used (left panel).}
    \label{fig:corner_lq}
\end{figure}

Finally, we can also investigate the R\'enyi index dependence for
non-integer values of $q$. In Fig.~\ref{fig:corner_lq} we report numerical linked expansions results 
for (a) O(2)~\cite{stoudenmire_corner_2014} and (b) O(3)~\cite{kallin_corner_2014} together with different estimates for the Ising fixed point, as well as our estimate for free scalar~\cite{laflorencie_spin-wave_2015}. The non-trivial $q$-dependences are qualitatively similar, and the ratio between O(2) and O(1) is well compatible with $2$, as well as for O(3) with a factor 3.

\subsection{Continuous symmetry breaking}
\label{sec:cs}
For condensed matter systems presenting a spontaneous breaking of continuous symmetry at zero temperature in the thermodynamic limit, such as Bose-condensed superfluids or ordered antiferromagnets, additive logarithmic corrections have been originally observed using modified SW theory~\cite{song_entanglement_2011-1} and numerical simulations~\cite{kallin_anomalies_2011} on $d=2$ quantum antiferromagnets. Shortly after, Metlitski and Grover~\cite{metlitski_entanglement_2011} proposed an analytical interpretation based on quantum rotor and non-linear sigma models where both SW excitations and the "tower of states" (TOS) due to the symmetry restoration in a finite volume~\cite{anderson_approximate_1952,lhuillier_frustrated_2005} are responsible for a logarithmic correction, proportional to the number of Nambu-Goldstone modes $n_G$ associated with the symmetry breaking, thus yielding
\be
S_q=a_qL^{d-1}+\frac{n_G}{2}\ln\left(\frac{\rho_{s}}{v} L^{d-1}\right) +\gamma_q^{\rm ord},
\label{eq:MG}
\ee
where $\rho_s$ is the stiffness, $v$ the SW velocity, and $\gamma_q^{\rm ord}$ a universal geometric constant (see below).
\subsubsection{Large $s$ approach for $d=2$}
\paragraph{Logarithmic corrections due to Nambu-Goldstone modes---}
As developped in Refs.~\cite{song_entanglement_2011-1,luitz_universal_2015,laflorencie_spin-wave_2015} for ordered phases where a continuous symmetry is broken, a SW (SW) treatment allows to capture subleading corrections in Eq.~\eqref{eq:MG}.
A canonical example is the so-called $J_1 - J_2$ spin-$s$ antiferromagnet~\cite{chandra_possible_1988}, governed on the square lattice by the Hamiltonian
\be
{\cal H}=J_1\sum_{\langle i j\rangle}{\vec{S}}_i\cdot {\vec{S}}_j + J_2\sum_{\langle\langle i j\rangle\rangle}{\vec{S}}_i\cdot {\vec{S}}_j
+h\sum_{i}(-1)^iS_i^z.
\label{eq:HJ1J2}
\ee
When the external staggered field $h=0$, this model exhibits N\'eel order with a spontaneous breaking of SU(2) symmetry in the thermodynamic limt at $T=0$ if $J_2<J_c(s)$~\cite{chandra_possible_1988}, with $J_c\to J_1/2$ for $s\gg 1$. Replacing spin-$s$ operators by Holstein-Primakoff bosonic deviations about the classically ordered moment allows to expand the above $J_1 - J_2$ model as ${\cal H}/s^2=E_{\rm cl.}+{\cal{H}}^{(2)}/s +{\cal{H}}^{(4)}/s^2 + \cdots$, where $E_{\rm cl.}$ is the classical energy of the ordered ground-state, ${\cal{H}}^{(2)}$ is a quadratic Hamiltonian in term of Holstein-Primakoff bosons, and ${\cal{H}}^{(4)}$ is quartic. Truncating the above expansion at the quadratic ($1/s$) level usually captures most of the low-energy physics of quantum antiferromagnets~\cite{manousakis_spin-textonehalf_1991}, and we further expect better accuracy for $s\gg 1$. Such a treatment leads to the quadratic bosonic model
\be
{\cal H}^{(2)}=\sum_{\bf k}{\rm{A}}_{\bf k}(b^{\dagger}_{\bf k}b^{\dagga}_{\bf k}+b^{\dagger}_{-\bf k}b^{\dagga}_{-\bf k})
+{\rm{B}}_{\bf k}(b^{\dagger}_{\bf k}b^{\dagger}_{-\bf k}+b^{\dagga}_{\bf k}b^{\dagga}_{-\bf k}),
\label{eq:Hk}
\ee
where ${\rm{A}}_{\bf k}=2s\left(J_2\cos k_x\cos k_y+J_1-J_2\right)+\frac{h}{2}$ and $
{\rm{B}}_{\bf k}=-sJ_1\left(\cos k_x + \cos k_y\right)$ that we solve
using a standard Bogoliubov transformation, yielding a quasi-particle dispersion $\Omega_{\bf k}=2\sqrt{{\rm{A}}_{\bf k}^2-{\rm{B}}_{\bf k}^2}$. In the vicinity of its two minima at ${\bf{k}}_0=(0,0)$ and $(\pi,\pi)$ the SW dispersion takes the relativistic form
\be
\Omega_{\bf{k}}\approx v\sqrt{|{\bf{k}}-{\bf{k}}_0|^2+m^2},
\ee
with a velocity $v=2\sqrt{2}s\sqrt{J_1(J_1-2J_2)}$ and a mass gap $m=\sqrt{\frac{h}{s(J_1-2J_2)}}$. Such a (small) gap plays a crucial role for finite lattice calculations. Indeed,
in order to correct the fact that spin rotational symmetry is broken on finite lattices within the SW framework, one can artificially proceed to a fictitious restoration by adjusting the staggered field $h$ in Eq.~\eqref{eq:HJ1J2} so that the (SW-corrected) sublattice magnetization vanishes, following Refs.~\cite{takahashi_modified_1989,hirsch_spin-wave_1989}. As discussed in Refs.~\cite{song_entanglement_2011-1,luitz_universal_2015,laflorencie_spin-wave_2015}, such a (size-dependent) artificial staggered field has to scale as $h^*\approx \frac{2J_1}{sL^4}$ for $L\times L$ lattices in the large $s$ limit. Interestingly, the induced gap has precisely the Anderson TOS scaling~\cite{anderson_approximate_1952,lhuillier_frustrated_2005}
\be
m^*(L)=\sqrt{\frac{2J_1}{s^2(J_1-2J_2)}}\frac{1}{L^2}.
\label{eq:mls}
\ee
\begin{figure}[t]
    \centering
    \includegraphics[width=.75\columnwidth]{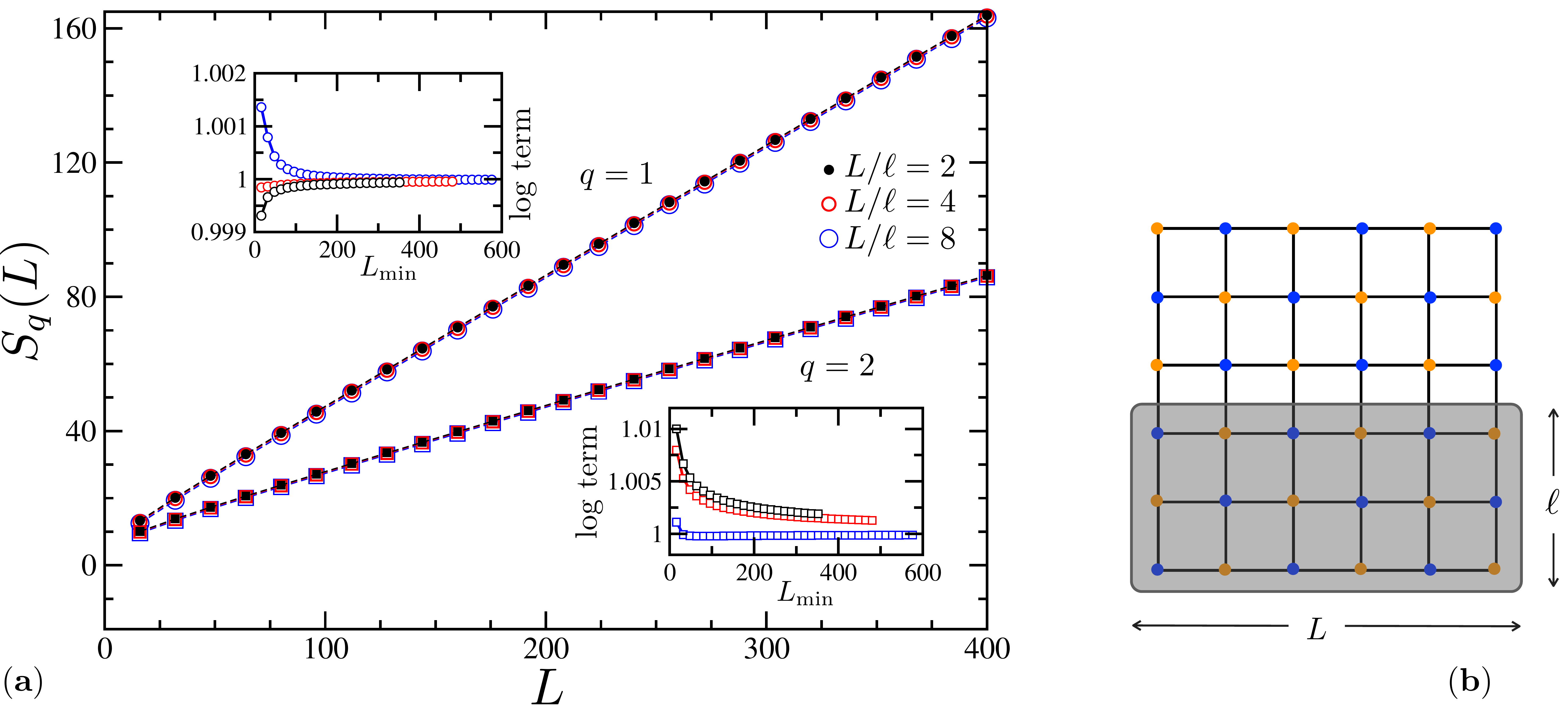}
    \caption{(a) SW results for the entanglement R\'enyi entropies ($q=1,2$) of the square lattice Heinseberg model Eq.~\eqref{eq:HJ1J2} with $s=100$ and $J_2=0$ for strip subsystems (panel b) with different aspect ratios
        $\ell/L$, plotted against $L$. The insets of (a) show fit results for the prefactor of the log correction as a function of the minimal system size $L_\text{min}$ included in the fit. The log prefactor unambiguously converges to 1, independently of $q$ and of the aspect ratio of the subsystem. Data taken from Ref.~\cite{laflorencie_spin-wave_2015}}
    \label{fig:S100}
\end{figure}

\begin{figure}[t]
    \centering
    \includegraphics[width=.55\columnwidth]{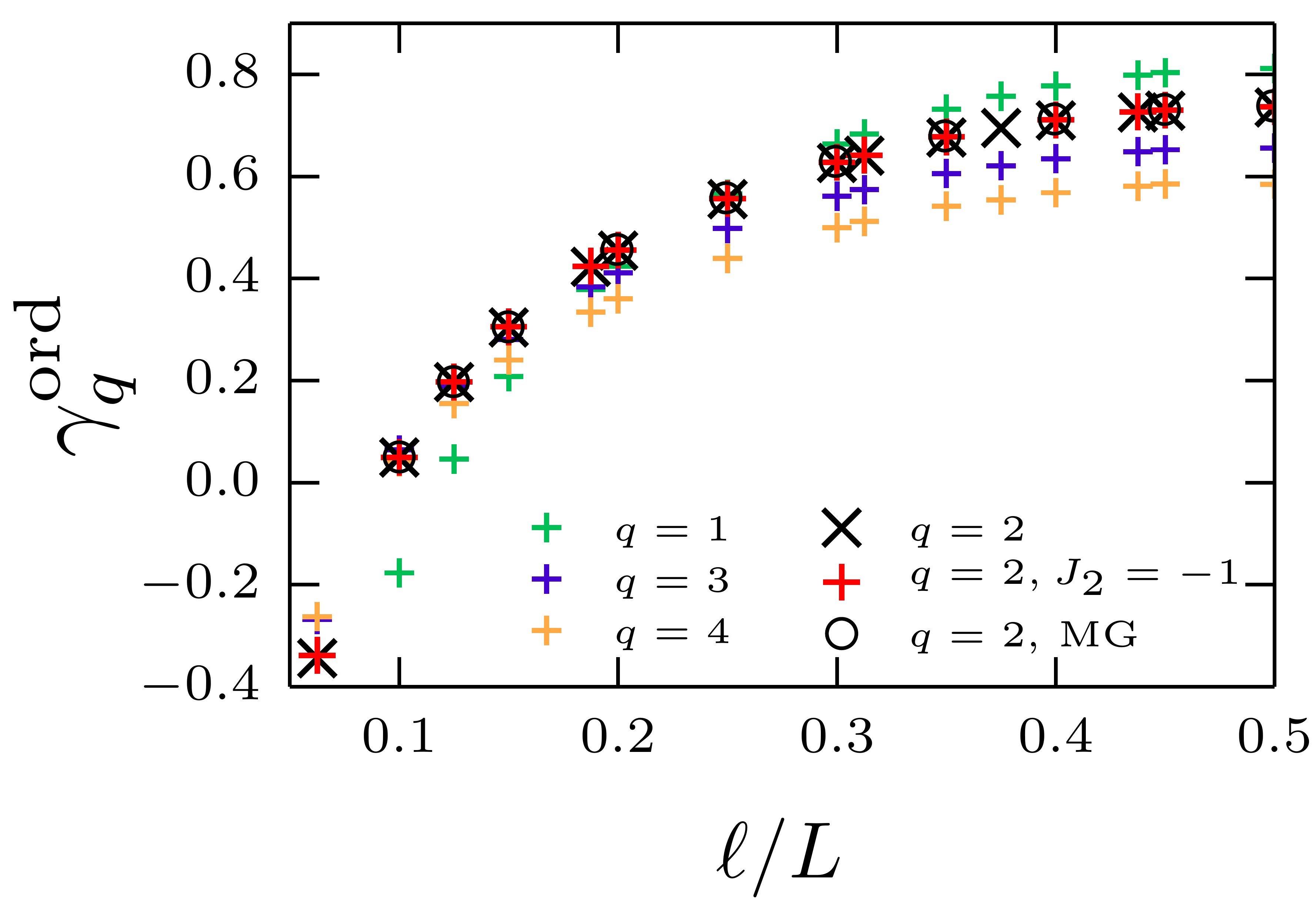}
    \caption{Additive constant $\gamma_q^{\rm ord}(\ell/L)$ for various R\'enyi parameters $q$ as a
    function of $\ell/L$ for $s=100$, and $J_2=0,-1$.
    Perfect agreement with Metlitski and Grover ($\circ$)~\cite{metlitski_entanglement_2011} is found for $q=2$.  Results for
    $J_2/J_1=-1$ at $q=2$ ({\color{red}{+}}) agree perfectly with 
$J_2=0$ (x), confirming universality. Reprinted from Ref.~\cite{laflorencie_spin-wave_2015}}
    \label{fig:gamma_ord}
\end{figure}

Such a finite size regularization allows to access R\'enyi entanglement entropies $S_q(A)$ for various shapes of subsystem $A$~\cite{luitz_universal_2015,laflorencie_spin-wave_2015} by means of numerical diagonalization using the correlation matrix technique~\cite{peschel_reduced_2009}. In Fig.~\ref{fig:S100} we show $S_q(A)$ results for $L\times\ell$ strip subsystems $A$ (panel b) embedded in $L\times L$ tori, up to $600\times 600$\footnote{Note that in order to access such large systems, translation symmetry of the subsystems was used for the strip geometry, thus optimizing the diagonalization procedure for large sizes.} with a large spin length $s=100$. This plot shows that (i) the area law term does not depend on the aspect ratio of the subsystem, and (ii) the logarithmic correction in Eq.~\eqref{eq:MG} is precisely governed by the prefactor $n_G/2=1$, independent of the R\'enyi index. 
The logarithmic corrections in Eq.~\eqref{eq:MG} have been also captured within a similar SW formalism for $d=3$ U(1) and SU(2) cases~\cite{frerot_area_2015}, and also using a Schwinger boson formalism~\cite{frerot_entanglement_2016}.
\paragraph{Universal additive constant $\gamma_{q}^{\rm ord}$---}
In the case of $d=2$ strip subsystems with an aspect ratio $\ell/L$, Metlitski and Grover have also predicted the existence of an additive universal constant $\gamma_q^{\rm ord}(\ell/L)$ in the scaling of the R\'enyi entropies in Eq.~\eqref{eq:MG}. Using the large-$s$ expression for the stiffness $\rho_s=s^2(J_1-2J_2)$ and the velocity $v=2\sqrt{2}s\sqrt{J_1(J_1-2J_2)}$, one can extract $\gamma_q^{\rm ord}(\ell/L)$ by fitting our numerical data to the above form Eq.~\eqref{eq:MG}, as displayed in Fig.~\ref{fig:gamma_ord} against the aspect ratio $\ell/L$. Results compare perfectly to thoses of Ref.~\cite{metlitski_entanglement_2011} for R\'enyi $q=2$. Universality is also confirmed by the fact that the results do not depend on the value of the second neighbor couplings $J_2$.

\paragraph{Connection to free scalar field theory---} An attempt to directly connect this result to free scalar field theory has also been made in Refs.~\cite{metlitski_entanglement_2011,laflorencie_spin-wave_2015}. On finite lattices, a diverging contribution from the zero mode has to be cut off by a small gap in the relativistic spectrum~\cite{barthel_entanglement_2006}. For a free scalar field, a small mass leads to a logarithmic correction to the entanglement entropy $\sim \ln\left(\frac{1}{mL}\right)$~\cite{metlitski_entanglement_2011}. However, an additional geometric constant $\gamma_{q}^{\rm free}(\ell/L)$, which depends on the R\'enyi coefficient $q$ and the aspect ratio $\ell/L$ of strip subsystems, is also expected~\cite{metlitski_entanglement_2011}, thus yielding
\be
S_q=a_q L-\frac{1}{2}\ln\left({mL}\right)+\gamma_{q}^{\rm free}(\ell/L).
\label{eq:freeb}
\ee
This behavior can be checked numerically for relativistic bosons on a square lattice, described by Eq.~\eqref{eq:Hk} with
${\rm{A}}_{\bf k}=2t-{t}\left(\cos k_x+\cos k_y\right)/2-{\Gamma}/{2}$ and ${\rm{B}}_{\bf k}={t}\left(\cos k_x+\cos k_y\right)/2$. 
Using a very small (albeit size-independent) mass gap $m=\sqrt{\Gamma/t}=10^{-9}$,
\begin{figure}[t]
    \centering
    \includegraphics[width=.7\columnwidth]{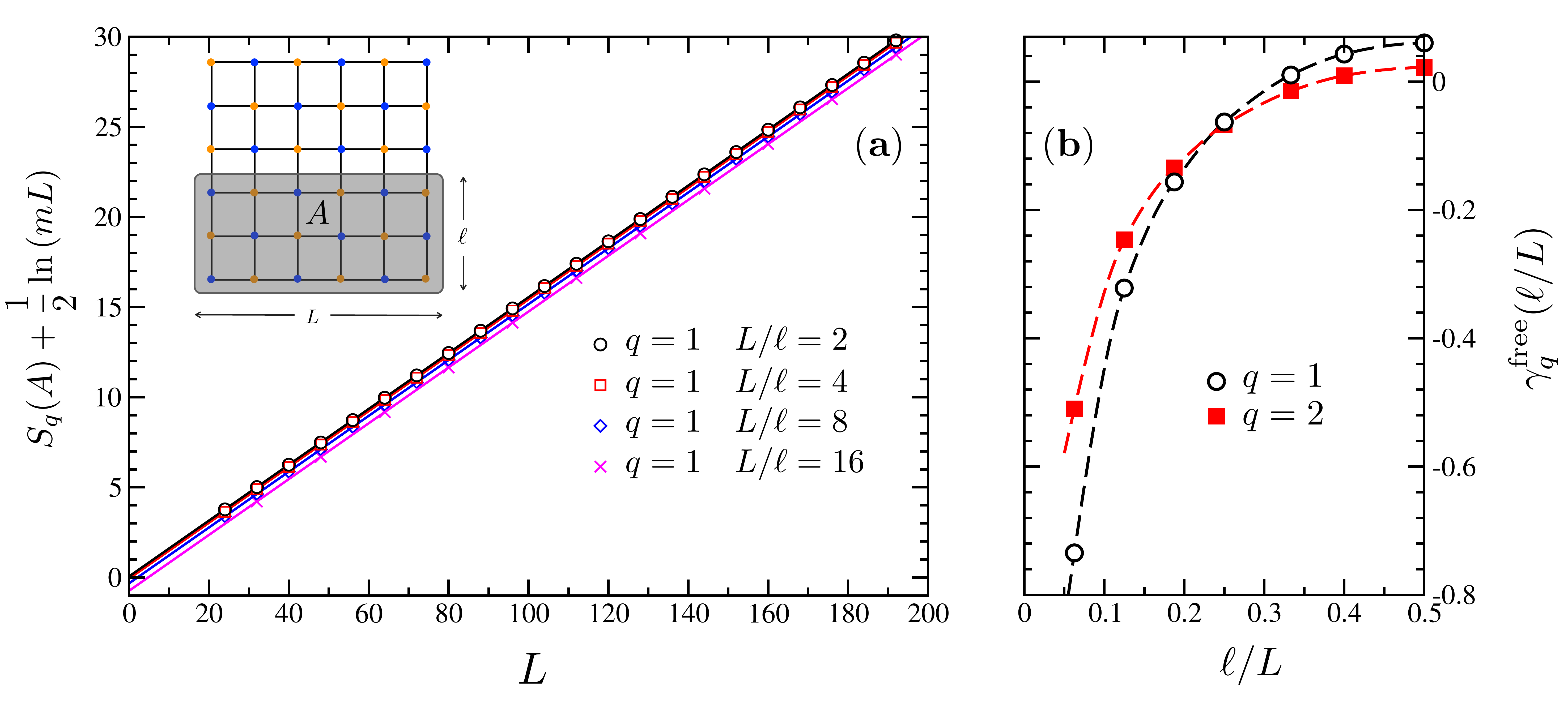}
    \caption{R\'enyi entanglement entropies of relativistic free bosons on the square lattice governed by the lattice Hamiltonian Eq.~\eqref{eq:Hk} with a small mass gap $m=10^{-9}$. Subsystem $A$ is a $L\times \ell$ strip (grey region in the inset). Panel (a) shows the area law scaling for $S_1(A)+\frac{1}{2}\ln(mL)$ for various aspect ratios. Panel (b) shows the geometric constant $\gamma_q^{\rm free}$ obtained from fits to the form Eq.~\eqref{eq:freeb} and plotted against $\ell/L$, the aspect ratio of subsystem $A$, for $q=1$ and $q=2$. Lines are guides to the eyes.}
    \label{fig:freeb}
\end{figure}
exact diagonalization results for $S_1+\frac{1}{2}\ln(mL)$ are dispalyed in panel (a) of Fig.~\ref{fig:freeb} where, on top of the clear area law scaling, one see the aspect ratio dependence of the intercept at $L\to 0$, thus giving  $\gamma_q^{\rm free}$, displayed in the panel (b) against $\ell/L$.

Building on the result Eq.~\eqref{eq:freeb} for a single mode, the correction part for SU(2) antiferromagnets with $n_G=2$ relativistic Glodstone modes and a TOS gap given by the mass term $m^*(L)$ in Eq.~\eqref{eq:mls},
can be expressed as
$
\Delta S_q=\frac{n_G}{2}\ln \left(\frac{\rho_s}{v}L\right) +n_G\left(\ln\sqrt{2}+\gamma_q^{\rm free}\right)
$, where we have used the large $s$ expression for the stiffness and the SW velocity. A direct comparison with Eq.~\eqref{eq:MG} from Metlitski and Grover~\cite{metlitski_entanglement_2011} yields
\be
\gamma_{q}^{\rm ord~su(2)}=2\gamma_q^{\rm free}+\ln 2,
\label{eq:gammasu2}
\ee
which agrees with them~\cite{metlitski_entanglement_2011}, but only at $q=2$.
Using similar arguments for the XY model with a single Goldstone mode~\cite{luitz_universal_2015}, one gets
\be
\gamma_{q}^{\rm ord~u(1)}=\gamma_{q}^{\rm free} +\frac{5}{4}\ln 2,
\label{eq:gammau1}
\ee
which compares well to QMC results of Kulchytskyy {\it{et al.}}~\cite{kulchytskyy_detecting_2015} (see below).
\subsubsection{Quantum Monte Carlo simulations}
\paragraph{Entanglement entropies---}

One of the most efficient numerical method to diagnose continuous symmetry breaking in the entanglement scaling for dimension $d\ge 2$ is based on QMC sampling, in particular after the work of Hastings {\it{et al.}}~\cite{hastings_measuring_2010}. Density Matrix Renormalization Group (DMRG) studies of $d=2$ quantum spin or bosonic models where SU(2)~\cite{jiang_spin_2012,kolley_entanglement_2013} or U(1)~\cite{alba_entanglement_2013} symmetry may be broken are also available, but they did not focus on logarithmic corrections.
The first QMC attempts~\cite{kallin_anomalies_2011,humeniuk_quantum_2012,
helmes_entanglement_2014} to check the Goldstone modes signature  $\frac{n_G}{2}\ln L$ in the
entropy~\cite{metlitski_entanglement_2011} have indeed found a logarithmic term for Heisenberg models, but estimated the prefactor in the range $0.5 - 0.8$, which is smaller that the predicted value of one (with $n_G=2$ for SU(2) symmetry).

\begin{figure}[b]
    \centering
    \includegraphics[width=.5\columnwidth,clip]{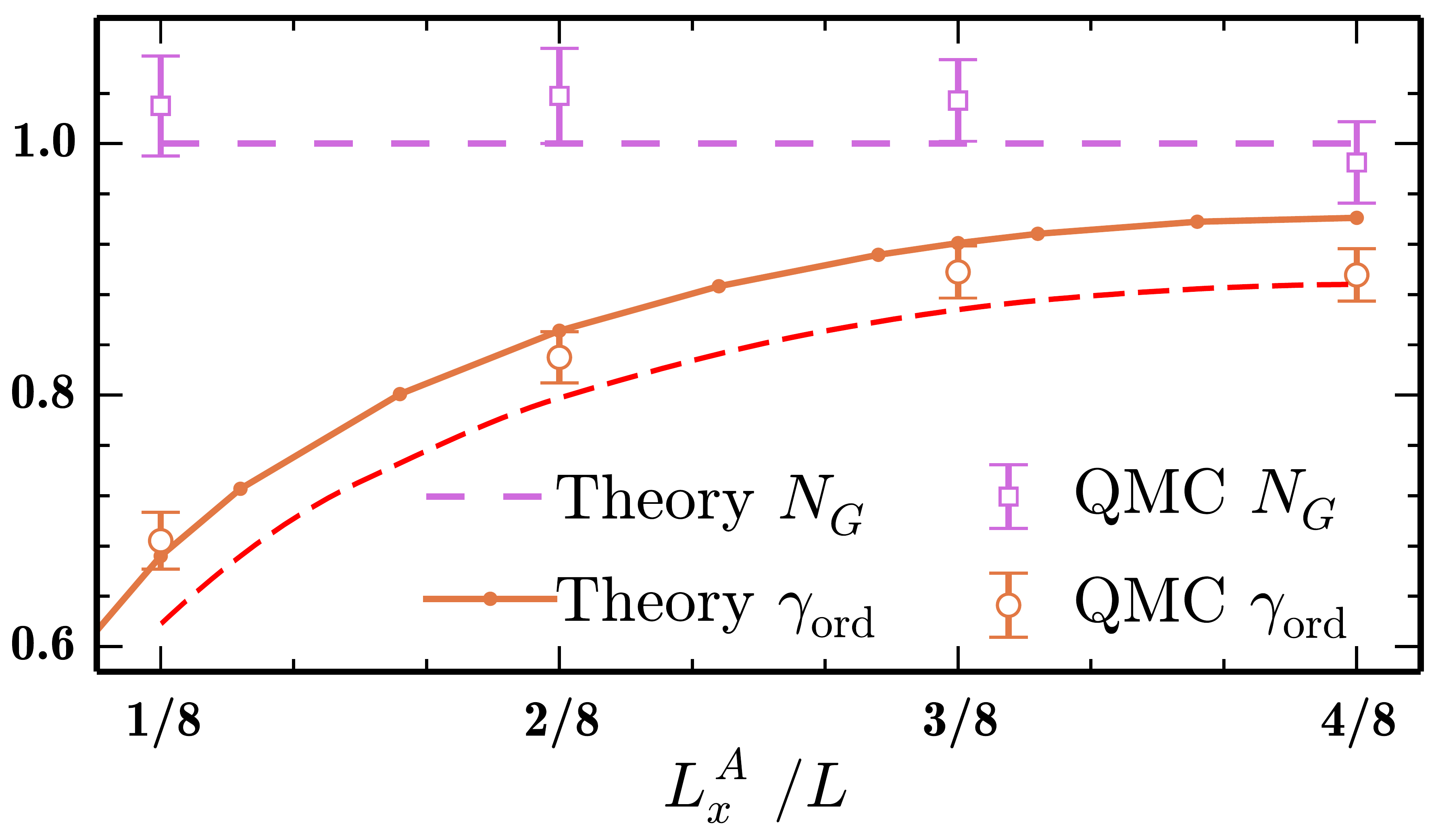}
    \caption{Subleading terms of the second R\'enyi entropy $\frac{n_G}{2}\ln L$ and $\gamma_2^{\rm ord}$ from Eq.~\eqref{eq:MG} obtained from QMC simulations of the $d=2$ $s=1/2$ XY model at $T=0$, up to $L=32$. The red dashed line is the large-$s$ prediction Eq.~\eqref{eq:gammau1}. Reprinted from Ref.~\cite{kulchytskyy_detecting_2015}.}
    \label{fig:detecting_melko}
\end{figure}

More recently, Kulchytskyy {\it{et al.}}~\cite{kulchytskyy_detecting_2015} have performed an improved measurement of the second R\'enyi entropy $S_2$ for the square lattice $s=1/2$ XY model from which they could estimate both the log prefactor (here $n_G=1$ for U(1) symmetry), as well as the constant in Eq.~\eqref{eq:MG} $\gamma^{\rm ord}$ for lattice up to $32\times 32$. Their results for both quantities, shown in Fig.~\ref{fig:detecting_melko} as a function of the aspect ratio of the strip subsystem, appear to be in good agreement the analytical prediction of Metlitski and Grover~\cite{metlitski_entanglement_2011} as well as with the large-$s$ prediction Eq.~\eqref{eq:gammau1} form Ref.~\cite{laflorencie_spin-wave_2015}.

\begin{figure}[t!]
    \centering
    \includegraphics[width=.6\columnwidth,clip]{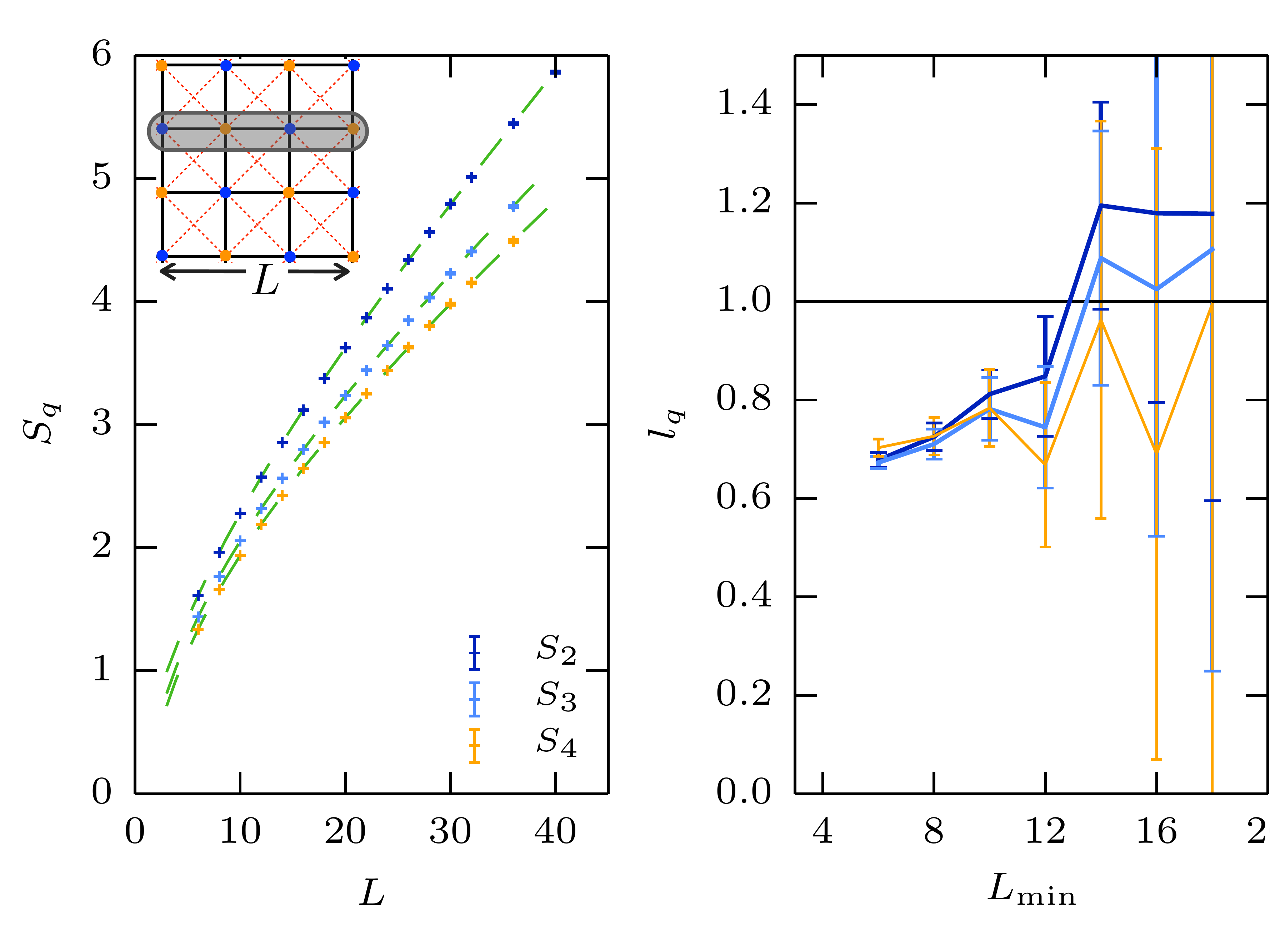}
    \caption{Left: QMC results for the
         entanglement R\'enyi entropies of the spin-$\frac{1}{2}$ $J_1-J_2$ Heisenberg model for $J_2/J_1=-1$, plotted against the length $L$ of the subsystem line (depicted in the inset). Right: Prefactor of the logarithmic scaling term obtained by fits to the form
 $S_q=a_q L + l_q \ln L + b_q + c_q/L$ over ranges $[L_{\rm{min}},L_{\rm{max}}]$ as a function of
 $L_{\rm{min}}$, with $L_{\rm{max}}=40$ for $q=2$ and $L_{\rm{max}}=36$ for $q=3,4$. Despite quite large error bars, these results are consistent with $l_q=1$ independent of $J_2$ and 
 $q$. Reprinted from Ref.~\cite{luitz_improving_2014}.}
    \label{fig:lineqmc}
\end{figure}

In order to go beyond $q=2$, building on the ratio trick proposed by Humeniuk and Roscilde~\cite{humeniuk_quantum_2012} one can improve the QMC estimate of R\'enyi (and also thermodynamic) entropies, as done in Ref.~\cite{luitz_improving_2014}. This method has been applied to the above $s=1/2$ $J_1 - J_2$ Heisenberg antiferromagnet on the square lattice Eq.~\eqref{eq:HJ1J2}, for which studying the R\'eny entropies at $q=2,3,4$ of the simplest corner-free subsystem - a periodic one-dimensional line (see Fig.~\ref{fig:lineqmc}) -  allowed to capture the logarithmic corrections with a prefactor close to unity, fully consistent with $n_G=2$ Goldstone bosons. The fact that universality can be captured for such a simple linear subsystem reinforces the idea that entanglement is dominated by boundary degrees of freedom.
\paragraph{Shannon-R\'enyi entropies---}
Spontaneous breaking of a continuous symmetry also appears in the (basis-dependent) so-called "Shannon-R\'enyi" (SR) entropy computed in the ground-state of finite size systems. These wave-function entropies, recently introduced in serie of works by St\'ephan and co-workers~\cite{stephan_shannon_2009,stephan_phase_2011,stephan_shannon_2014}, and further sudied by other groups~\cite{zaletel_logarithmic_2011,atas_multifractality_2012,alcaraz_universal_2013,luitz_universal_2014,luitz_shannon-renyi_2014,luitz_participation_2014,alcaraz_universal_2014,monthus_pure_2015,alcaraz_generalized_2015}, can efficiently capture universal properties, {\it{e.g.}} Luttinger liquid physics, quantum criticality, symmetry breaking phases. 

\begin{figure}[t]
    \centering
    \includegraphics[width=.6\columnwidth,clip]{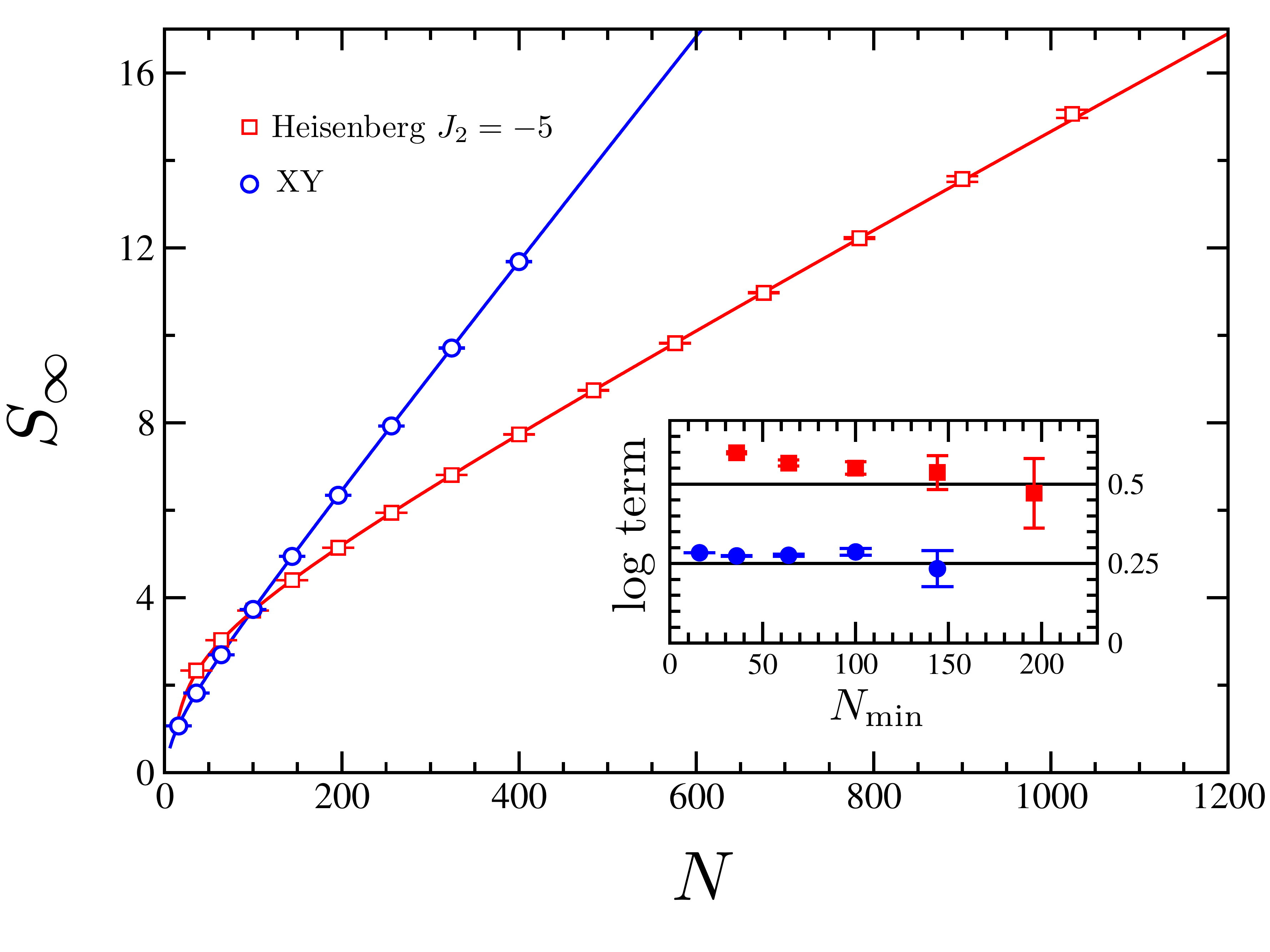}
    \caption{QMC results for the Shannon R\'enyi entropy $S_\infty$ of two-dimensional symmetry breaking phases. $T=0$ QMC data for SU(2) Heisenberg antiferromagnet with second neighbor ferromagnetic coupling $J_2/J_1=-5$ (red squares) and U(1) symmetric XY model (blue circles) are shown both for $s=1/2$ versus the total number of lattice sites $N$. Inset: the prefactors of the logarithmic correction in Eq.~\eqref{eq:SR}, obtained after fitting over windows $[N_{\rm min},N_{\rm max}]$, are plotted for the two models against $N_{\rm min}$. Data from \cite{luitz_universal_2014}.}
    \label{fig:shannon}
\end{figure}

Expanding a given (ground-) state in a computational discrete orthonormal basis $|\Phi\rangle=\sum_i a_i|i\rangle$, the SR entropy of this wave-function is defined by
\be
S_q^{\rm SR}=\frac{1}{1-q}\ln\left(\sum_i |a_i|^{2q}\right),
\label{eq:SRE}
\ee
where $|a_i|^2$ is simply interpreted as the probability of occupying state $|i\rangle$, provided $\sum_i |a_i|^2=1$. Contrary to single particle problems such as the Anderson localization~\cite{evers_anderson_2008} where the SR entropy (related to the inverse participation ratio) does not grow with the number of sites, many-body states occupy a finite portion of the Hilbert space and therefore the leading term grows with the volume. However, as observed numerically using large scale QMC simulations~\cite{luitz_universal_2014}, the first corrections to such a volume law are logarithmic for SU(2) and U(1) broken phases. This is examplified in Fig.~\ref{fig:shannon} where QMC data~\cite{luitz_universal_2014} for $q=\infty$ (corresponding to the coefficient of the most propable state in the basis expansion) are shown for the ground-state of two different spin-$\frac{1}{2}$ models on a square lattice: the SU(2) Heisenberg antiferromagnet Eq.~\eqref{eq:HJ1J2} with $J_2/J_1=-5$ and the U(1) XY model, both computed in the $\{S^z\}$ basis. One clearly observes a volume law scaling as well as an additive logarithmic correction with a prefactor whose value depends on the symmetry.

An analytical explanation has been proposed recently by Misguich and co-workers~\cite{misguich_shannon-renyi_2015}, using a massless free-field description of the SW modes supplemented by a phase space argument treating the rotational symmetry in finite volume. They arrived at the following correction for $q>1$
\be
\Delta S_{q}^{SR}=\frac{n_G}{4}\left(\frac{q}{q-1}\right)\ln N,
\label{eq:SR}
\ee
which appears consistent with the numerics~\cite{luitz_universal_2014}.

\subsubsection{Summary and outlook}
Relativistic Nambu-Goldstone modes associated with continuous symmetry breaking show up as logarithmic corrections in the R\'enyi entanglement entropies, as written in Eq.~\eqref{eq:MG}. This was clearly verified using large-$s$ calculations, and QMC simulations for strip and line subsystems. Interestingly, the coefficients of the ground-state wave function expressed in a local computational basis also contain such a correction Eq.~\eqref{eq:SR}, as verified with large scale QMC.

Regarding the additive geometric contant $\gamma_{\rm ord}$ in Eq.~\eqref{eq:MG}, the discrepancy between the prediction of Metlitski and Grover and the large-$s$ result Eqs.~\eqref{eq:gammasu2}\eqref{eq:gammau1} remains to be understood. A QMC study for $q>2$, while notoriously difficult, could perhaps resolve this issue.
Another potentially interesting check would be to study other continuous symmetry breaking states, such as SU(N) for instance using QMC~\cite{beach_$textsun$_2009} or flavor-wave theory~\cite{toth_three-sublattice_2010}. Spin nematic ordered states~\cite{smerald_theory_2015} are also interesting exotic candidates where similar logarithmic corrections should occur.

\subsection{Long-range entanglement in topologically ordered phases}
\label{sec:to}
The current understanding of topological ordered phases has benefited from intensive works during the past 25 years~\cite{wen_mean-field_1991,nayak_non-abelian_2008}. While a precise definition of topological order is still an active field of research~\cite{wen_theory_2016}, one can simply see it as a zero temperature disordered gapped state ({\it{i.e.}} with only short-range order of any local operators) which does not break any symmetry, and whose degeneracy is robust against local perturbations and depends on the topology of the space. This dependence implies that infinitely far boundaries can influence the ground-state properties, which seems to contradict the short-range nature of the correlations. Such a "robust hidden long-range structure"~\cite{wen_topological_1990} has been diagnosed as "long-range entanglement"~\cite{chen_local_2010} (even though a topological ordered state is not necessarily more entangled than other states of matter).

\subsubsection{Topological entanglement entropy}
\label{sec:toee}
One of the simplest way to characterize topological order relies on the entanglement entropy which displays a sub-leading constant beyond the conventional area law term, thus yielding
\be
S=aL^{d-1}-\gamma,
\ee
as first identified by Hamma {\it{et al.}}~\cite{hamma_ground_2005}, Kitaev and Preskill~\cite{kitaev_topological_2006}, and Levin and Wen~\cite{levin_detecting_2006}. This topological entropy $\gamma$ is a universal number, characteristic of the topological order, which depends non-trivially on the quantum dimension of emergent quasi-particles above the degenerate ground-state~\cite{kitaev_topological_2006,levin_detecting_2006}:
\be \gamma=\ln \cal D,\label{eq:D}\ee where $\cal D$ is the total quantum dimension. 

Let us give a few examples: for $\mathbb Z_2$ liquids, such as the toric code~\cite{kitaev_anyons_2006}, $\gamma=\ln 2$, while for the chiral spin liquid state~\cite{kalmeyer_equivalence_1987,wen_chiral_1989} $\gamma=\ln\sqrt 2$. For $\nu=1/m$ quantum Hall states, $\gamma=\ln \sqrt m$ for the Laughlin fractional wave-function~\cite{laughlin_anomalous_1983}, while for the Moore-Read state~\cite{moore_nonabelions_1991} $\gamma=\ln\sqrt{4m}$. A very useful interpretation of this topological constant has been proposed by St\'ephan and co-workers in a series of works~\cite{stephan_shannon_2009,stephan_phase_2011,stephan_renyi_2012} where they showed using CFT a correspondence between entanglement entropies of $d=2$ Rokhsar-Kivelson (RK) states and the (basis-dependent) Shannon-R\'enyi entropies associated to the coefficient of the wave-function of spin chains models. An interesting example is the $d=2$ eight-vertex RK state~\cite{ardonne_topological_2004} which has $\mathbb Z_2$ topological order, and corresponds to the Ising chain in transverse field in the ferromagnetic phase where the Shannon-R\'enyi entropies follows $aL-\ln 2$.

\subsubsection{Numerical results}
\paragraph{${\mathbb Z}_2$ spin liquids---}

\begin{figure}[b]
    \centering
    \includegraphics[width=.6\columnwidth,clip]{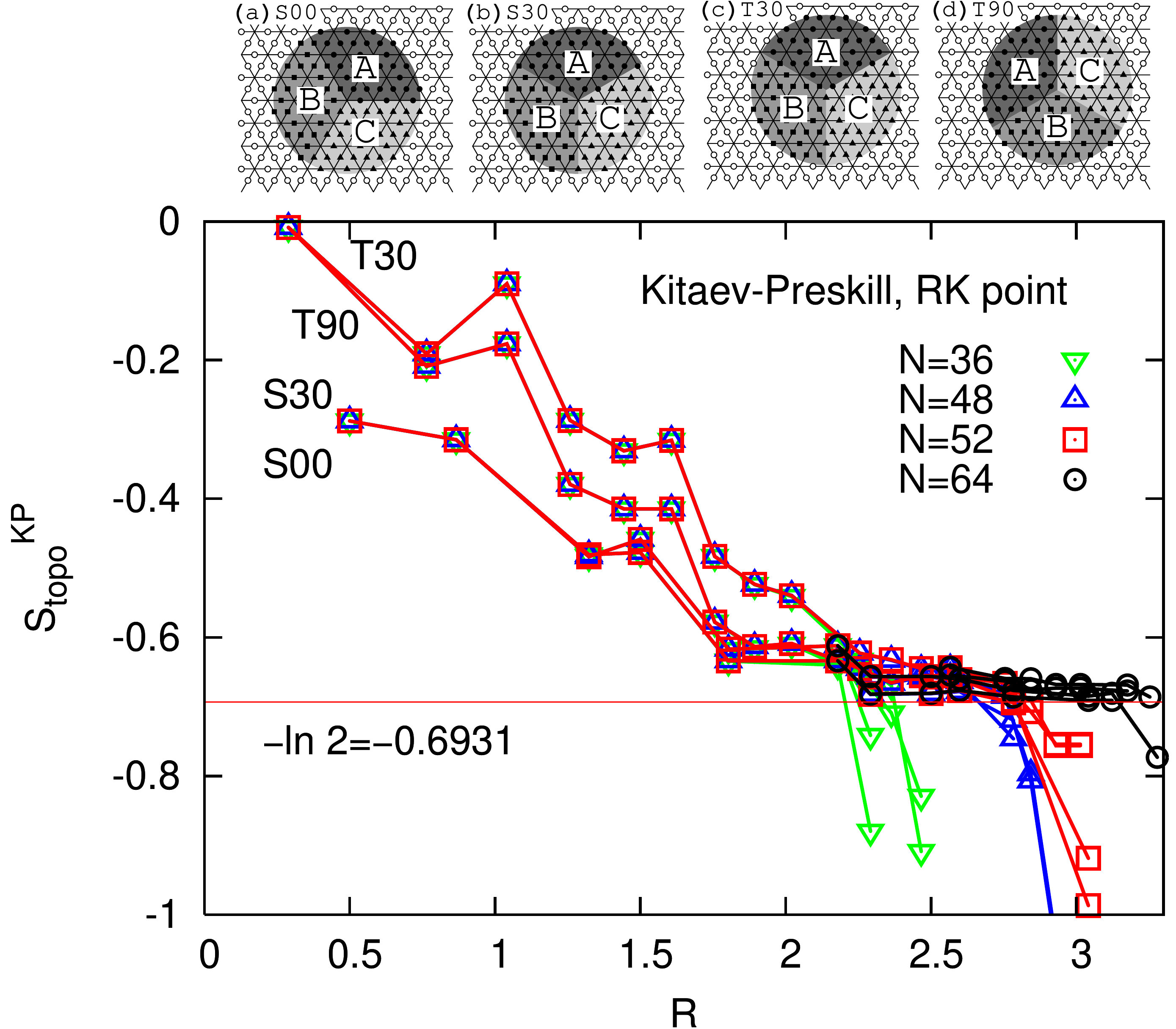}
    \caption{Topological entanglement entropy for the $\mathbb Z_2$ spin liquid phase of the quantum dimer model on the triangular lattice at the RK point computed using exact diagonalization by Furukawa and Misguich~\cite{furukawa_topological_2007}, plotted as a function of the circle radius $R$. The 4 insets (top) show the 4 different Levin-Wen constructions used to extract $\gamma$ which converges to the expected $\ln 2$. Reprinted from \cite{furukawa_topological_2007}.\label{fig:topo_gregoire}}
    \end{figure}
Considered to be the smoking gun of topological order, non-zero topological entanglement entropy  may be hard to measure in numerical simulations. A construction in real space was proposed~\cite{kitaev_topological_2006,levin_detecting_2006} in order to cancel boundary and corner effects and get a better access to $\gamma$. This is illustrated in Fig.~\ref{fig:topo_gregoire} for the $\mathbb Z_2$ spin liquid phase of the quantum dimer model on the triangular lattice~\cite{moessner_resonating_2001} at the RK point~\cite{rokhsar_superconductivity_1988}, computed using exact diagonalization by Furukawa and Misguich~\cite{furukawa_topological_2007}. 

\begin{figure}[h]
    \centering
        \includegraphics[width=.8\columnwidth,clip]{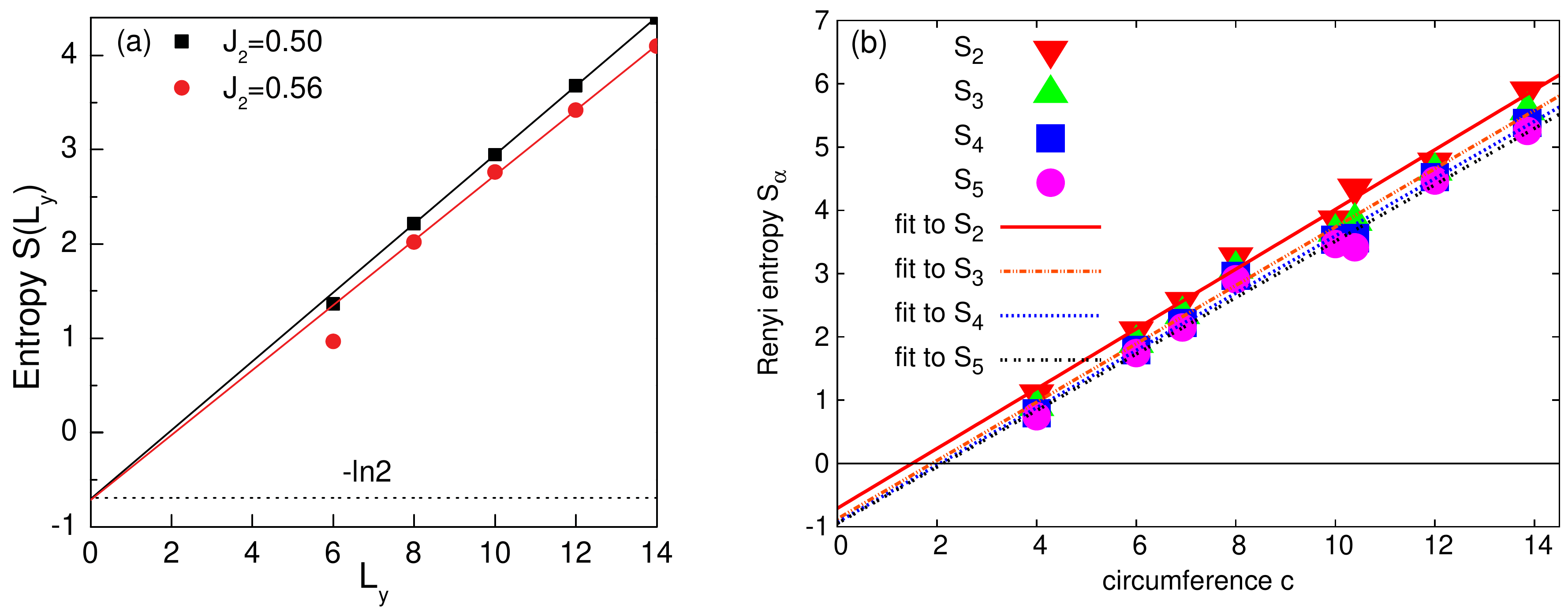}
    \caption{Entanglement entropy for $\mathbb Z_2$ topological spin liquid states computed with DMRG on long cylinders of finite widths. Panel (a), reprinted from \cite{jiang_spin_2012}, shows the von Neuman entropy of the $J_1 - J_2$ antiferromagnet on the square lattice as function of the perimeter of the cylinder $L_y$  with an intercept $\gamma=-\ln 2 $ for $L_y\to 0$. Panel (b), reprinted from \cite{depenbrock_nature_2012}, shows various R\'enyi entropies $S_q$ for the Kagom\'e antiferromagnet as function of the perimeter of the cylinder $c$ with an intercept $\gamma\to -\log_2 (2)=-1$ for increasing values of $q$.}
    \label{fig:tee_z2sl}
\end{figure}
This subtraction trick has been used by several authors to measure the topological entanglement entropy~\cite{furukawa_topological_2007,zhang_topological_2011,isakov_topological_2011,stephan_renyi_2012,wildeboer_entanglement_2015}, but it turns out that it is easier and more accurate to work with a cylinder geometry, as done for instance in Refs.~\cite{lauchli_entanglement_2010,stephan_renyi_2012,jiang_identifying_2012,jiang_spin_2012,zhang_quasiparticle_2012,poilblanc_topological_2012,depenbrock_nature_2012,cincio_characterizing_2013,poilblanc_simplex_2013,jiang_accuracy_2013,wildeboer_entanglement_2015}. In Fig.~\ref{fig:tee_z2sl}, we report $d=2$ DMRG results for $\mathbb Z_2$ spin liquid states on (a) the frustrated $J_1 - J_2$ spin-$\frac{1}{2}$ Heisenberg antiferromagnet on the square lattice Eq.~\eqref{eq:HJ1J2} from \cite{jiang_spin_2012}, and (b) the Kagom\'e Heisenberg antiferromagnet\footnote{Note that there is still some debates regarding the nature of the ground-state
due to the existence of various competing low-energy states. Among the three candidates, the gapped ${\mathbb Z}_2$ spin liquid~\cite{lu_$mathbbz_2$_2011}, the U(1) Dirac spin liquid~\cite{iqbal_gapless_2013}, and the 36-site unit cell valence-bond solid~\cite{singh_ground_2007}, the most recent $d=2$ DMRG studies concluded for a ${\mathbb Z}_2$ spin liquid~\cite{yan_spin-liquid_2011,depenbrock_nature_2012}, with ${\mathbb Z}_2$ topological order. For recent discussions, see also Refs.~\cite{gong_global_2015,kolley_phase_2015}.} from \cite{depenbrock_nature_2012}. In both cases one reads the topological entanglement entropy as the intercept when the cylinder circumference vanishes, in perfect agreement with the predicted 
quantum dimension ${\cal D}=2$ from Eq.~\eqref{eq:D}. Note the better accord for the Kagom\'e model in Fig.~\ref{fig:tee_z2sl} at larger R\'enyi parameters $q$, a consequence of the less good accuracy when computing lower weight entanglement modes. Surprisingly this numerical limitation seems to be at variance with the results obtained in Ref.~\cite{jiang_accuracy_2013}.

\paragraph{Fractional quantum Hall states---}

\begin{figure}[h]
    \centering
    \includegraphics[width=.7\columnwidth,clip]{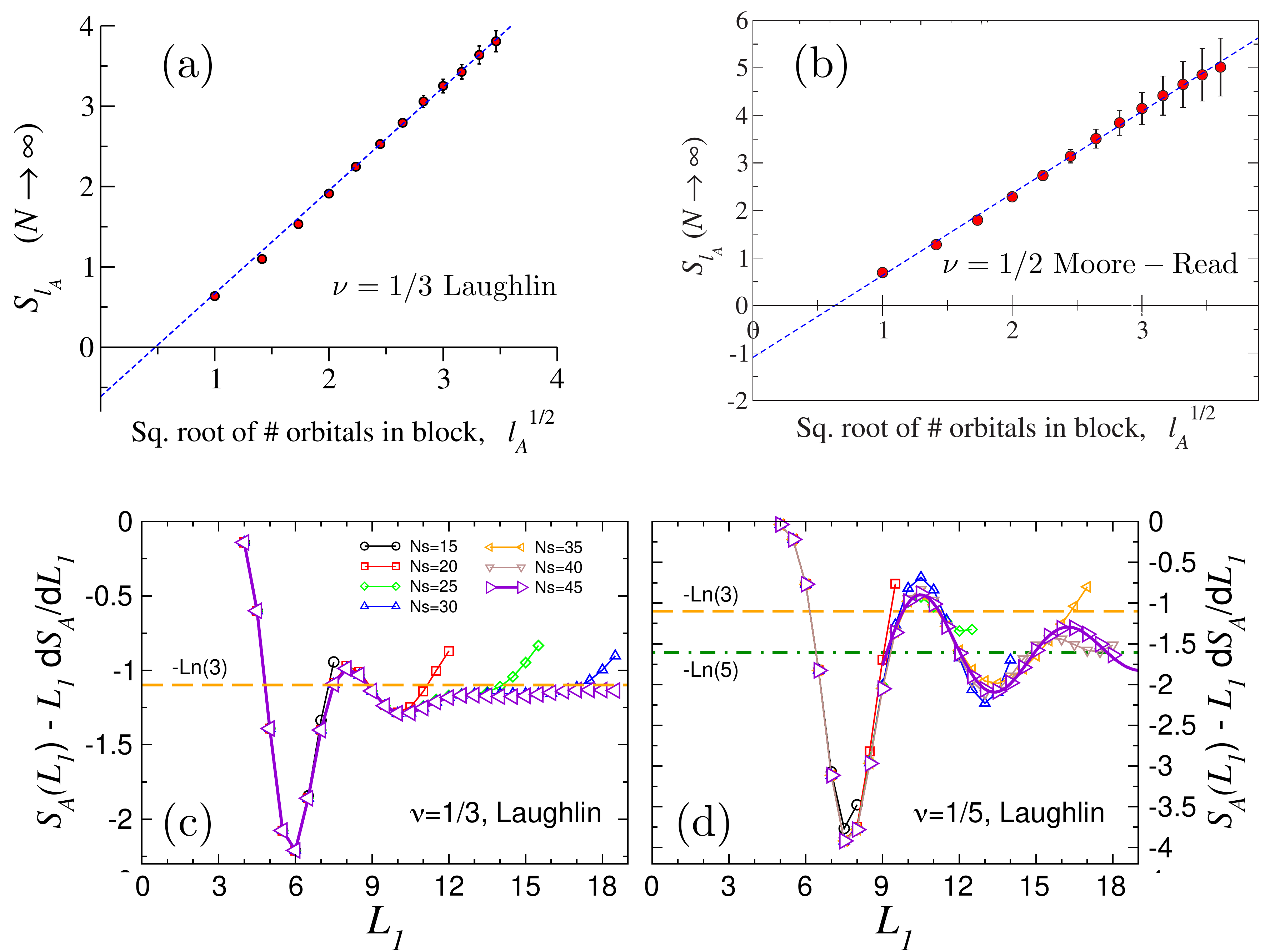}
    \caption{Entanglement entropies for fractional quantum Hall states computed (a,b) on a sphere with orbital partitions and (c,d) on a torus with real space cuts. (a) For the $\nu=1/3$ Laughlin wave-function the topological term is found in perfect agreement with $\gamma=-\ln\sqrt{3}\simeq -0.55$, reprinted from~\cite{haque_entanglement_2007}. (b) For the $\nu=1/2$ Moore-Read state $\gamma=-\ln{\sqrt 8}\simeq -1.04$, reprinted from \cite{zozulya_bipartite_2007}.  For real space bipartitions on a torus (c,d), reprinted from \cite{lauchli_entanglement_2010}, $2\times \gamma$ is plotted against the torus perimeter $L_1$ for various magnetic flux quanta $N_s=N/\nu$ with $N$ electrons for $\nu=1/3$ and $1/5$ Laughlin wave-functions. A good agreement is found with $2\gamma=\ln \nu$.}
    \label{fig:tee_hall}
\end{figure}

The topological entanglement entropy has also been measured with a great accuracy for Laughlin wave function at $\nu=1/m$~\cite{iblisdir_entropy_2007,haque_entanglement_2007,lauchli_entanglement_2010,zhang_topological_2011} as well as for the non-abelian Moore-Read state~\cite{zozulya_bipartite_2007,wildeboer_spin_2015}. In Fig.~\ref{fig:tee_hall} we report numerical results from Refs.~\cite{haque_entanglement_2007,zozulya_bipartite_2007,lauchli_entanglement_2010}.
More recently, non-abelian states such as fractional quantum Hall states at $\nu=13/5$ and $\nu=12/5$, apparently captured by the $k=3$ parafermion Read-Rezayi state~\cite{read_beyond_1999}, have been studied in Refs.~\cite{mong_fibonacci_2015,zhu_fractional_2015}. The topological entanglement entropy of such a state $\gamma=\frac{1}{2}\ln(5+5\phi^2)\simeq 1.45$ was successfully captured using large scale DMRG simulations.
Note also the non-abelian phases in the two-component $\nu=2/3$ fractional quantum Hall states on a bilayer, with the emergence of Fibonacci anyons~\cite{liu_non-abelian_2015}.

\paragraph{Other examples---} We now give a very short (and clearly non-exhaustive) list of recently discussed topological ordered phases characterized by their topological entropy. Using $d=2$ DMRG for the $s=1/2$ kagom\'e antiferromagnet in an external field, a magnetization plateau at filling $1/9$ of saturation has been claimed to be a ${\mathbb Z}_3$ spin liquid state~\cite{nishimoto_controlling_2013}. The existence of this exotic plateau state had not been reported previously~\cite{capponi_numerical_2013}, but was diagnosed in a more recent study~\cite{picot_spin-$s$_2016} to be a valence bond crystal coexisting with spin order, a phase where topological order is absent.
Still on the kagom\'e lattice, hard-core bosons at $\nu=1/3$ filling have revealed an interesting ${\mathbb Z}_2$ topological liquid state~\cite{roychowdhury_$mathbbz_2$_2015}. In Ref.~\cite{nielsen_local_2013} a frustrated spin-$\frac{1}{2}$ model on a square lattice was found to exhibit the same topological order as the spin chiral state. Note also the topological color code model~\cite{bombin_topological_2006} which exhibits a ${\mathbb Z}_2\times {\mathbb Z}_2$ spin liquid ground-state with $\gamma=\ln 4$, as studied numerically in \cite{jahromi_full_2015}.

For extension to $d> 2$, we refer to the work of Grover {\it{et al.}}~\cite{grover_entanglement_2011} (see also \cite{kimchi_three-dimensional_2014} for three-dimensional quantum spin liquids in iridate materials and\cite{pretko_entanglement_2015}  for U(1) quantum spin liquids).
Finally, for topological order and topological entropy in classical systems, see {\it{e.g.}} Refs.~\cite{castelnovo_topological_2007,castelnovo_entanglement_2007,helmes_renyi_2015}.

\section{Entanglement spectroscopy}
\label{sec:ES}
Soon after the discovery of the DMRG algorithm for one dimensional quantum systems~\cite{white_density_1992, schollwock_density-matrix_2005}, some attempts have been made to extend it to $d=2$~\cite{liang_approximate_1994,nishino_density_1995,white_density_1998}. However, it was then quickly realized that such an extension is a quite difficult task, as observed through the study of the spectrum of the RDM, the entanglement spectrum (ES), for two dimensional non-interacting quantum systems~\cite{chung_density-matrix_2000}. Although such studies were first motivated by performance issues regarding the DMRG algorithm, some deeper investigations have later started to become a topic of increasing interest, in particular for quantum chains~\cite{peschel_density-matrix_1999,chung_density-matrix_2001,cheong_many-body_2004,cheong_operator-based_2004,eisler_fluctuations_2006}.

A breakthrough came in 2008 with the work of Li and Haldane~\cite{li_entanglement_2008} where they showed for $\nu=5/2$ fractional quantum Hall states that the low lying part of the RDM spectrum contains universal features regarding topological properties. This has triggered a huge interest, in particular for topological order that we discuss below in Section~\ref{sec:esto}. Roughly at the same time, Calabrese and Lef{\`e}vre~\cite{calabrese_entanglement_2008} focused on the eigenvalues distribution of the RDM for critical spin chains described by a CFT, which also raised a large inquisitiveness, as we review in Section~\ref{sec:es}.
Among numerous fascinating properties of entanglement spectroscopy, the notion of entanglement Hamitonian has been also intensively debated, as we discuss in Section~\ref{sec:eh}.

\subsection{Entanglement spectrum in one dimension and beyond}
\label{sec:es}
\subsubsection{Quantum spin chains and ladders}
\paragraph{Quantum spin chains}
The ground-state of the spin $s=1/2$ XXZ chain, defined on a chain of $L$ sites by the following Hamiltonian
\be
{\cal{H}}_{xxz}=\sum_{i=1}^{L}\left(S_{i}^{x}S_{i+1}^{x}+S_{i}^{y}S_{i+1}^{y}+\Delta S_{i}^{z}S_{i+1}^{z}\right),
\label{eq:XXZ}
\ee
displays critical spin-spin correlations for $-1<\Delta\le 1$ with a continuously varying Luttinger liquid parameter~\cite{giamarchi_quantum_2003} 
\be K=\frac{1}{2\arccos(-\Delta)/\pi}.\label{eq:K}
\ee
This critical regime is described by a CFT~\cite{alcaraz_surface_1987} with central charge $c=1$, yielding the logarithmic growth~\cite{calabrese_entanglement_2004} of the R\'eny entanglement entropies Eq.~\eqref{eq:S1d}. Making a real-space bipartition, which defines a subsystem $A$ (depicted in the inset of Fig.~\ref{fig:dist_eig}), its RDM $\rho_A$ has eigenvalues $\lambda_i$ which gives the ES $\xi_i=-\ln\lambda_i$.
The eigenvalues distribution was first studied by Calabrese and Lef\`evre~\cite{calabrese_entanglement_2008} using CFT and exact diagonalization of Eq.~\eqref{eq:XXZ} at $\Delta=0$. Exploiting the fact that $ {\rm{Tr}}\,\rho^q_A\sim L^{-\frac{c}{6}\left(q-\frac{1}{q}\right)}$,
 they derived an approximate expression for the mean number of eigenvalues larger than a given $\lambda$:
\begin{figure}
    \centering
    \includegraphics[width=.5\columnwidth,clip]{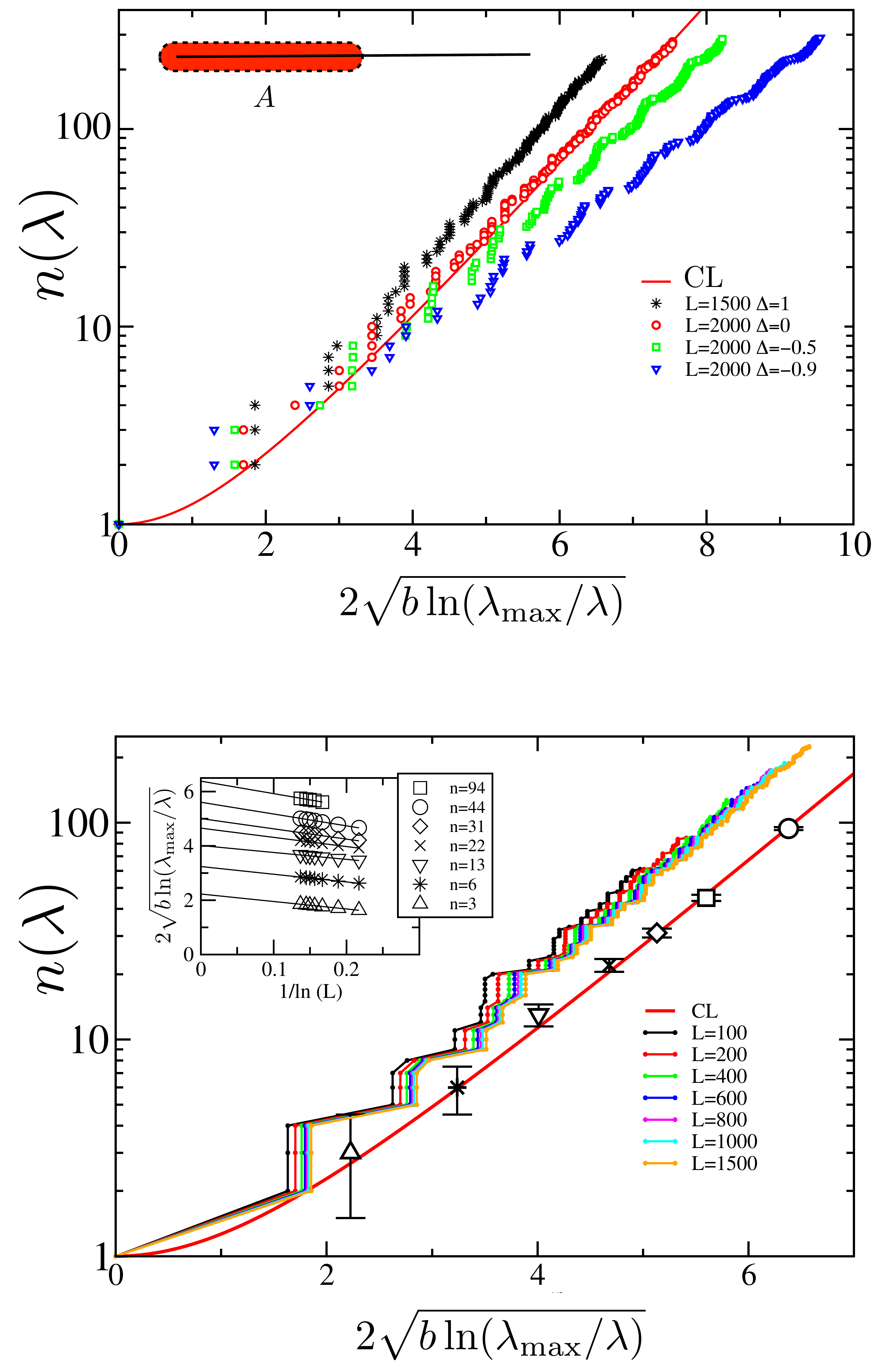}
    \caption{DMRG results for the entanglement eigenvalue distribution for open spin-$\frac{1}{2}$ XXZ chains with a real-space bipartition at half-chains $L/2$ (inset). Top: Numerical results for $n(\lambda)$ at various anisotropies $\Delta$ for system sizes $L\ge 1500$, compared to the analytical prediction Eq.~\eqref{eq:nlambda} from Calabrese and Lef\`evre (CL)~\cite{calabrese_entanglement_2008}. Bottom: Finite size convergence of $n(\lambda)$ towards the CL formula Eq.~\eqref{eq:nlambda}. DMRG data for $\Delta=1$. Inset: logarithmic convergence to the thermodynamic limit. Figure reprinted from Ref.~\cite{laflorencie_spin-resolved_2014}.}
    \label{fig:dist_eig}
\end{figure}
\be
n(\lambda)=I_0\Bigl[b\ln\left(\frac{\lambda_{\text{max}}}{\lambda}\right)\Bigr],
\label{eq:nlambda}
\ee
where $I_0$ is the modified Bessel function of first kind, $\lambda_{\text{max}}$ is the largest eigenvalue, and $b=-\ln\lambda_{\rm max}$, nicely confirmed numerically for the XX point~\cite{calabrese_entanglement_2008}. A good agreement was also found later by Pollmann and Moore using infinite time-evolved block decimation~\cite{pollmann_entanglement_2010} for various XXZ anisotropies, albeit a fitting prefactor was used in Eq.~\eqref{eq:nlambda}. In the critical ferromagnetic regime $-1<\Delta<0$, Alba {\it{et al.}}~\cite{alba_entanglement_2012} also observed a sizeable deviation form prediction Eq.~\eqref{eq:nlambda}. Note that the theory of finite-entanglement scaling~\cite{tagliacozzo_scaling_2008} precisely builds on such a universal distribution~\cite{pollmann_theory_2009}.

Using large scale DMRG we have explored~\cite{laflorencie_spin-resolved_2014} the distribution of $\lambda_i$ for various anisotropy parameters $\Delta$ along the critical regime of the XXZ model.
Numerical results for $n(\lambda)$ with very large systems, up to $L=2000$ lattice sites and open boundary conditions, are shown in Fig.~\ref{fig:dist_eig}. There, we see that the prediction Eq.~\eqref{eq:nlambda} works remarkably well at the XX point (free fermions), in agreement with Calabrese-Lef\`evre, but with significant deviations appear for finite interaction $\Delta\neq 0$. Interestingly, attractive $\Delta<0$ and repulsive $\Delta>0$ interactions display opposite deviations, with a sign change at $\Delta=0$. A finite size analysis, shown in Fig.~\ref{fig:dist_eig} (Bottom) for the isotropic point $\Delta=1$, reveals that finite size effects are responsible for the observed disagreement. Indeed, convergence to the asymptotic form Eq.~\eqref{eq:nlambda} is slowed by logarithmic corrections $\sim 1/\ln L$, with a prefactor which change sign with the anisotropy $\Delta$\footnote{These corrections exist on the entire critical regime and are not restricted to $\Delta=1$~\cite{laflorencie_spin-resolved_2014}.}. So far there is no analytical understanding for such finite size effects.

\begin{figure}[t]
    \centering
    \includegraphics[width=.85\columnwidth,clip]{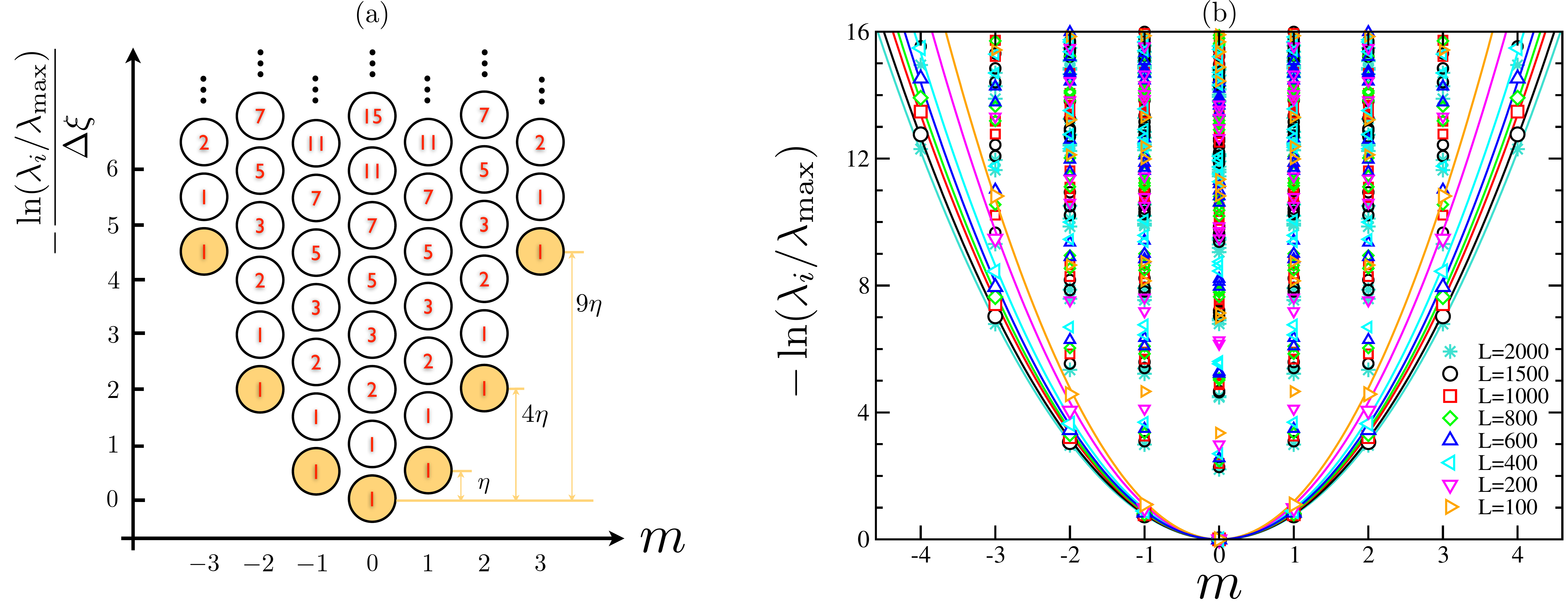}
    \caption{(a) Schematic representation of the CFT energy spectrum of a compactified boson $(\eta=\frac{1}{2K}$) with open boundary conditions, reprinted from~\cite{lauchli_operator_2013}.
(b) Spin-resolved ES from DMRG calculations: results for open XXZ chains at $\Delta=-0.5$ and various lengths $L$, as indicated on the plot. The lower part of the spectrum is fitted to the quadratic form Eq.~\eqref{eq:Sm}, reprinted from~\cite{laflorencie_spin-resolved_2014}.}
    \label{fig:esD}
\end{figure}
\begin{figure}[hb]
    \centering
    \includegraphics[width=.815\columnwidth,clip]{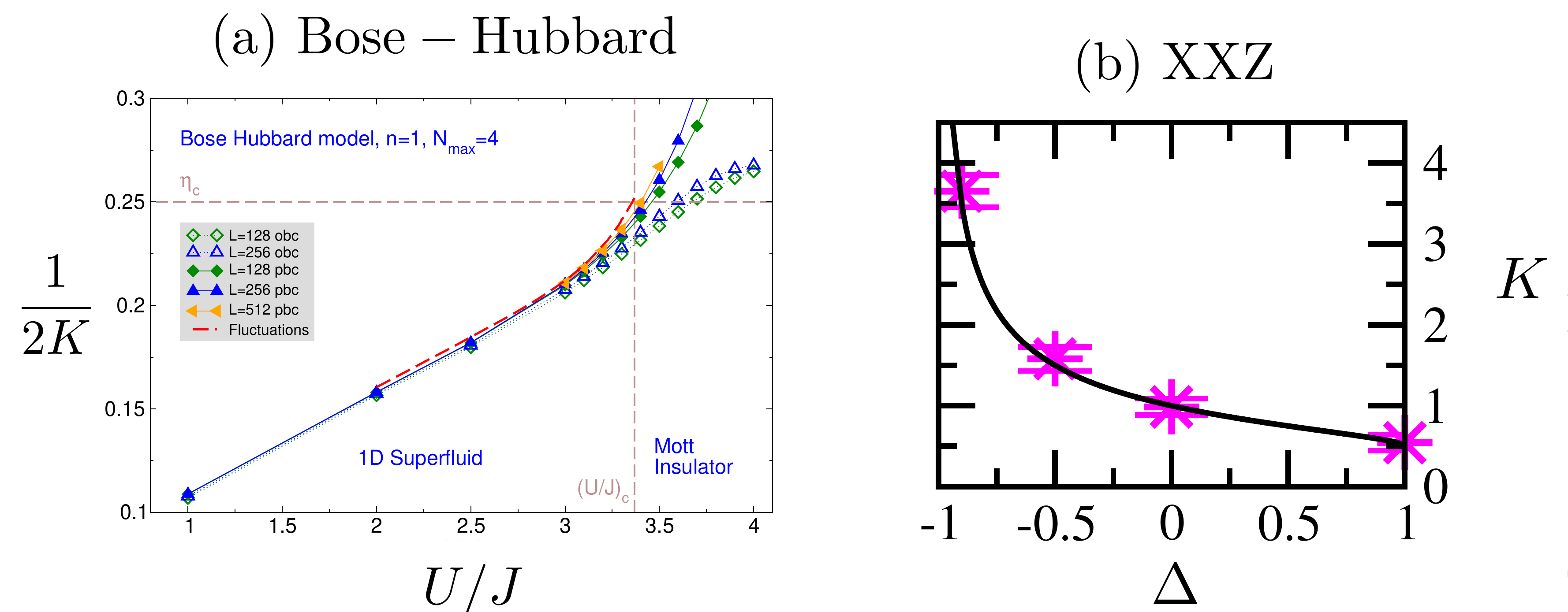}
    \caption{Luttinger $K$ parameter estimated from the curvature of the low-energy part of the ES described by Eq.~\eqref{eq:Sm}. (a) Bose-Hubbard chains at integer filling, reprinted from~\cite{lauchli_operator_2013}. (b) $s=1/2$ XXZ chains compared to exact Bethe ansatz (full line) Eq.~\eqref{eq:K}, reprinted from~\cite{laflorencie_spin-resolved_2014}.}
    \label{fig:K0}
\end{figure}

The CFT description is also encoded in the microscopic structure of the entanglement levels, as discussed in Refs.~\cite{lauchli_operator_2013,laflorencie_spin-resolved_2014}. Indeed, for critical XXZ~\cite{lauchli_operator_2013,laflorencie_spin-resolved_2014}, as well as for critical Bose-Hubbard chains~\cite{deng_entanglement_2011,lauchli_operator_2013}, the low lying part of the ES for subsystems of length $\ell_A$ corresponds to the energy spectrum of open chains of length $\ell_A$ with the same Luttinger parameter as the full system. This is shown in Fig.~\ref{fig:esD} (b) for XXZ chains at $\Delta=-0.5$ where the low "energy" entanglement levels are perfectly described by the CFT prediction~\cite{alcaraz_surface_1987} for the low-energy spectrum of a critical open chain of $\ell_A$ sites
\be
E_0^m-E_0^0=\frac{\pi u}{2K\ell_A}m^2,
\label{eq:Em}
\ee
where $m$ is the $S^z$ quantum number of the subsystem, $E_0^m$ is the ground-state energy in a given magnetization sector $m$, $u$ the velocity of excitations, and $K$ the Luttinger parameter. The entanglement levels in Fig.~\ref{fig:esD} can be identified with Eq.~\eqref{eq:Em} using the correct "entanglement temperature"~\cite{laflorencie_spin-resolved_2014}  (see also below in Section~\ref{sec:eh}), thus yielding
\be
-\ln\left(\lambda^{(m)}_{i}/\lambda^{(m)}_{\rm max}\right)=\frac{\pi^2}{K\ln(\ell_A/\ell_0)}m^2,
\label{eq:Sm}
\ee
where $\ell_0$ is a length scale of order 1. 
As a consistency check, one can extract the Luttinger parameter $K$ from the quadratic enveloppe Eq.~\eqref{eq:Sm}, which is shown in Fig.~\ref{fig:K0} (a) for Bose-Hubbard chains~\cite{lauchli_operator_2013}, and (b) for XXZ chains~\cite{laflorencie_spin-resolved_2014}.

One should also mention some earlier results obtained by Peschel~\cite{peschel_reduced_2004} on free electronic chains which suggest that the effective Hamiltonian is an open free fermion chain, but with non-homogeneous hopping terms vanishing at the boundaries. Such microscopic details should not change the above picture for quantum critical chains. However, in the gapped Ising regime of the XXZ chain ($\Delta\gg 1$), the boundary-local nature of the ES has been clearly identified~\cite{alba_boundary-locality_2012}. Also one could interpret the inhomogeneity in a local thermodynamic with a spatially varying local temperature~\cite{wong_entanglement_2013,swingle_area_2016} decaying away from the boundary.

In Refs.~\cite{thomale_nonlocal_2010,de_chiara_entanglement_2012,lepori_scaling_2013,giampaolo_universal_2013}, the entanglement gap between the first two largest eingenvalues (the Schmidt gap) has been diagnosed as an order parameter to locate quantum phase transitions for quantum spin chain models, an idea further applied to quantum impurity problems~\cite{bayat_order_2014}. One should however note that universal features captured by entanglement spectroscopy have been recently questioned when looking at momentum-space entanglement~\cite{balasubramanian_momentum-space_2012,lundgren_momentum-space_2014} where the closure of the Schmidt gap may not occur at the physical critical point. These second thoughts regarding universality are also supported by the recent discussion in Ref.~\cite{chandran_how_2014}.

\paragraph{Ladders---} 
\begin{figure}[t]
    \centering
    \includegraphics[width=.7\columnwidth,clip]{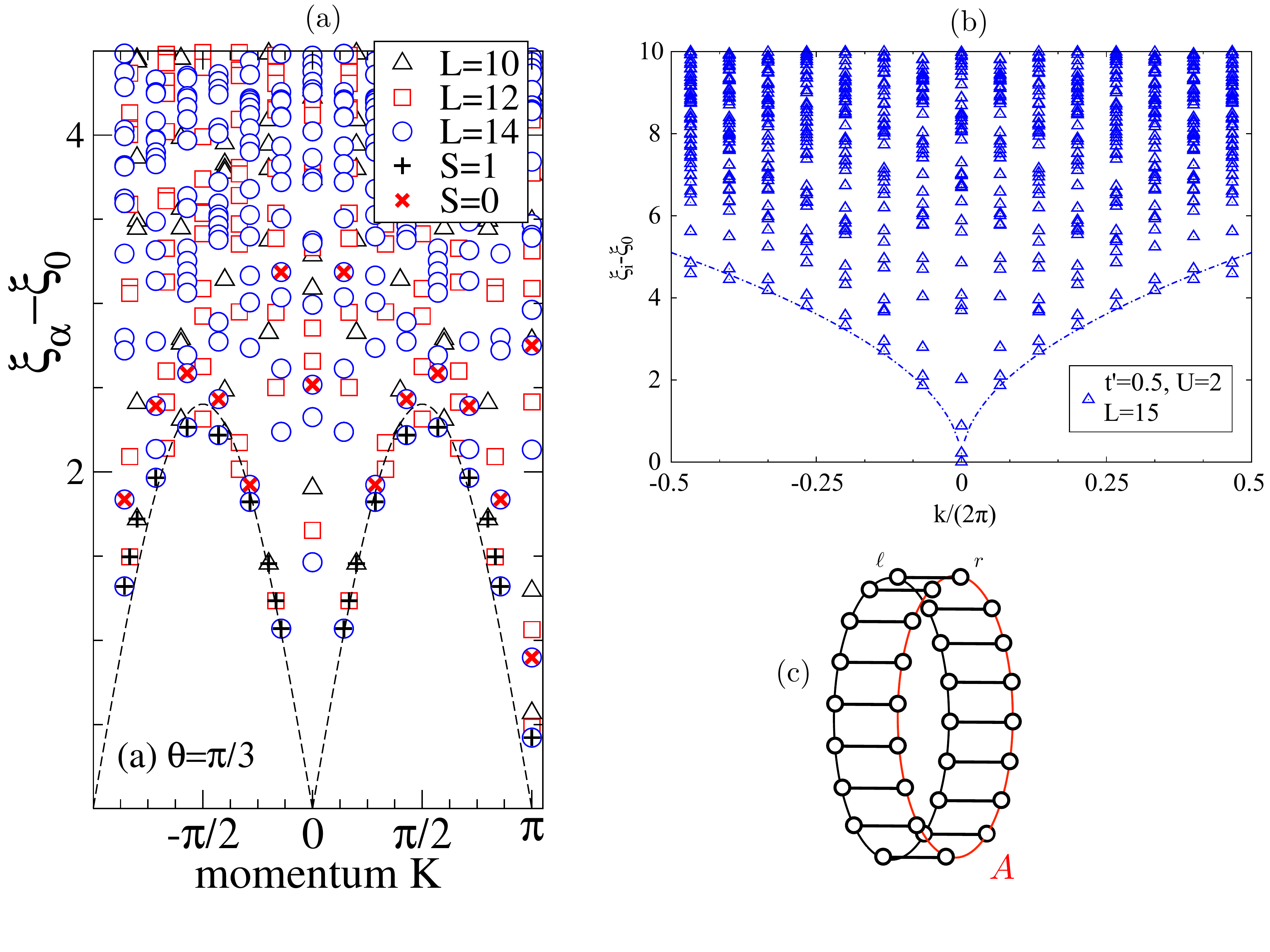}
    \caption{Exact diagonalization results for the momentum-resolved ES of 2-leg ladders, obtained after a real-space bipartition with one leg as susbsystem $A$ (c).
    (a) $s=1/2$ Heisenberg model which has a gapped ground-state, the dashed line being the de Cloiseaux - Pearson gapless dispersion (reprinted from~\cite{poilblanc_entanglement_2010}). (b) Interacting hard-core boson model with a single gapless mode, the dashed line being a $\sqrt{k}$ dispersion (reprinted from~\cite{lundgren_entanglement_2013}).}
    \label{fig:es_ladders}
\end{figure}
Ladder materials, originally introduced to study two-dimensional systems from a quasi-$d=1$ perspective~\cite{dagotto_surprises_1996} are now well recognized to harbour many fascinating phenomena, and numerous experimental achievements~\cite{abbamonte_crystallization_2004,klanjsek_controlling_2008,ruegg_thermodynamics_2008}. A paradigmatic example is the two-leg ladder geometry (depicted in panel (c) of Fig.~\ref{fig:es_ladders}) for which the antiferromagnetic Heisenberg model presents a singlet ground-state separated from the first excited triplet state by a finite energy gap. Neverthless, as first shown by Poilblanc~\cite{poilblanc_entanglement_2010} by tracing out a single leg (subsystem $A$ in panel (c) of Fig.~\ref{fig:es_ladders}), the ES reveals an unexpected gapless feature when studied against the momentum quantum number $k$ of the chain. Visible in Fig.~\ref{fig:es_ladders} (a), the eigenvalues $\lambda$ of the RDM, plotted as $\xi=-\ln \lambda$ against the momentum $k$ show a "low-energy" structure similar to the gapless des Cloizeaux - Pearson spectrum of a single Heisenberg chain~\cite{des_cloizeaux_spin-wave_1962}.
It is quite remarkable that the ES appears to reflect the low-energy spectrum of each individual edge. This observation has then triggered further extensions and studies of quantum ladder systems
\cite{cirac_entanglement_2011,peschel_relation_2011,furukawa_entanglement_2011,lauchli_entanglement_2012,schliemann_entanglement_2012,lundgren_entanglement_2013,chen_quantum_2013,luitz_improving_2014}. In particular, the authors of Ref.~\cite{lundgren_entanglement_2013} analyzed in details, both analytically and numerically, the case of an interacting two-leg ladder of hard-core bosons with one gapless mode. The resulting momentum-resolved ES, displayed in Fig.~\ref{fig:es_ladders} (b), features a $\sqrt{k}$ low-energy dispersion, in contrast with the gapped ladder case~\cite{poilblanc_entanglement_2010,cirac_entanglement_2011,lauchli_entanglement_2012,schliemann_entanglement_2012}, thus reflecting the long-range nature of the boundary Hamiltonian (see also below Section~\ref{sec:eh}).

\paragraph{Quantum Monte Carlo approaches for entanglement spectroscopy---}
While exact diagonalization, DMRG, and tensor network approaches have been the most useful numerical techniques able to access ES of strongly correlated quantum systems, very recent developments have been made using QMC methods~\cite{assaad_entanglement_2014,chung_entanglement_2014,luitz_improving_2014,assaad_stable_2015}. For fermionic systems, Assaad and co-workers proposed a method based on the computation of spectral functions along the imaginary time~\cite{assaad_entanglement_2014,assaad_stable_2015}. Another route was suggested for bosonic and quantum spin models~\cite{chung_entanglement_2014,luitz_improving_2014}, using the fact that one can reconstruct the ES from higher-order R\'enyi entropies. This method, introduced in Ref.~\cite{song_bipartite_2012}, and implemented in a QMC framework for Bose-Hubbard chains~\cite{chung_entanglement_2014} and quantum spin ladders~\cite{luitz_improving_2014}, relies on the Newton-Girard identities, which links the coefficients of a polynomial to the
   power sums of its roots.

   \begin{figure}
       \centering
       \includegraphics[width=.8\columnwidth,clip]{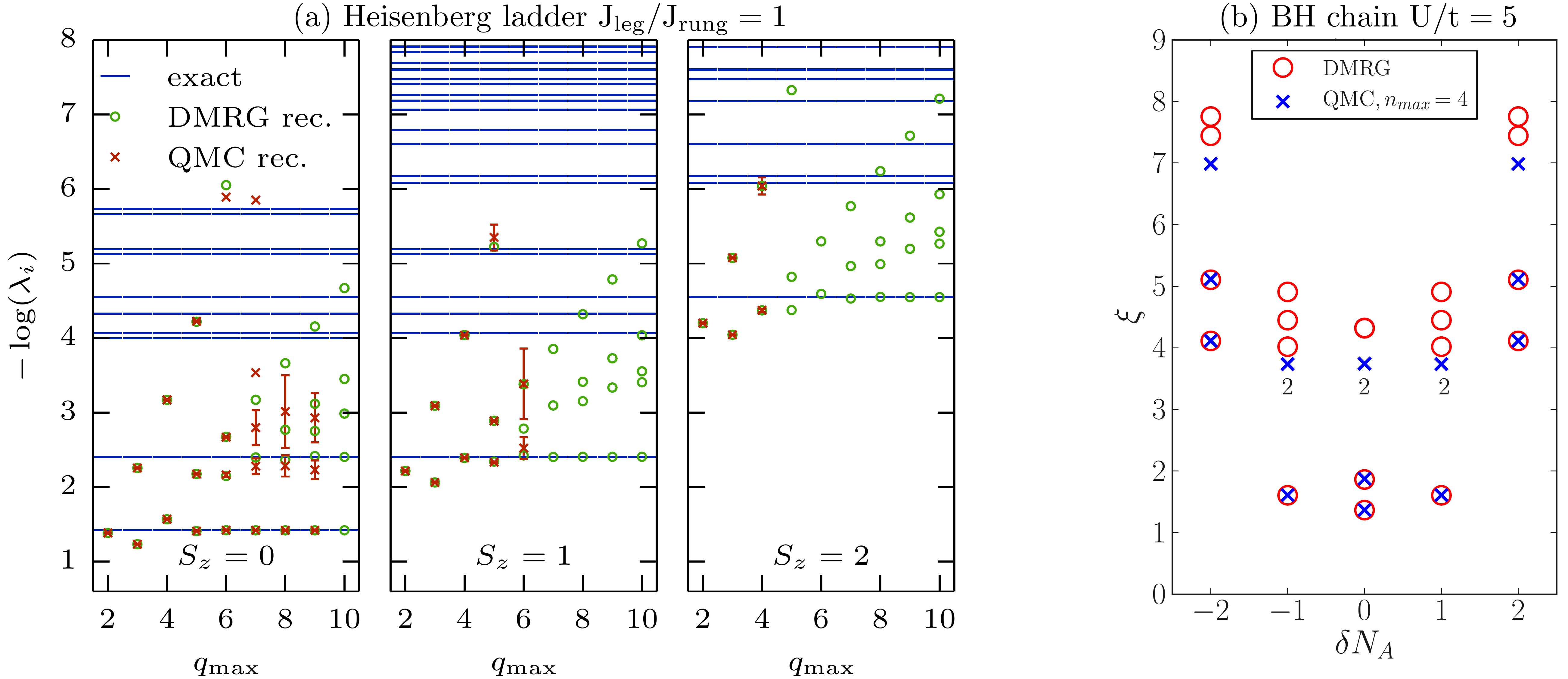}
       \caption{Reconstructed ES from QMC data. (a) Results for the $2\times 10$ $s=1/2$ Heisenberg ladder with $J_{\rm leg}=J_{\rm rung}$ plotted against $q_{\rm{max}}$, and separated in spin sectors $S_z$. Exact and reconstructed DMRG data are also shown for comparison. Reprinted from~\cite{luitz_improving_2014}. (b) Results for a $L=48$ extented Bose - Hubbard chain at unit filling  in the Mott insulating phase. Reprinted from~\cite{chung_entanglement_2014}.}
       \label{fig:ladderspec}
   \end{figure}

   However, in any practical QMC simulation, the knowledge of R\'enyi entanglement entropies is limited to some finite precison, and to not too large value of the R\'enyi parameter $q\le q_{\rm max}$.
   For the extended
   Bose-Hubbard chain at unit filling, Chung {\it{et al.}}~\cite{chung_entanglement_2014} have been able reconstruct the low lying levels of the ES using $q_{\rm{max}}=4$, as plotted in Fig.~\ref{fig:ladderspec} (b). In Ref.~\cite{luitz_improving_2014} we have studied the isotropic $s=1/2$ Heisenberg ladder with $J_{\rm leg}=J_{\rm rung}$ up to $q_{\rm max}=9$. QMC results are shown in Fig.~\ref{fig:ladderspec} (a) for a $2\times 10$ system (with the same bipartition as discussed previously in Fig.~\ref{fig:es_ladders}), and compared to exact DMRG results, as well as to the same reconstruction trick using DMRG data for $S_{q\le 10}$.
   The convergence of the lowest entanglement level (in the
   $S_z=0$ sector) is very good, thus providing an efficient estimate of the
   single copy entanglement $S_\infty=-\ln \lambda_{\rm max}$~\cite{eisert_single-copy_2005,peschel_single-copy_2005} 
   for which no direct QMC estimate is available, as it would rely on a Monte Carlo sampling of an infinite number of replicas. 
   Nevertheless, despite the numerical gain using the spin-resolved structure of the RDM, we clearly see the limitations of this technique to access higher entanglement levels, even for exact DMRG data, while the reconstructed QMC spectrum carries additional errors due
   to the statistical uncertainty of the R\'enyi entanglement entropies.

Before switching to $d=2$, one should mention an alternative spectroscopic tool relying on the participation spectrum~\cite{luitz_shannon-renyi_2014,luitz_participation_2014}, which can be computed very efficiently within a QMC framework. This will be discussed below, when addressing the issue of entanglement Hamiltonian in Section~\ref{sec:eh}.

\subsubsection{Conventional ordered and gapped states in $d> 1$}
\paragraph{Gapped phases---}
Accessing the ES of quantum interacting systems beyond $d=1$ is a challenging task. While entanglement spectroscopy was popularised after the work of Li and Haldane~\cite{li_entanglement_2008} as a smoking gun of topological order for $d=2$ non-abelian fractional quantum Hall states (see below in Section~\ref{sec:esto}), somehow surprisingly, less is known regarding traditional $d=2$ phases of matter, {\it{e.g.}} symmetry breaking states or conventional quantum critical points. Indeed, 
only a few analytical works are available, exploring for example ES for broken continuous symmetry states~\cite{metlitski_entanglement_2011}, complex paired superfluids~\cite{dubail_entanglement_2011}, valence-bond solid states~\cite{santos_bulk-edge_2013}, or the quantum Ising model~\cite{chandran_how_2014}. On the numerical side, most of the simulation results have been obtained using diagonalization techniques~\cite{lou_entanglement_2011}, tensor network approaches~\cite{cirac_entanglement_2011,pizorn_tree_2013}, or $d=2$ DMRG~\cite{james_understanding_2013,alba_entanglement_2013,kolley_entanglement_2013}.

   \begin{figure}[b]
       \centering
       \includegraphics[width=.6\columnwidth,clip]{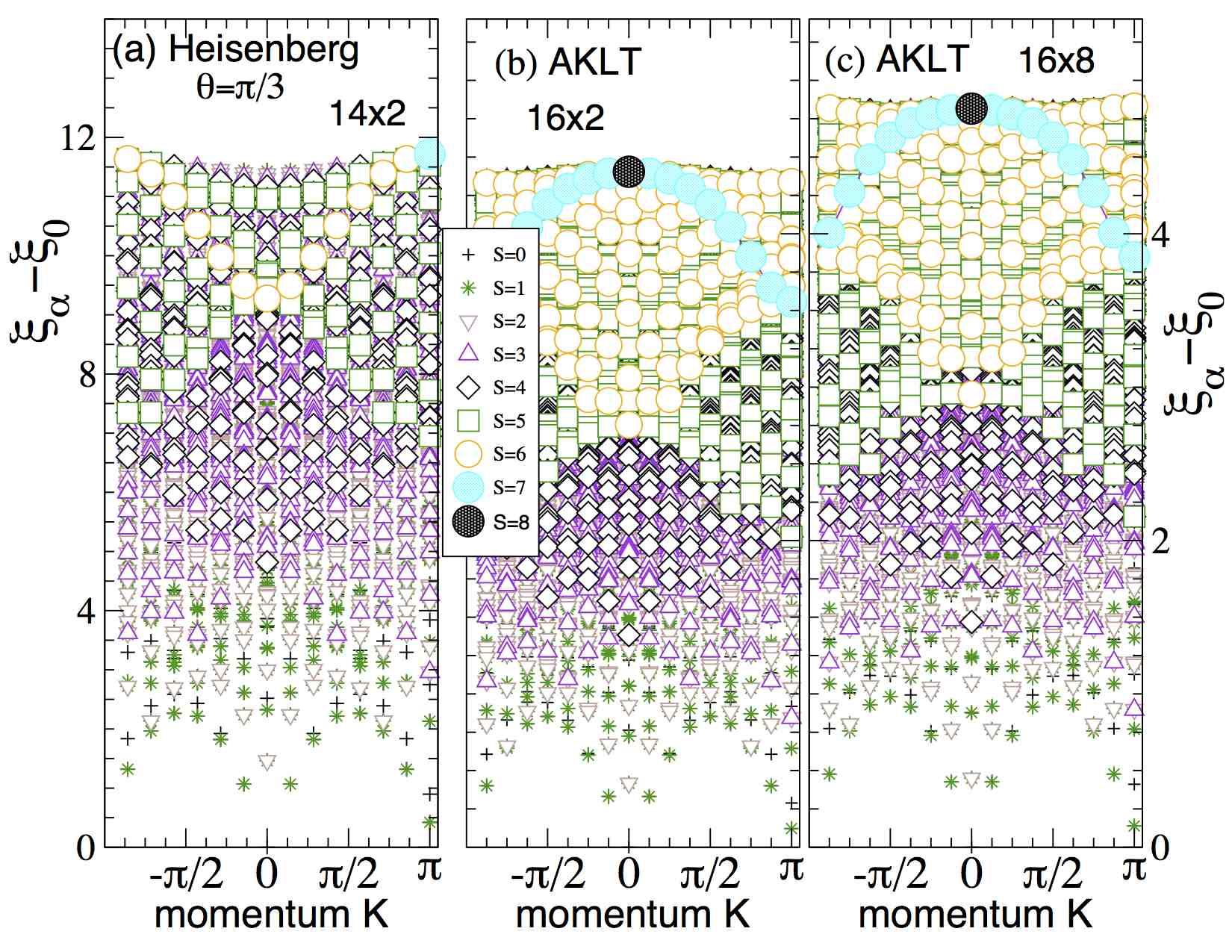}
       \caption{ES $\xi_\alpha=-\ln\lambda_\alpha$ of Heisenebrg and AKLT ladders (a,b) and $d=2$ AKLT cylinder (c) shown against the total momentum K along the  direction of the cut. Eigenvalues are labelled according to their total spin quantum number using different symbols displayed on the plots. Figure reprinted from Ref.~\cite{cirac_entanglement_2011}.}
       \label{fig:AKLT_PEPS}
   \end{figure}

For gapped phases, the structure of the ES reflects the physics at the boundary, similarly to the Heisenberg ladder case~\cite{poilblanc_entanglement_2010} discussed above, with a short-range entanglement Hamiltonian following a correpondence between the ES and the edge physics~\cite{cirac_entanglement_2011,santos_bulk-edge_2013}. This is illustrated in Fig.~\ref{fig:AKLT_PEPS} from Ref.~\cite{cirac_entanglement_2011} where one sees that the AKLT state~\cite{affleck_rigorous_1987} features the same ES in $d=2$ and on a 2-leg ladder.

\begin{figure}
   \centering
       \includegraphics[width=.6\columnwidth,clip]{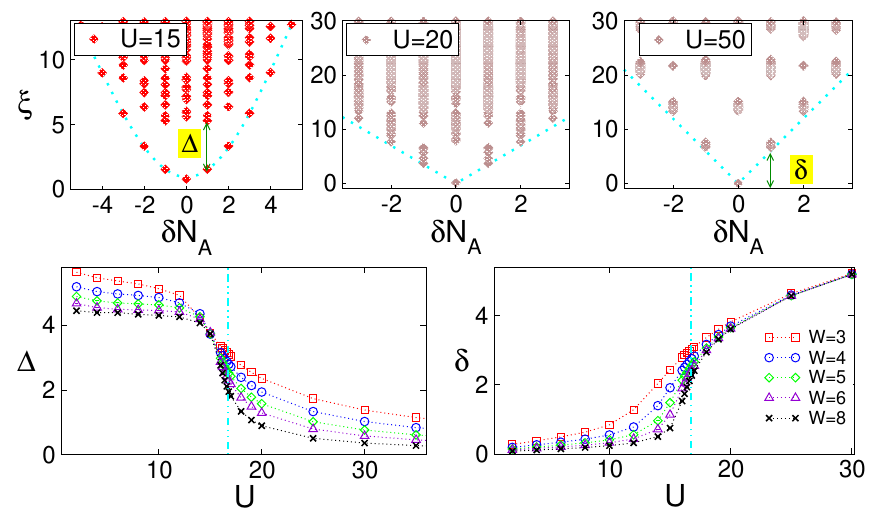}
       \caption{DMRG results for the $d=2$ Bose-Hubbard model from Alba {\it{et al.}}~\cite{alba_entanglement_2013}. Top: ES across the superfluid - Mott insulator transition. Bottom: $\Delta$ and $\delta$ (defined in top panels) plotted against the on-site repulsion $U$ for several values of the boundary length $W$. The vertical line denotes the critical repulsion $U_c\simeq 16.74$. Reprinted from Ref.~\cite{alba_entanglement_2013}}
       \label{fig:BH2D}
   \end{figure}

For the $d=2$ Bose-Hubbard model studied by Alba {\it{et al.}}~\cite{alba_entanglement_2013}, in the gapped Mott insulating phase at large on-site repulsion, the ES acquires the boundary structure of a simple tight-biding chain with short-range hopping of the excess particles on top of the insulating state, (see Fig.~\ref{fig:BH2D}).

\paragraph{Long-range order---}
\label{sec:ESLRO}
The case of long-range ordered states where a continuous symmetry is broken, such as the U(1) for a superfluid~\cite{alba_entanglement_2013} or SU(2) for N\'eel ordered antiferromagnets~\cite{kolley_entanglement_2013} is more subtle. Indeed, the ES has to reflect the broken symmetry and therefore cannot be a simple edge spectrum of an effective model with short-range interactions like in the gapped case discussed previously. As suggested by the field theory approach of Metlitski and Grover~\cite{metlitski_entanglement_2011}, and observed numerically using DMRG for the superfluid regime of the Bose-Hubbard on the square lattice~\cite{alba_entanglement_2013} and the N\'eel phase of the Heisenberg model on various lattices~\cite{kolley_entanglement_2013}, the ES has to contain both Anderson TOS~\cite{anderson_approximate_1952} and oscillators (SW) structures. This is indeed what we can see for the U(1) broken superfluid in Fig.~\ref{fig:BH2D} where the TOS yields a quadratic envelope in term of the particle number $\sim \delta\times (\delta N_A)^2$, with a "TOS gap" $\delta$ vanishing with the linear size $W$ of the subsystem as $1/W$ (with possible logarithmic corrections~\cite{metlitski_entanglement_2011}). The SW part has a finite gap $\Delta$ (or possibly slowly vanishing $\sim 1/\ln W$~\cite{metlitski_entanglement_2011}) in the ordered regime. As we discuss more below in Section~\ref{sec:eh}, when comparing ES to the physical spectrum of a given "entanglement Hamiltonian", one should be careful about the entanglement temperature which renormalizes the ES, here with a factor $T_{\rm eff}\sim 1/W$.

\subsection{Entanglement spectroscopy of topological order}
\label{sec:esto}
Quantum states of matter which exhibit topological properties, {\it{e.g.}} fractional quantum Hall effect (FQHE)~\cite{tsui_two-dimensional_1982}, toplogical insulators~\cite{hasan_textitcolloquium_2010}, chiral or $\mathbb Z_2$ spin liquids~\cite{balents_spin_2010}, AKLT states~\cite{affleck_rigorous_1987}\ldots~have been conjectured by Li and Haldane~\cite{li_entanglement_2008} to display egde excitations in their ground-state ES. This idea has then been extensively pursued for 
FQHE~\cite{regnault_topological_2009,zozulya_entanglement_2009,lauchli_disentangling_2010,thomale_entanglement_2010,schliemann_entanglement_2011,hermanns_haldane_2011,chandran_bulk-edge_2011,qi_general_2012,liu_edge-mode_2012,sterdyniak_real-space_2012,
dubail_real-space_2012,rodriguez_evaluation_2012,dubail_edge-state_2012,liu_fractional_2013,zaletel_topological_2013,hsieh_bulk_2014,mong_fibonacci_2015,he_bosonic_2015,regnault_entanglement_2015}, 
topological insulators~\cite{ryu_entanglement_2006,turner_entanglement_2010,fidkowski_entanglement_2010,fidkowski_topological_2011,alexandradinata_trace_2011,regnault_fractional_2011,turner_topological_2011,fang_entanglement_2013,hermanns_entanglement_2014}, 
quantum spin liquids~\cite{poilblanc_simplex_2013,he_chiral_2014,bauer_chiral_2014,gong_emergent_2014,poilblanc_chiral_2015}, 
valence bond spin states~\cite{pollmann_entanglement_2010-1,lou_entanglement_2011,santos_entanglement_2012,santos_entanglement_2012-1,huang_topological_2011,santos_bulk-edge_2013,santos_bulk_2015}, 
or the toric code~\cite{yao_entanglement_2010,cirac_entanglement_2011}. While this field is still relatively young, there has been an impressive number of works since Li and Haldane proposal. Therefore, providing an exhaustive review of such recent developments is clearly out of the scope of the present paper. Below we try to give a short overview, focusing on a few aspects such as the identification of topological order through the ES, the edge-ES correspondence, and the recent advances in topological properties of quantum spin liquids.
\subsubsection{Identify topological order with entanglement spectra}

%
\begin{figure}[b]
\begin{center}
\includegraphics[width=.48\columnwidth,clip]{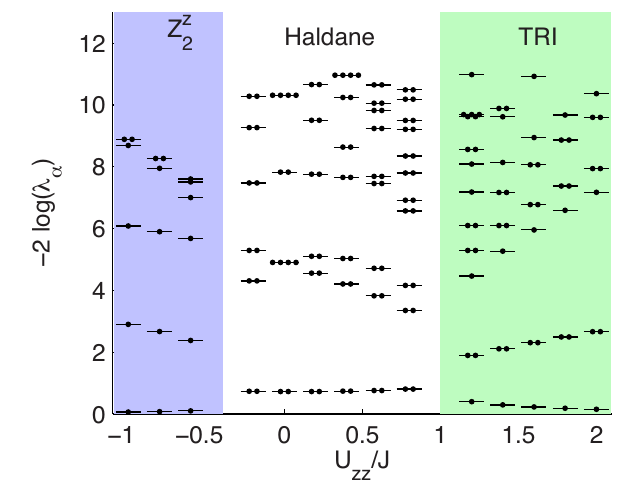}
\caption{Lower part of the entanglement spectrum of the spin $s=1$ chain model Eq.~\eqref{eq:pollmann}. The multiple dots show the degeneracy of the eigenvalues, even in the Haldane regime, and odd otherwise.
Reprinted from Ref.~\cite{pollmann_entanglement_2010-1}.}
\label{fig:haldane}
\end{center}
\end{figure}

Before discussing FQHE, one can first illustrate the power of entanglement spectroscopy for characterizing non-trivial states, such as the Haldane phase of open spin $s=1$ Heisenberg chains, governed by
\be
{\cal{H}}=J\sum_{i=1}^{L-1}{\vec{S}}_i\cdot{\vec{S}}_{i+1} +U_{zz}\sum_{i=1}^{L}\left(S_i^z\right)^2 +\cdots
\label{eq:pollmann}
\ee
For $U_{zz}/J\in [-0.4,1]$ the ground-state is connected to its parent VBS phase of the AKLT model~\cite{affleck_rigorous_1987}, and is a "symmetry protected topological phase", as also found in Bose-Hubbard
\cite{dalla_torre_hidden_2006} or multicomponent fermionic chains~\cite{nonne_haldane_2010}. Note that such VBS phase was previously addressed from an entanglement point of view using the concept of localizable entanglement~\cite{verstraete_entanglement_2004,verstraete_diverging_2004,popp_localizable_2005}.

Despite the lack of local order parameter, and the fragility  regarding small perturbations of the string (non-local) order and of the edge states, Pollmann and co-workers have shown~\cite{pollmann_entanglement_2010-1} 
that provided an appropriate set of symmetries remains preserved, the Haldane phase is stable. The signature of this topological state has to be read in the lower part of the ES, as visible in Fig.~\ref{fig:haldane} where one sees 2 degenerate non-zero eigenvalues, mimicking the edge spectrum of a system with a true physical boundary. Such a stability can be used as an operational definition of the topologically protected Haldane phase~\cite{pollmann_symmetry_2012}. Note this was previously discussed by Ryu and Hatsugai~\cite{ryu_entanglement_2006} for a Chern insulator.

\begin{figure}[h]
\begin{center}
\includegraphics[width=.785\columnwidth,clip]{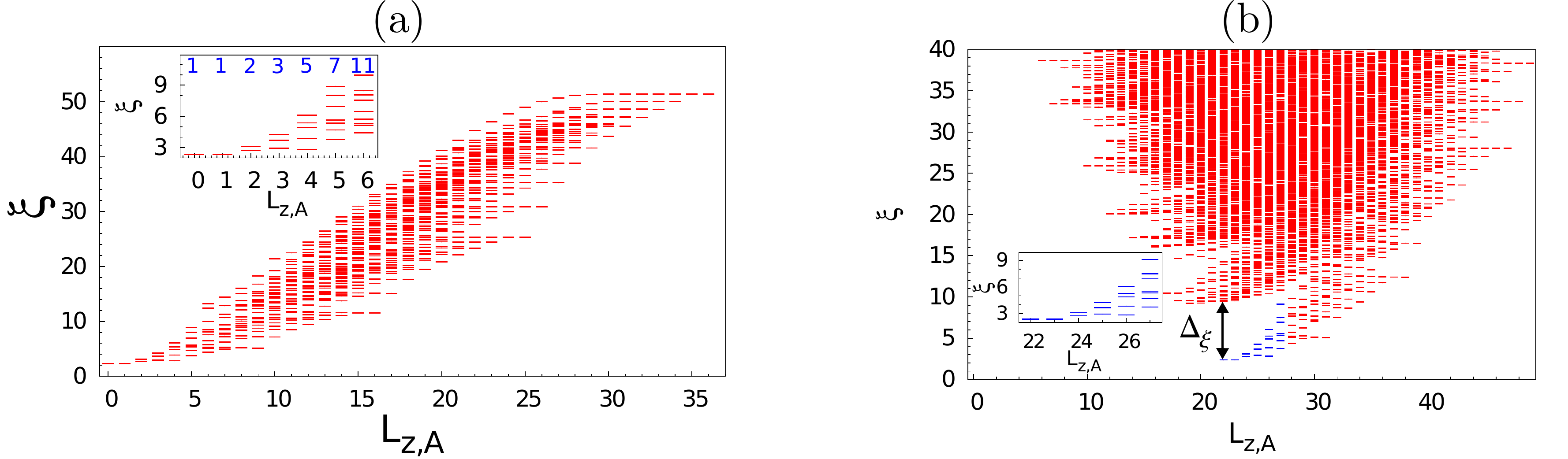}
\caption{Orbital ES for (a) the Laughlin wave-function at $\nu=1/3$ and (b) the Coulomb interaction, both on a sphere geometry. Results from Regnault~\cite{regnault_entanglement_2013} obtained with 12 fermions (in the sector with 6 particles), keeping 17 orbitals. The inset (a) demonstrates the direct correspondence between the lowest part of the spectrum and the chiral U(1) bosonic edge mode, with the correct counting 1, 1, 2, 3, 5, 7, 11. In (b) the blue levels are the edge modes similar to (a). The entanglement gap $\Delta_\xi$ separates the universal edge part from the non-universal high eenergy part. Reprinted from~\cite{regnault_entanglement_2013}.}
\label{fig:regnault}
\end{center}
\end{figure}

For the FQHE Li and Haldane, building on Ref.~\cite{haque_entanglement_2007}, suggested~\cite{li_entanglement_2008} an orbital bipartition which, while not equivalent, may imitate a real space cut. They conjectured that such an orbital ES should reflect the edge mode. This has effectively been observed in several cases, either for Laughlin wave-functions describing $\nu=1/m$ states, the Moore-Read state at $\nu=5/2$, and also beyond model wave-functions through more realistic Hamiltonians including Coulomb interaction. In Fig.~\ref{fig:regnault} we illustrate this with the result obtained by Regnault~\cite{regnault_entanglement_2013} for the Laughlin model wave-function and the Coulomb case for $\nu=1/3$ which both display the correct U(1) edge mode counting. For the more realistic spectrum obtained with Coulomb interactions (panel (b) of Fig.~\ref{fig:regnault}), the universal structure, similar to the Laughling model state, is protected from the non-universal high-energy part by a finite entanglement gap $\Delta_\xi$. The universal part is described by CFT~\cite{li_entanglement_2008}. When approaching quantum Hall phase transitions, this entanglement gap has been found to vanish~\cite{zozulya_entanglement_2009,thomale_entanglement_2010,liu_entanglement_2011}.

\subsubsection{Correspondence between edge and entanglement spectra}

A natural question which immediately arises concerns the universal character of the correspondence between the low lying part of the ES and the edge states, as posed by Chandran and co-workers~\cite{chandran_bulk-edge_2011} for several FQH states. They found a direct correspondence between mode countings of particle ES, the orbital ES studied by Li and Haldane~\cite{li_entanglement_2008}, and 
bulk quasi-holes using CFT. The latter being equal to the counting of edge modes at physical hard-cut on the sample, they interpret their results as a bulk-edge correspondence~\cite{halperin_quantized_1982} for the ES.

\begin{figure}[b]
\begin{center}
\includegraphics[width=.57\columnwidth,clip]{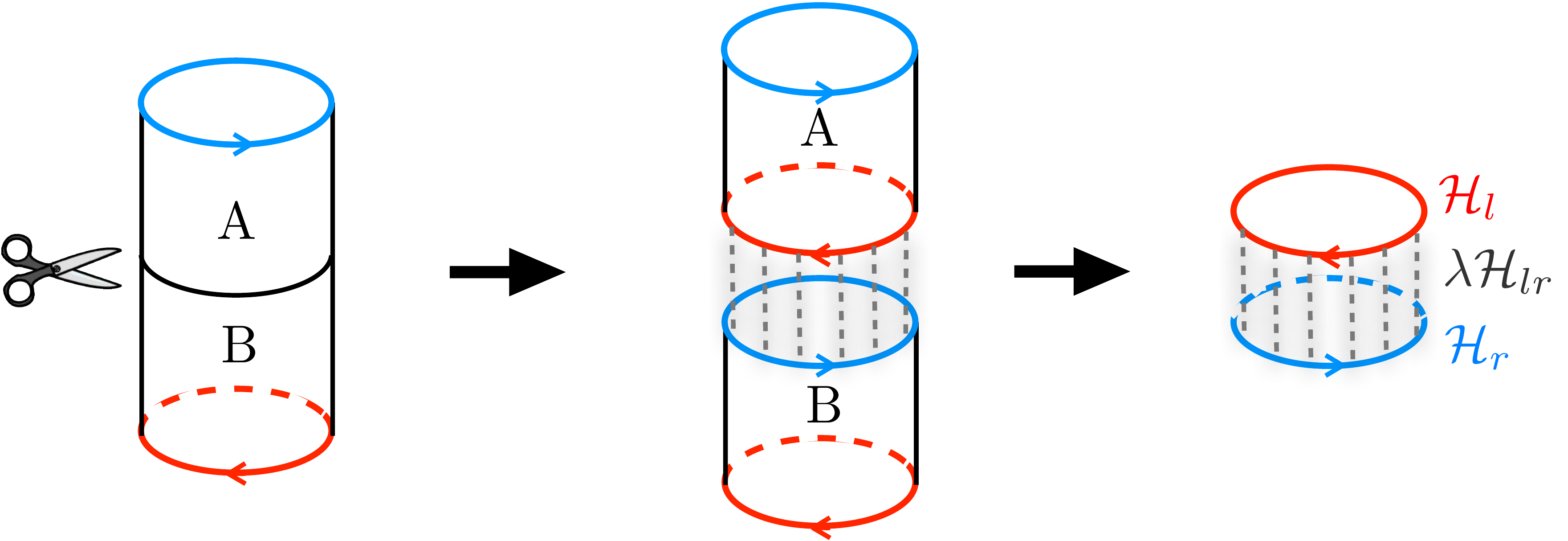}
\caption{ÒCut and glueÓ approach~\cite{qi_general_2012} applied to a quantum Hall state on a cylinder. The system is cut into A and B, obtaining gapless modes described by chiral Luttinger liquids, ${\cal{H}}_r$ and ${\cal H}_l$ living on the new edges. We then glue them along the edges by switching on an interaction $\lambda{\cal H}_{lr}$ that couples A and B. The entanglement problem of the quantum Hall state can then be reduced to the problem of entanglement between the two coupled chiral Luttinger liquids~\cite{lundgren_entanglement_2013}.
Inspired from Ref.~\cite{qi_general_2012}.}
\label{fig:cutglue}
\end{center}
\end{figure}

Using CFT, a general proof for this correspondence has been proposed by Qi, Katsura and Ludwig~\cite{qi_general_2012} for $d=2+1$ topological states with $d=1+1$ chiral edges using the Òcut and glueÓ approach, as illustrated in Fig.~\ref{fig:cutglue}. Starting by cutting a cylinder in two subsystems A and B which both support gapless chiral edge states, they are then glued together along the edges by switching an interaction between A and B. The cut being considered as a sudden quantum quench, it was proved using boundary CFT that both chiral edge and entanglement Hamiltonians are equivalent. Let us also mention other proofs that have been proposed, also based on CFT~\cite{swingle_geometric_2012,dubail_edge-state_2012}.

From a numerical point of view, the tensor networks PEPS formalism~\cite{orus_practical_2014} gives new insights on this question by directly constructing an "holographic"~\cite{bousso_holographic_2002} correspondence between the ES and a boundary Hamiltonian, as initiated by Cirac and co-workers~\cite{cirac_entanglement_2011,schuch_topological_2013} where it was found that for a gapped bulk the ES corresponds to a short-range gapless Hamiltonian~\cite{lou_entanglement_2011,santos_bulk-edge_2013}. In the case of non-chiral topological phases where there are no protected
gapless edge modes, such as the
Kitaev toric code model which has ${\mathbb Z}_2$ topological order, it has been shown~\cite{ho_edge-entanglement_2015} that there is no such a correspondence between the ES and edge physics.
\subsubsection{Chiral spin liquid state}
Spin liquids~\cite{balents_spin_2010} are very good candidates for testing the correspondence between edge modes appearing along a physical cut and the entanglement modes along the virtual edge  after bipartition. Quantum spin liquid states have been intensively studied using entanglement spectroscopy~\cite{poilblanc_topological_2012,poilblanc_simplex_2013,he_chiral_2014,bauer_chiral_2014,gong_emergent_2014,poilblanc_chiral_2015}.

\begin{figure}[b]
\begin{center}
\includegraphics[width=.8\columnwidth,clip]{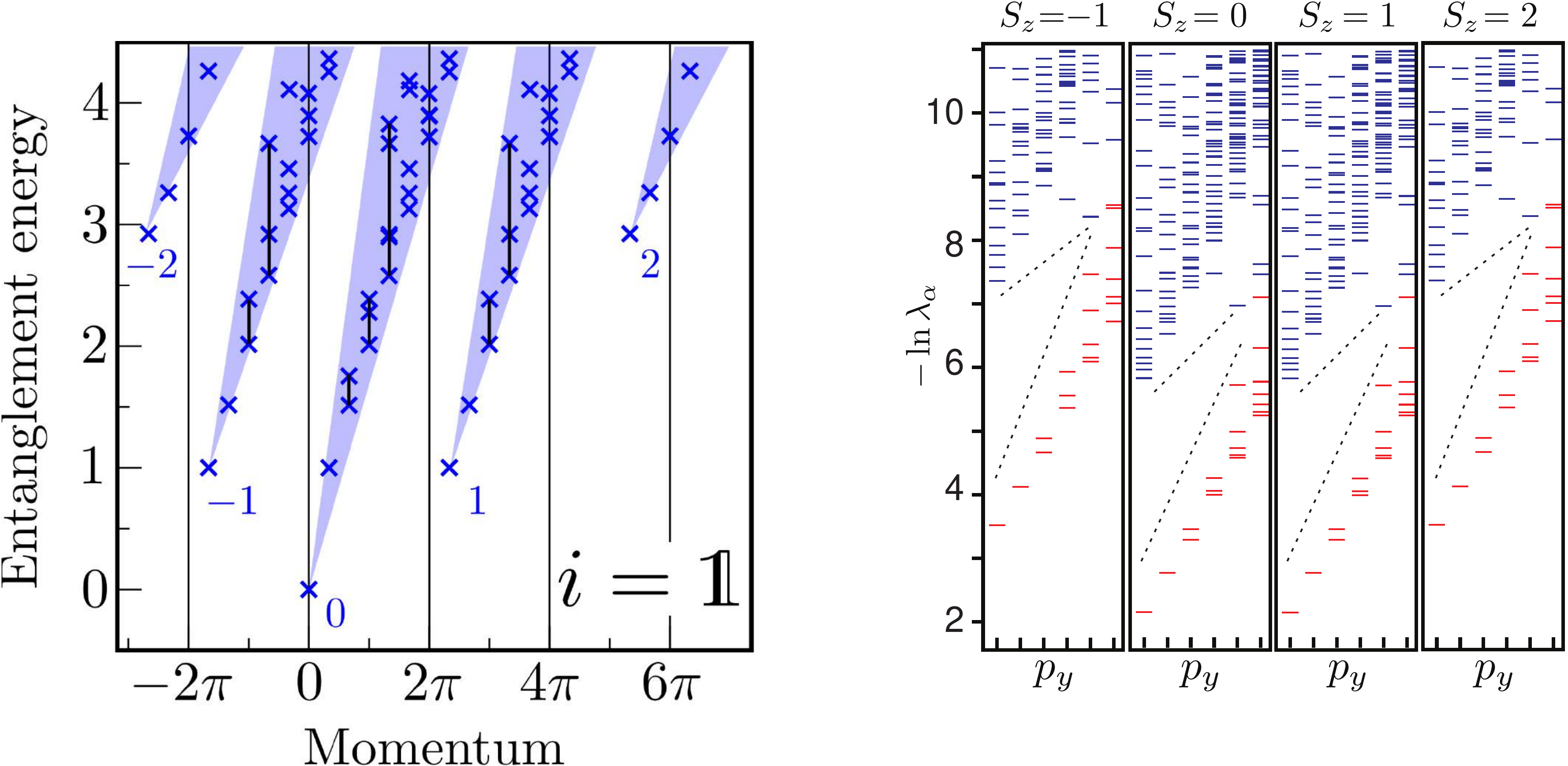}
\caption{ES of $s=1/2$ models on the kagom\'e lattice Eq.~\eqref{eq:chiralH} from Ref.~\cite{bauer_chiral_2014} (Left) and Eq.~\eqref{eq:J123} from Refs.~\cite{gong_emergent_2014,he_chiral_2014} (Right). Reprinted from Refs.~\cite{bauer_chiral_2014,he_chiral_2014}.}
\label{fig:escsl}
\end{center}
\end{figure}

Quite recently, the kagom\'e lattice has again revealed a very rich physics exhibiting chiral phases~\cite{messio_kagome_2012,messio_time_2013,he_chiral_2014,bauer_chiral_2014,gong_emergent_2014,poilblanc_chiral_2015,cincio_semion_2015,yang_chiral_2015,poilblanc_chiral_2015,bieri_gapless_2015,wietek_nature_2015}.
Indeed, two types of Heisenberg models have been found to realize the so-called chiral spin liquid state~\cite{kalmeyer_equivalence_1987,wen_chiral_1989}, expected to have time-reversal and parity symmetries (spontaneously or explicitly) broken, with a non-zero chiral order. The $J_1 - J_2 - J_3$ model
\be
{\cal H}=J_1\sum_{\langle i, j\rangle} {\vec{S}}_i\cdot{\vec{S}}_j\quad+ \quad J_2\sum_{\langle\langle i, j\rangle\rangle} {\vec{S}}_i\cdot{\vec{S}}_j\quad+\quad J_3\sum_{\langle\langle\langle i, j\rangle\rangle\rangle}{\vec{S}}_i\cdot{\vec{S}}_j,
\label{eq:J123}
\ee
which preserves both symmetries, studied in Refs.~\cite{gong_emergent_2014,he_chiral_2014,wietek_nature_2015}, and the $J - \chi$ model~\cite{bauer_chiral_2014,wietek_nature_2015}, which explicitly break them
\be
    {\cal H} = J \sum_{\langle i,j\rangle}\vec{S}_i \cdot \vec{S}_j \quad+\quad    {\chi}\sum_{i,j,k \in \bigtriangleup,\bigtriangledown} \vec{S}_i\cdot(\vec{S}_j\times \vec{S}_k).
      \label{eq:chiralH}
\ee
ES, plotted in Fig.~\ref{fig:escsl} for both models, unambiguously show the chiral nature of the edge excitations in momentum space.
%

\subsection{Entanglement Hamiltonian}
\label{sec:eh}
The notion of  entanglement Hamiltonian ${\cal H}_E$, simply defined by
\be
\rho_A=\exp(-{\cal H}_E),
\label{eq:HE}
\ee 
has been first discussed in the context of non-interacting systems~\cite{peschel_reduced_2009}, for which Peschel and Chung~\cite{peschel_density_1999,chung_density-matrix_2000,chung_density-matrix_2001} have shown that ${\cal H}_E$ has the same free-particle form as the original Hamiltonian. After Li and Haldane work~\cite{li_entanglement_2008} for fractional quantum Hall states, the idea that  ${\cal H}_E$ contains some universal properties of the virtual edge induced by the bipartition was 
vastly popularized for various condensed matter systems~\cite{poilblanc_entanglement_2010,peschel_relation_2011,lou_entanglement_2011,cirac_entanglement_2011,lauchli_entanglement_2012,schliemann_entanglement_2012,alba_entanglement_2013,chen_quantum_2013,kolley_entanglement_2013,lundgren_entanglement_2013,schliemann_entanglement_2014,luitz_participation_2014,hermanns_entanglement_2014,poilblanc_entanglement_2014,klich_entanglement_2015}. Below we discuss a few representative examples of gapped states, such as spin ladders or $d=2$ coupled dimers for which the entanglement Hamiltonian has a short-range nature. We also investigate gapless states where the situation is more involved when correlations get long ranged.
\subsubsection{Boundary theories for gapped phases}
\paragraph{Perturbative approach for spin ladders---}
After the observation~\cite{poilblanc_entanglement_2010,cirac_entanglement_2011} of the correspondence for (gapped) 2-leg spin ladders between the ES and the energy spectrum of a single (gapless) Heisenberg chain (see also Fig.~\ref{fig:es_ladders}), an analytical perturbative approach was proposed~\cite{peschel_relation_2011,lauchli_entanglement_2012,chen_quantum_2013} for a real space cut along the rungs (as depicted in the panel (c) of Fig.~\ref{fig:es_ladders}).  For a $s=1/2$ Heinsenberg ladder, governed by the Hamiltonian
\be
{\cal{H}}=\sum_{i} \Bigl(J_2 {\vec{S}}_{i,1}\cdot {\vec{S}}_{i,2}+\sum_{\ell=1,2} J_1{\vec{S}}_{i,\ell}\cdot{\vec{S}}_{i+1,\ell} \Bigr),
\ee
in the strong rung coupling limit $J_2\gg J_1$, the entanglement Hamiltonian derived using $2^{\rm nd}$ order perturbation is a $s=1/2$ unfrustrated short-range Heisenberg model\footnote{Only pair-wise interactions are considered here, while small multiple $2p$-spin interactions are generated at $p^{\rm th}$-order perturbation~\cite{cirac_entanglement_2011,lauchli_entanglement_2012}. For larger spins $s\ge 1$, the entanglement Hamiltonian is still well described by a spin-$s$ chain with a nearest-neighbor coupling independent of $s$, whereas longer range couplings do depend on $s$~\cite{schliemann_entanglement_2012}. For XXZ chains, the ${\cal H}_E$ has a renormalized Ising anisotropy~\cite{lauchli_entanglement_2012,schliemann_entanglement_2012}, for instance for $s=1/2$ $\Delta_{\rm eff}=\frac{\Delta}{2}(1+\Delta)$, which keeps the same value only for free-fermions $\Delta=0$ and Heisenberg $\Delta=1$ points.}
\be
{\cal{H}}_E=\frac{2J_1}{J_2}\sum_i {\vec{S}}_{i}\cdot{\vec{S}}_{i+1}  +\frac{1}{2}\left(\frac{J_1}{J_2}\right)^2\sum_i\left({\vec{S}}_{i}\cdot{\vec{S}}_{i+1}-{\vec{S}}_{i}\cdot{\vec{S}}_{i+2}\right)+O\left(\frac{J_1}{J_2}\right)^3.
\ee
One can infer the following generic unfrustrated form for the entanglement Hamiltonian, with exponentially decaying exchanges
\be
\rho_A=\frac{1}{Z}\exp\left(-\beta_{\rm eff}\sum_i\sum_r (-1)^{r-1}J_1{\rm e}^{-\frac{r-1}{\xi_E}}{\vec{S}}_{i}\cdot{\vec{S}}_{i+r}\right)
\label{eq:HE2}
\ee
The effective inverse temperature 
\be
\beta_{\rm eff}=\frac{2}{J_2}\left(1+\frac{J_1}{4J_2}\right),
\label{eq:beta_eff}
\ee
and $\xi_E$ is the length scale governing the coupling range. Based on $2^{\rm nd}$ order perturbation~\cite{lauchli_entanglement_2012,schliemann_entanglement_2012} we get
\be
\xi_E=\frac{1}{\ln(J_2/J_1)+\ln(4+J_1/J_2)}\propto \frac{1}{\ln(J_2/J_1)},\quad J_2\gg J_1,
\label{eq:xie}
\ee
in good agreement with the fact that pair-wise effective interactions at distance $r$ are generated at $r^{\rm th}$-order perturbation with amplitude $\sim (J_1/J_2)^r$~\cite{lauchli_entanglement_2012}. Interestingly this entanglement length $\xi_E$ is qualitatively different from the true correlation length of the spin ladder $\xi_{\rm corr}\sim J_1/J_2$~\cite{greven_monte_1996}. While multi-spin (beyond pair-wise) interactions may also play a role~\cite{cirac_entanglement_2011}, the above bilinear Heisenberg chain model Eq.~\eqref{eq:HE2} is already a very good approximation for the entanglement Hamiltonian in the rung singlet phase of the Heisenberg ladder, as we demonstrate below using QMC.

\paragraph{Quantum-thermal mapping---}
$T=0$ R\'enyi entanglement entropies can be exactly identified with thermal entropies (see for instance Ref.~\cite{casini_towards_2011} for an holographic approach) at temperature $\beta=1$ for the effective model ${\cal H}_E$ defined by Eq.~\eqref{eq:HE}. One can also interpret the RDM as a thermal density matrix at inverse temperature $\beta_{\rm eff}$ of an effective model whose energy scale is set to $J_1$ (the energy scale of the underlying Hamiltonian), as in Eq.~\eqref{eq:HE2}.
For a short-range interacting $s=1/2$ Heisenberg chain of $\ell$ sites, CFT predicts~\cite{affleck_universal_1986} for the low temperature regime $1\gg T/J_1\gg 1/\ell$ a thermal entropy $S_q^{\rm th}(T)=\frac{c}{3}(1+\frac{1}{q})\ell T/J_1$. Critical Luttinger liquids (Section~\ref{sec:es}) display an effective temperature which is vanishing with the subsystem size $\sim (\ln \ell)/\ell$~\cite{korepin_universality_2004,laflorencie_spin-resolved_2014}. For Heisenberg chains $T/J_1=\frac{1}{2}\ln(\ell/a)/\ell$, thus yielding the Calabrese-Cardy scaling Eq.~\eqref{eq:S1d}.

\begin{figure}[ht!]
\begin{center}
\includegraphics[width=.5\columnwidth,clip]{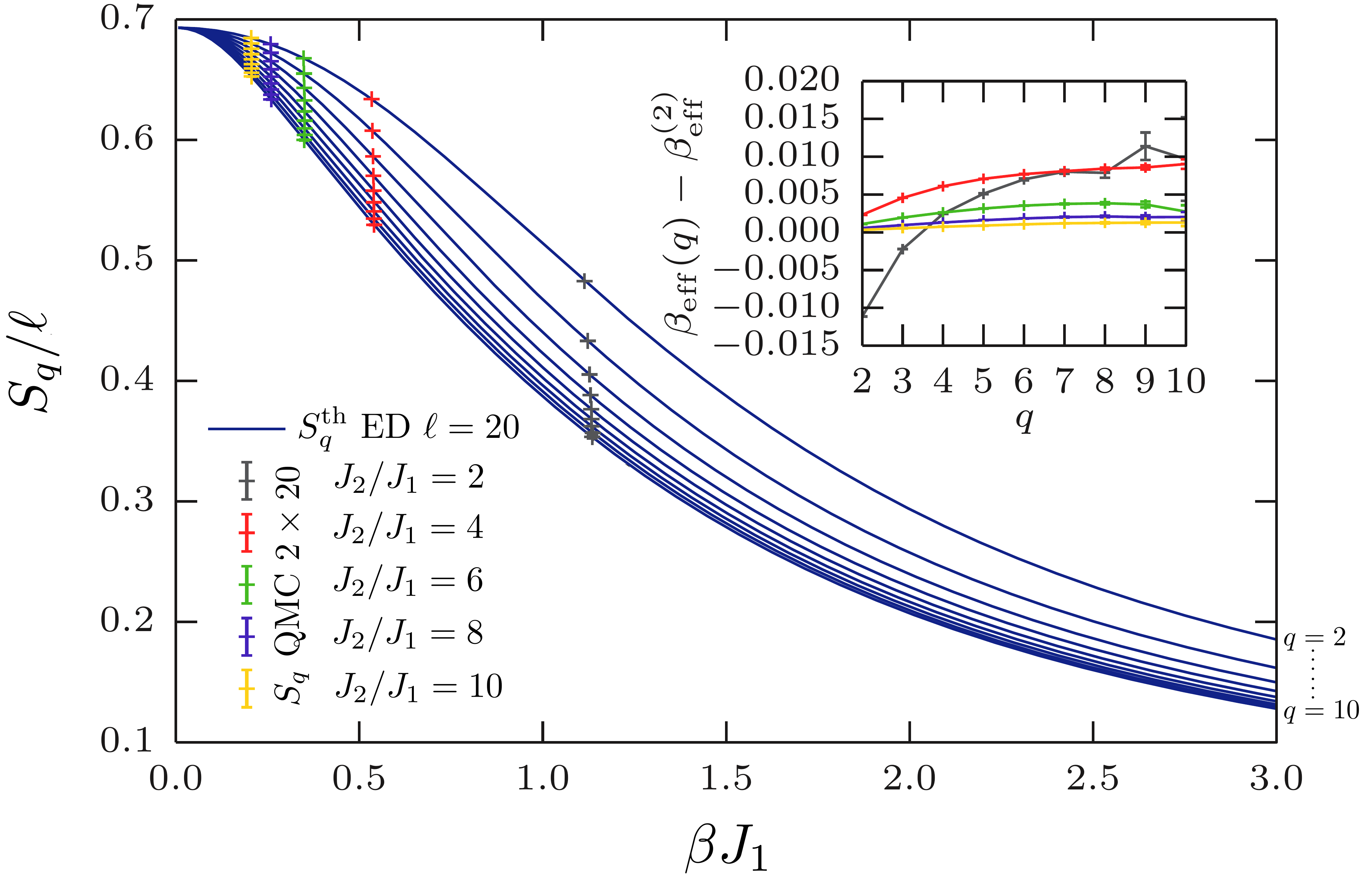}
\caption{Comparison of the thermal entropies $S_q^{\rm th}$ of a $\ell=20$ Heisenberg chain with entanglement entropies $S_q$ of Heisenberg ladders with various rung couplings $J_2/J_1$ and $q=2,\ldots,10$. Inverse temperatures $\beta_{\rm eff}$ at which  $S_q$ matches  $S_q^{\rm th}$ of the Heisenberg chain is compared to the
      $2^{\rm nd}$ order result Eq.~\eqref{eq:beta_eff}
      against $q$ (inset). 
From \cite{luitz_improving_2014}.}
\label{fig:ladder_thermo}
\end{center}
\end{figure}

For gapped spin ladders the effective temperature is of $O(1)$, a regime  where there is no analytical expression for $S_q^{\rm th}$. We have therefore performed numerical simulations~\cite{luitz_improving_2014} to compute $S_q^{\rm th}(T)$ of a $s=1/2$ Heisenberg chain (with nearest-neighbor only) at high temperature using exact diagonalization for $\ell=20$ sites, which was then compared to the R\'enyi-entanglement entropy for $2\times 20$ Heisenberg ladders at various R\'enyi indices and rung couplings $J_2$, computed using the improved QMC estimate developed in Ref.~\cite{luitz_improving_2014}. These results are displayed in Fig.~\ref{fig:ladder_thermo}  for varying rung couplings 
    $J_2/J_1=2,~4,~6,~8,~10$, where comparing with the simplest nearest-neighbor entanglement Hamiltonian ${\cal H}_E$ ($\xi_E=0$), we extract the effective inverse temperature for which the two quantities match. If
    the entanglement Hamiltonian is correct, the effective inverse temperature has to be
    independent of $q$. This is clearly the case if $J_2/J_1$ becomes large, as visible in the main panel and inset of Fig.~\ref{fig:ladder_thermo}, where the deviation from Eq.~\eqref{eq:beta_eff} is getting smaller and flatter (as a function of $q$) when $J_2/J_1$ increases.

    \subsubsection{Participation spectroscopy} A very powerful tool to further investigate this quantum - thermal mapping is the participation spectroscopy, introduced in~\cite{luitz_participation_2014}. The participation spectrum is defined in a given computational basis $\{|i\rangle\}$ by the diagonal of the RDM
    \be
    \zeta_i^A=-\ln\Bigl(\langle i| \rho_A |i\rangle\Bigr),
    \ee
    with $i=1,\ldots,2^N$ for a subsystem $A$ of $N$ spin $1/2$ for example. Using the entanglement Hamiltonian in Eq.~\eqref{eq:HE2}, the effective participation spectrum
\be
\zeta_i^{E}=\ln Z-\ln\Bigl(\langle i|\exp(-\beta_{\rm eff} {{\cal H}}_{E}) |i\rangle\Bigr),
\ee
must fulfil, for all levels $i$, $\zeta_i^E=\zeta_i^A$, provided ${\cal H}_{E}$ is the correct entanglement Hamiltonian and $\beta_{\rm eff}$ the effective inverse temperature\footnote{This condition is necessary but not sufficient.}.
Interestingly, participation levels $\{\zeta_i\}$ are easily measurable with QMC, contrary to the ES, except for the very low lying part which might be accessible\footnote{Note however that the low "energy" part of the ES may not capture universal properties since we are not interested in ground-state but finite temperature properties of the entanglement Hamiltonian.} in some cases~\cite{chung_entanglement_2014,luitz_improving_2014,assaad_entanglement_2014}.
\begin{figure}[ht]
\begin{center}
\includegraphics[width=.6\columnwidth,clip]{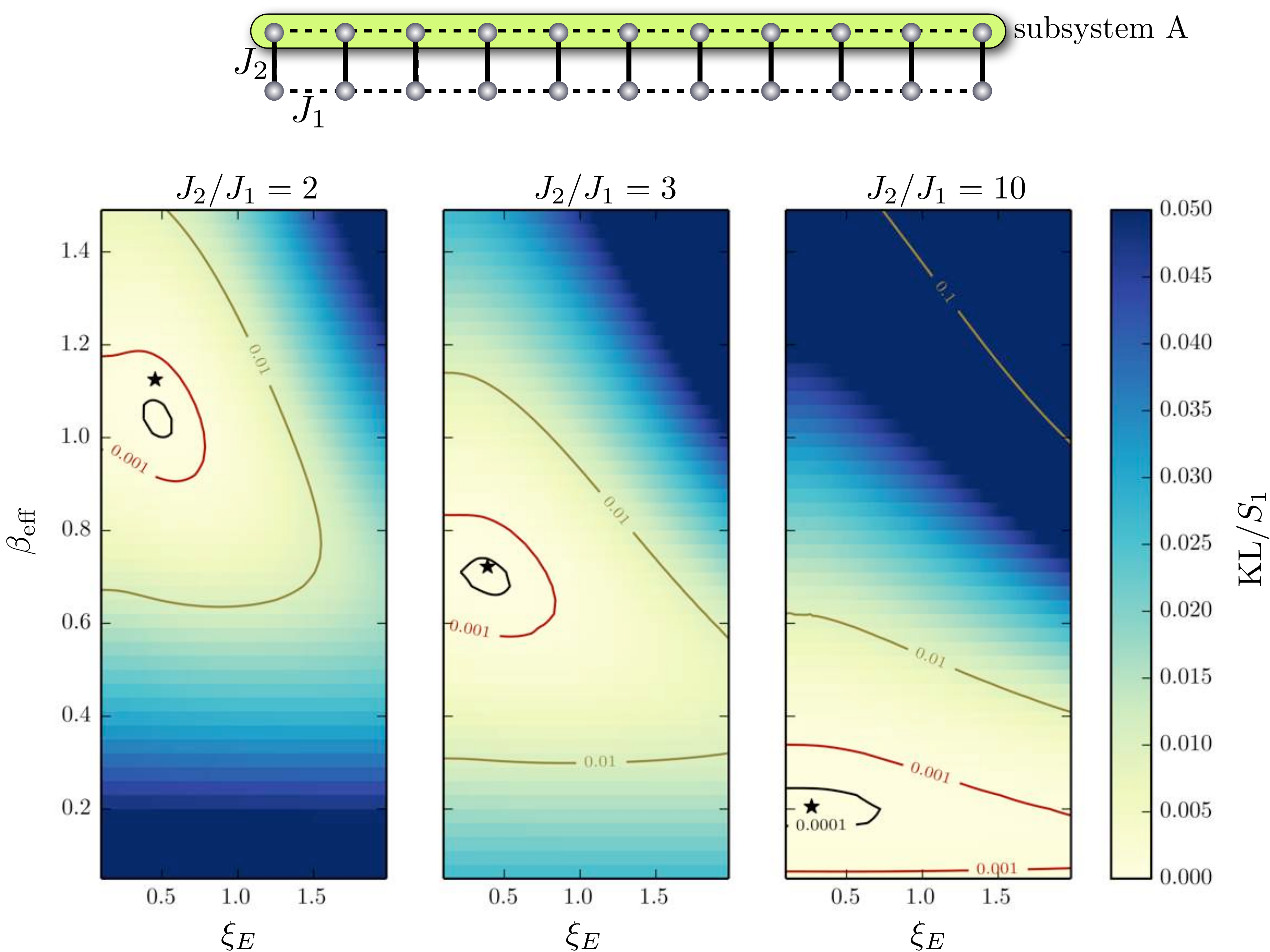}
\caption{Relative KL divergence ${\rm KL}/S_1$ Eq.~\eqref{eq:KL} of subsystem A (see schematic picture) and
     the effective short-range Hamiltonian Eq.~\eqref{eq:HE2} for different inverse temperatures $\beta_{\rm eff}$
     and entanglement lengths $\xi_E$. QMC results for $2\times 16$ ladders. The black stars show the location of the second-order perturbative result Eqs.~\eqref{eq:beta_eff}\eqref{eq:xie}.}
\label{fig:PS_LADDERS}
\end{center}
\end{figure}

A quantitative way to compare two spectra relies on the so-called Kullback-Leibler divergence KL~\cite{kullback_information_1951} which measures the difference between the two probability distributions ${\rm{e}}^{-\zeta^A}$ and ${\rm{e}}^{-\zeta^E}$. When normalized by the Shannon entropy Eq.~\eqref{eq:SRE} at $q=1$, it reads\footnote{The KL divergence and its R\'enyified version~\cite{renyi_measures_1961} $I_q=\frac{1}{1-q}\ln\left[\sum_i{\rm{e}}^{-\zeta_i^A}{\rm{e}}^{-(q-1)\left(\zeta_i^A-\zeta_i^E\right)}\right]$ compare two spectra {\textit state by state}, including possible degeneracies. Below we display results for $q=1$ (KL) which
allow to compare the spectra across their entire range, contrary to $I_{q\gg 1}$ which increases the
weight in the low "energy" part. 
However, it is important to emphasize that we have always checked the stability under
variations of $q$.}:

\be
{\rm KL}/S_1=\sum_i (\zeta_i^A-\zeta_i^E) {\rm{e}}^{-\zeta_i^A}/\sum_i\zeta_i^A{\rm{e}}^{-\zeta_i^A}.
\label{eq:KL}
\ee
In Fig.~\ref{fig:PS_LADDERS} we show a color map of such a KL divergence Eq.~\eqref{eq:KL} comparing the participation spectra of half ladders for various rung couplings $J_2/J_1$ against the effective short-range Heisenberg chain model Eq.~\eqref{eq:HE2} in the parameter space spanned by the effective inverse temperature $\beta_{\rm eff}$ and the entanglement length $\xi_E$. We clearly see a minimum of the KL divergence with a very small magnitude $<10^{-4}$, in good agreement with the perturbative result  Eqs.~\eqref{eq:beta_eff}\eqref{eq:xie}. The KL divergence provides a very precise qualitative tool to test the validity of a given entanglement Hamiltonian.

\begin{figure}
\begin{center}
\includegraphics[width=.35\columnwidth,clip]{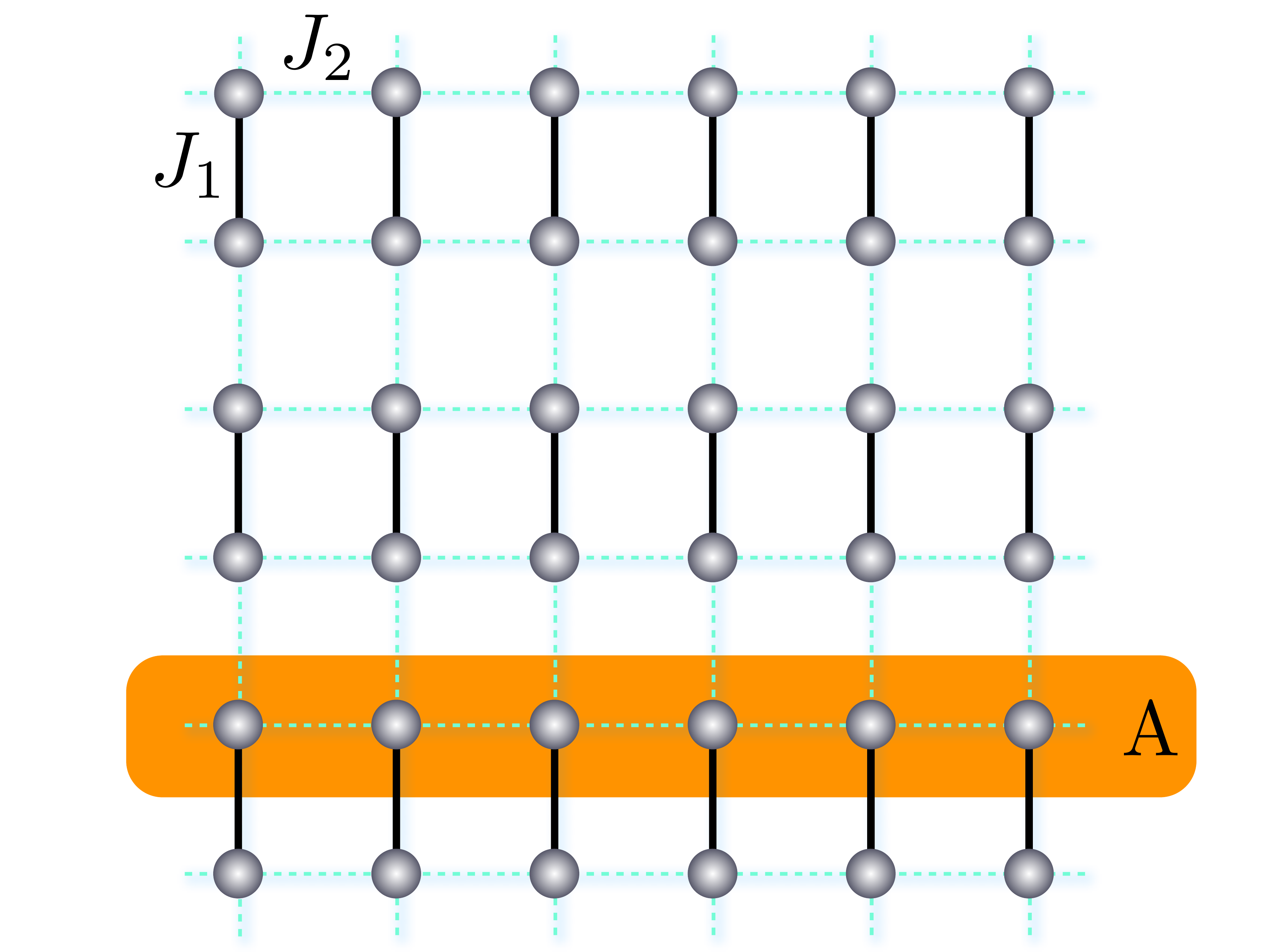}
\caption{Schematic picture for the square lattice $s=\frac{1}{2}$ dimerized Heisenberg model Eq.~\eqref{eq:HD}. The one-dimensional substystem A is also schematized.}
\label{fig:latt_dim}
\end{center}
\end{figure}

\begin{figure}
\begin{center}
\includegraphics[width=.675\columnwidth,clip]{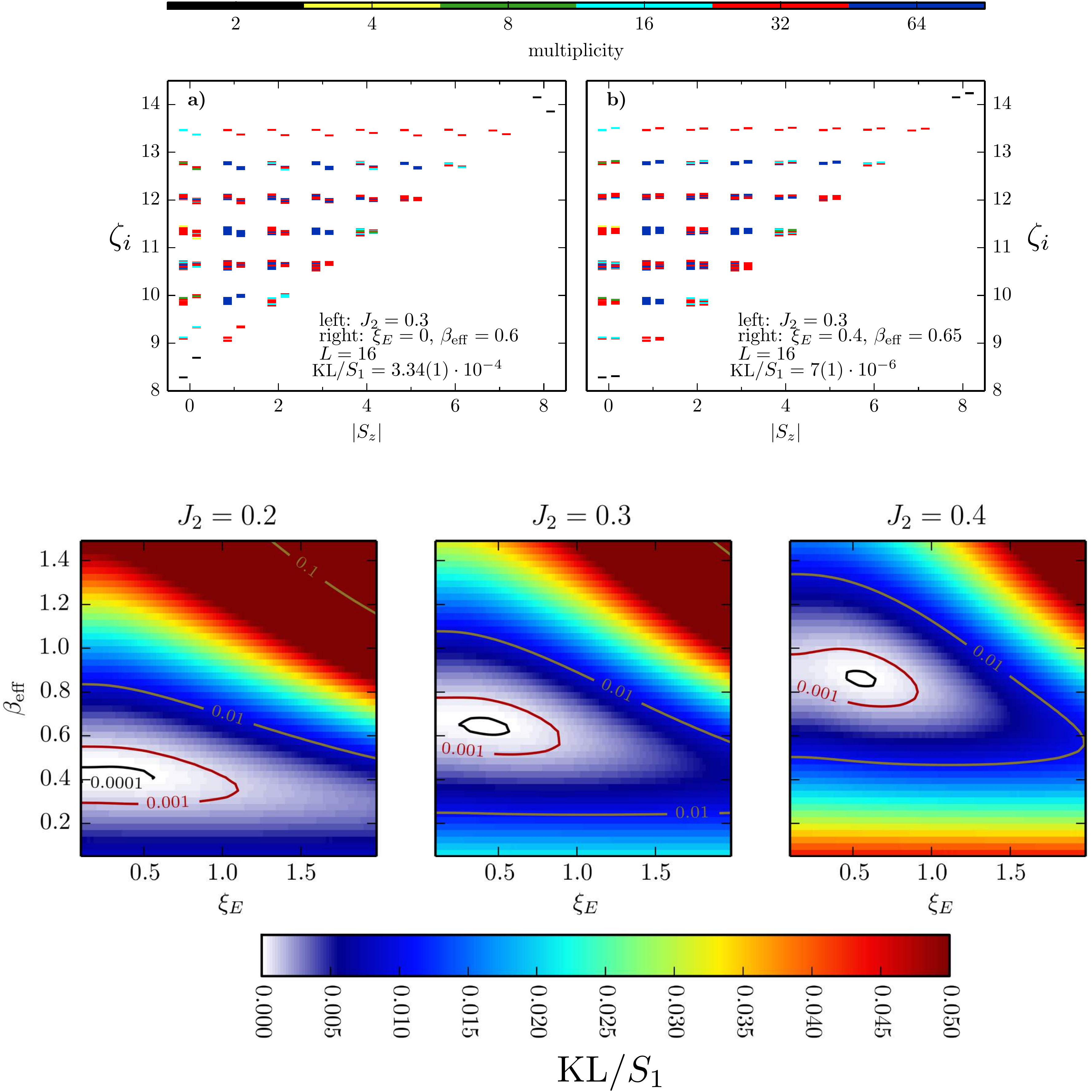}
\caption{QMC results for the $16\times 16$ coupled dimer Hamiltonian Eq.~\eqref{eq:HD}. Top: Comparison of participation spectra of the line subsystem for $J_2/J_1=0.3$
         with the effective model~\eqref{eq:HE2}. For each spin $|S_z|$ sector, two spectra are displayed (left: line subsystem, right: effective model). Panel (a) shows the effective model is the Heisenberg chain with nearest-neighbor only ($\xi_{E}=0$) at $\beta_{\rm{eff}}=0.6$. Panel (b) shows the improved effective model with $\xi_E=0.4$
 and $\beta_{\rm{eff}}=0.65$. Bottom: Color map of the relative KL divergence ${\rm KL}/S_1$ of the 1d subsystem spectrum and
     the effective short-range Hamiltonian Eq.~\eqref{eq:HE2} for different inverse temperatures $\beta_{\rm eff}$
     and entanglement lengths $\xi_E$. From~\cite{luitz_participation_2014}.}
\label{fig:psxi2d}
\end{center}
\end{figure}

This approach has also been used for a $d=2$ quantum spin model~\cite{luitz_participation_2014}, defined by coupled $s=1/2$ dimers (Fig.~\ref{fig:latt_dim})
\be 
{\cal H}_{\rm dim}= J_1 \sum_{{\rm dimers}} \vec{S}_i \cdot\vec{S}_j +J_2 \sum_{\rm{links}} \vec{S}_i \cdot \vec{S}_j, 
\label{eq:HD}
\ee
with $J_1,J_2\geq 0$. Considering only $g=J_2/J_1\leq1$ ($g=1$ being the square lattice Heisenberg antiferromagnet), this model, intensively studied at zero
temperature~\cite{troyer_critical_1997,matsumoto_ground-state_2001,wang_high-precision_2006,albuquerque_quantum_2008,wenzel_comprehensive_2009,sandvik_computational_2010}, exhibits at $g_c=0.52370(1)$~\cite{sandvik_computational_2010} a $2+1$ $O(3)$ quantum critical point separating a disordered gapped regime for $g<g_c$  from an antiferromagnetic N\'eel ordered phase at $g>g_c$, with a spontaneous breaking of the SU(2) symmetry. The participation spectrum of a one-dimensional subsystem (see Fig.~\ref{fig:latt_dim}) in the gapped regime is plotted in Fig.~\ref{fig:psxi2d} and compared with the one-dimensional effective Hamiltonian Eq.~\eqref{eq:HE2}. Deep in the gapped regime, the results are very similar to the 2-leg ladder case, with a well-defined minimum in the KL divergences.

%

%
%

\subsubsection{Gapless states}
\begin{figure}[h]
\begin{center}
\includegraphics[width=.65\columnwidth,clip]{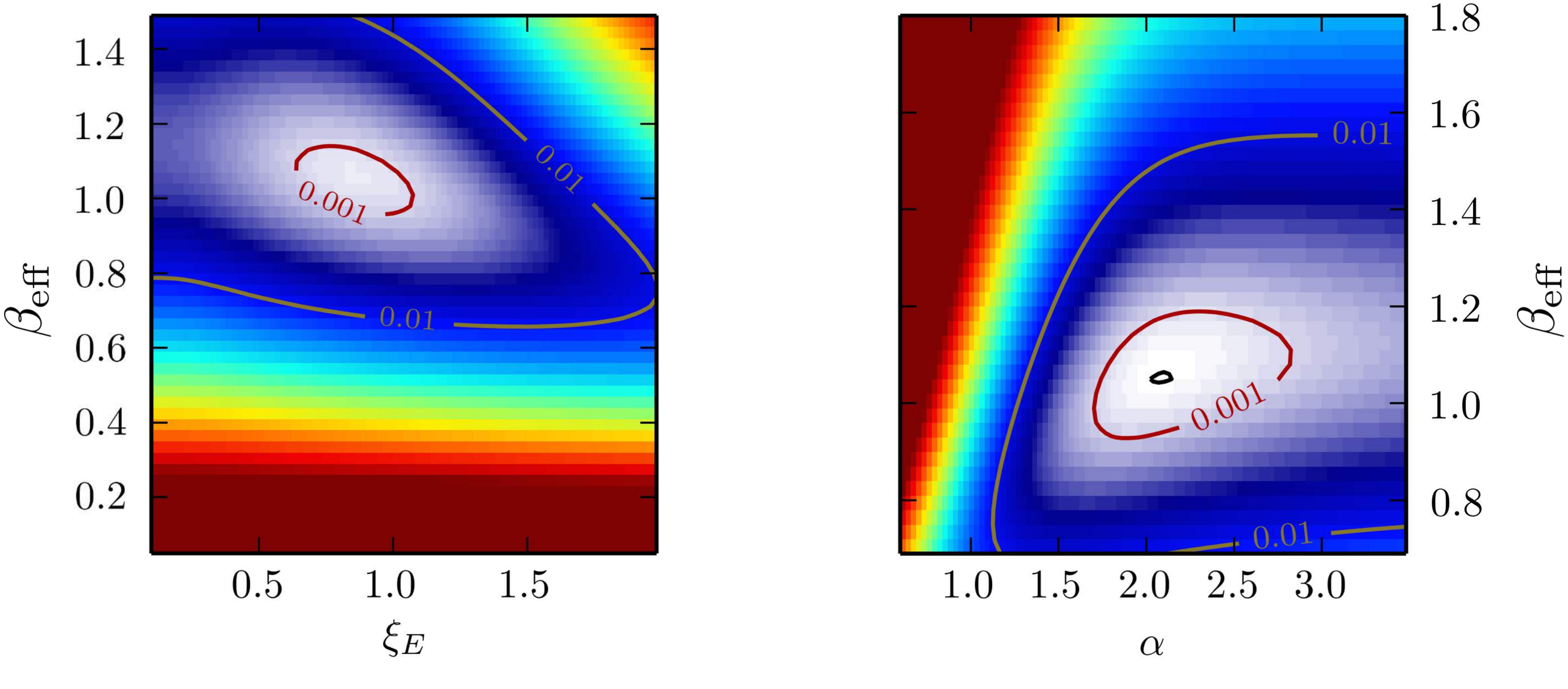}
\caption{QMC results for the participation spectra of a one-dimensional subsystem in the $16\times 16$ coupled dimer Hamiltonian Eq.~\eqref{eq:HD} at $g=J_2/J_1=0.5$, with the same color code as in Fig.~\ref{fig:psxi2d}. Relative KL divergences ${\rm KL}/S_1$ Eq.~\eqref{eq:KL} with the short-range interacting Heisenberg chain model Eq.~\eqref{eq:HE2} (left) and the power-law interacting Hamiltonian Eq.~\eqref{eq:alpha} (right) for which the minimum is clearly smaller. Reprinted from~\cite{luitz_participation_2014}.}
\label{fig:J205}
\end{center}
\end{figure}

%
\paragraph{Quantum critical point---}
When the $O(3)$ quantum critical point is approached, the entanglement length increases, while the effective temperature slightly decreases but remains of order one. As discussed by Cirac {\it et al.}~\cite{cirac_entanglement_2011}, the short-range nature of the entanglement Hamiltonian  qualitatively changes when approaching criticality. Based on the participation spectroscopy analysis~\cite{luitz_participation_2014}, the best effective model which captures criticality for the line subsystem  is no longer the short-range Hamiltonian Eq.~\eqref{eq:HE2} (which cannot display algebraic correlations at finite temperature), but the following unfrustrated power-law decaying model~\cite{yusuf_spin_2004,laflorencie_critical_2005,beach_fractal_2007}
\be
{\cal H}_E(\alpha)=J_1\sum_{i,j}\frac{(-1)^{r_{ij}}}{r_{ij}^{\alpha}}\vec{S}_i \cdot
\vec{S}_{j}.
\label{eq:alpha}
\ee
In Fig.~\ref{fig:J205} we provide a direct comparison between the short-range model Eq.~\eqref{eq:HE2} and the one with power-law interactions Eq.~\eqref{eq:alpha} for $g=0.5$, {\it{i.e.}} very close to the quantum critical coupling of the $d=2$ dimerized Hamiltonian Eq.~\eqref{eq:HD}. The latter model clearly displas a much better agreement, with an effective inverse temperature $\beta_{\rm eff}J_1\simeq 1$, and a power $\alpha\simeq 2$.

\paragraph{N\'eel ordered regime---}

\begin{figure}
\begin{center}
\includegraphics[width=.65\columnwidth,clip]{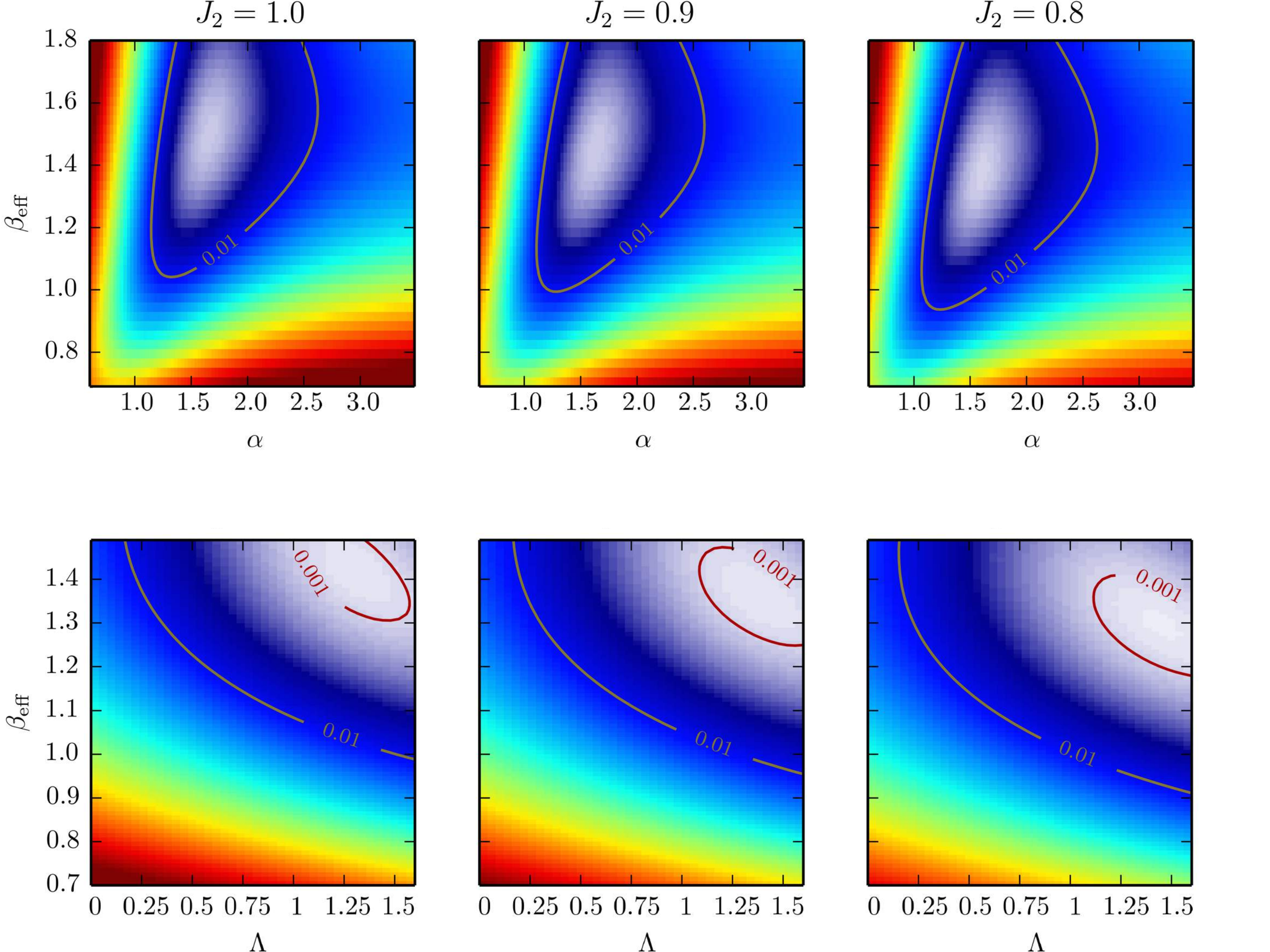}
\caption{QMC results for the participation spectra of a one-dimensional subsystem in the $16\times 16$ Hamiltonian Eq.~\eqref{eq:HD} at $J_2/J_1=0.8,~0.9,~1$ (N\'eel phase), with the same color code as in Fig.~\ref{fig:psxi2d}. Relative KL divergences ${\rm KL}/S_1$ Eq.~\eqref{eq:KL} with the power-law effective Hamiltonian Eq.~\eqref{eq:alpha} (top) and the Lieb-Matis  + power-law interacting Hamiltonian Eq.~\eqref{eq:HLambda} (bottom) for which the minimum is smaller. Reprinted from~\cite{luitz_participation_2014}.}
\label{fig:psneel}
\end{center}
\end{figure}

Keeping the same one-dimensional setup for subsystem A, one can cross the critical point and explore the N\'eel ordered side for $g>g_c$, as done in Refs.~\cite{luitz_shannon-renyi_2014,luitz_participation_2014}. In this regime, while the power-law model Eq.~\eqref{eq:alpha} gives reasonable KL (Fig.~\ref{fig:psneel} top) for parameters ($\alpha-\beta_{\rm eff}$) where finite temperature order is expected, a better agreement between participation spectra is found for the model
\be
{\cal H}_E(\Lambda)=J_1\sum_{i,j}{(-1)^{r_{ij}}} \left(\frac{\Lambda}{L}+\frac{1}{r_{ij}^{3}}\right)\vec{S}_i \cdot
\vec{S}_{j},
\label{eq:HLambda}
\ee
\noindent as shown in Fig.~\ref{fig:psneel} for three values of $J_2$ in the N\'eel regime.
 
%
 Following Refs.~\cite{metlitski_entanglement_2011,alba_entanglement_2013,kolley_entanglement_2013,rademaker_tower_2015}, we expect the entanglement Hamiltonian to have both TOS and SW structures, as already discussed above for the superfluid regime of the $d=2$ Bose-Hubbard model, see Section~\ref{sec:ESLRO}. In Eq.~\eqref{eq:HLambda}, the $\Lambda$ term (constant at all distances, with a $1/L$ normalisation to ensure extensivity) has the Lieb-Mattis~\cite{lieb_ordering_1962} form which has TOS but does not sustain relativistic SW excitations, whereas the power-law
component $\sim 1/r^3$ is expected to bring $\Omega_{\rm sw} \sim k$ excitations and $1/r$ decaying spin correlation functions~\cite{laflorencie_critical_2005}.  

Entanglement spectroscopy for the N\'eel state has also been studied using PEPS calculations~\cite{poilblanc_entanglement_2014} where it was interpreted using a mapping onto a bosonic $t - J$ chain with long distance hopping terms.

\newpage 
\section{Impurity and disorder effects}
\label{sec:dis}
Disorder and quantum fluctuations have the common tendency to destabilize classical order. Whether
intrinsically present in materials, chemically controlled via doping, or
explicitly introduced via a random potential (as in ultra-cold atomic systems) or by varying the thickness in superconducting films,
randomness can lead to dramatic changes in physical properties of condensed
matter systems, as experienced for instance with the Anderson localization~\cite{anderson_absence_1958,evers_anderson_2008}, the Kondo effect~\cite{kondo_resistance_1964,hewson_kondo_1993}, glassy physics~\cite{binder_spin_1986}, etc.

In such a context, entanglement witnesses have provided new tools to investigate various disordered systems as we review below, focusing on three representative examples. In section~\ref{sec:disqs}, we discuss disordered quantum spin systems which despite some similarities with their clean counterparts, display major differences. Another paradigmatic example of impurity physics is the Kondo effect for which the putative "Kondo screening cloud" can be probed using the concept of impurity entanglement entropy, as we present in Section~\ref{sec:kondo}. Finally, in Section~\ref{sec:mbl} we discuss recent advances in the context on Anderson localization in the presence of interactions, the so-called "many-body localization" from entanglement perspectives, either at and out of equilibrium.

\subsection{Disordered quantum spin systems}
\label{sec:disqs}
Impurity and disorder in quantum spin systems have been intensively investigated for several decades, in particular in the context of spin-glass physics~\cite{edwards_theory_1975,kirkpatrick_infinite-ranged_1978,binder_spin_1986}. For random field and random exchange quantum magnets, the strong disorder 
decimation method in real space~\cite{ma_random_1979,fisher_critical_1995,igloi_strong_2005} have been used quite intensively, with the celebrated example of the infinite randomness fixed point (IRFP) analytically described by Fisher in a serie of seminal papers for random spin chains~\cite{fisher_random_1994,fisher_critical_1995}. Later these idea have been extended to $d>1$ where a numerical approach is inevitable~\cite{motrunich_infinite-randomness_2000,lin_low-energy_2003,kovacs_infinite-disorder_2011}.
\subsubsection{Entanglement in random spin chains}
\begin{figure}[h!]
\begin{center}
\includegraphics[width=\columnwidth,clip]{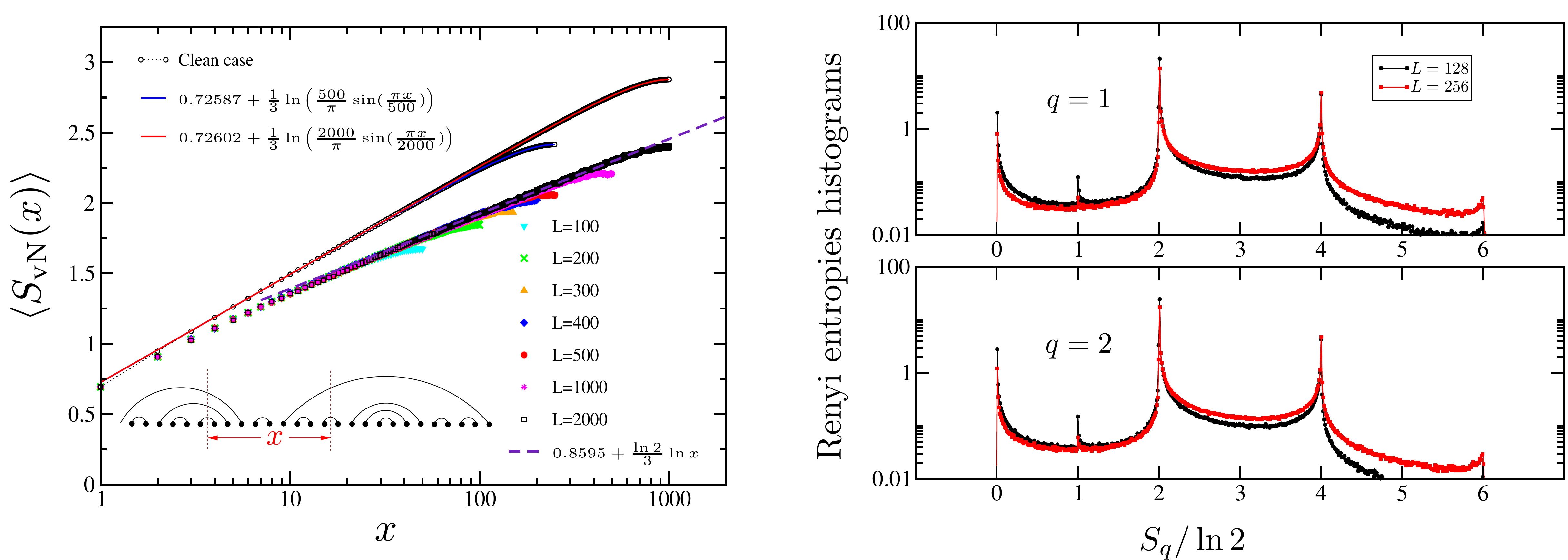}
\caption{Exact diagonalization results for $s=1/2$ XX chains with random bonds. Left (reprinted from~\cite{laflorencie_scaling_2005}): Entanglement entropy of a subsystem of size
  $x$ embedded in a closed ring of size $L$, shown against $x$ in a
  log-linear plot. 
For clean systems with $L=500,~2000$ (open circles), $S_{\rm vN}$ is perfectly described by Eq.~(\ref{eq:S1d}) (red and blue curves).
Data for the random case (with a uniform box distribution of couplings $J\in [0,1]$) have been averaged over $10^4$ samples for $L=500,~1000,~2000$ and $2\times 10^4$ samples for $100\le L \le 400$.
The form $0.8595+\frac{\ln 2}{3}\ln x$ (dashed line) fits the data in the regime where finite size effects are absent.
Right: Histograms of R\'enyi entropies $q=1,~2$ collected at half-chain over $\sim 10^6$ independent samples, with $L=128,~256$  sites at  strong disorder $P(J)\propto{J^{-1+1/D}}$ with $D=5$. The peaks, independent on the R\'enyi parameter, clearly show the random singlet structure.}
\label{fig:RSP}
\end{center}
\end{figure}

\paragraph{Random singlet state for random bonds spin $s=1/2$ chains---} Building on the decimation solution of the $s=1/2$ random exchange XXZ chain~\cite{fisher_random_1994} and of the random transverse field Ising chain~\cite{fisher_critical_1995}, Refael and Moore~\cite{refael_entanglement_2004} have shown that at such an IRFP, the disorder-average von-Neumann entropy follows a similar logarithmic growth with sub-system size $x$ as in the clean case Eq.~\eqref{eq:S1d}, but with a smaller prefactor $c\ln 2$, where $c$ is the central charge of the critical clean system ($c=1/2$ for Ising and $c=1$ for XXZ). This result can be simply understood as a direct consequence of the perturbative structure of the ground-state of the XXZ chain in the limit of strong disorder, the so-called random-singlet state~\cite{fisher_random_1994} (depicted in Fig.~\ref{fig:RSP}) where the probability of finding a singlet of length $x$ is $\simeq 1/(3x^{2})$~\cite{hoyos_correlation_2007}. The quantum Ising case follows immediately using a direct relationship with the XX chain~\cite{igloi_exact_2008} which can be seen as two decoupled quantum Ising chains~\cite{turban_exactly_1984}.

This surprising analytical prediction was rapidly confirmed numerically~\cite{laflorencie_scaling_2005} using exact diagonalization at the XX point where free-fermion techniques~\cite{peschel_reduced_2004} allow to reach quite large chains. These results are shown in Fig.~\ref{fig:RSP} (left panel) where the disorder average von-Neuman entropy displays the expected $\frac{\ln 2}{3} \ln x$ growth. This was later verified by other groups~\cite{hoyos_correlation_2007,igloi_finite-size_2008,fagotti_entanglement_2011,pouranvari_entanglement_2013}. The random singlet structure also led to the notion of valence bond entanglement entropy~\cite{alet_valence_2007,chhajlany_topological_2007,
mambrini_hard-core_2008,jacobsen_exact_2008,tran_valence_2011} which is asymptotically equivalent to the usual entanglement entropy at the IRFP~\cite{alet_valence_2007,tran_valence_2011}.

For higher order R\'enyi indices, Fagotti and co-workers~\cite{fagotti_entanglement_2011} have shown that the situation is more subtle. Indeed, depending how the disorder averaging $\langle \ldots\rangle$ is performed, they found a different prefactor for the logarithmic scaling:
\be
\langle{S}_q\rangle=\frac{1}{1-q}\langle\ln {\rm Tr}\rho_A^q\rangle=\frac{\ln 2}{3}\ln L+{\rm{const}}(q),
\label{eq:Sq}
\ee
while
\be
{\widetilde{\langle S_q\rangle}}=\frac{1}{1-q}\ln\langle {\rm Tr}\rho_A^q\rangle=f_q\frac{\ln 2}{3}\ln L+{\rm{const'}}(q),
\label{eq:Sqtilde}
\ee
with a non-trivial prefactor $f_q=\frac{3\left(\sqrt{5+2^{3-q}}-3\right)}{2\ln 2 (1-q)}\le 1$, vanishing at large $q$ and recovering $f_q\to 1 $ when $q\to 1$. This ensemble averaging dependence is the hallmark of infinite randomness physics, as first identified by Fisher for correlations functions in Ref.~\cite{fisher_critical_1995} where typical and average have qualitatively different scalings. 

\begin{figure}[t]
\begin{center}
\includegraphics[width=.65\columnwidth,clip]{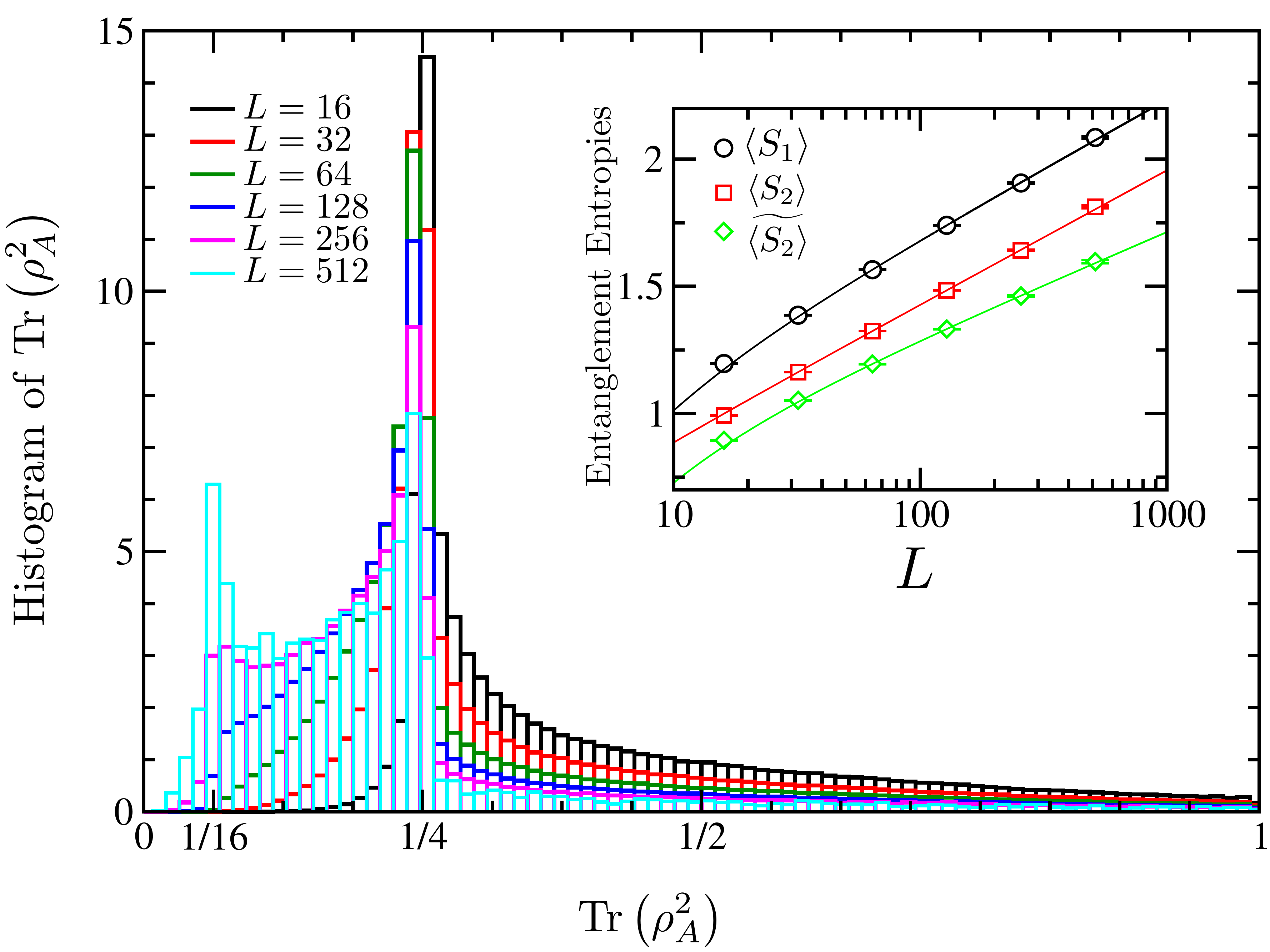}
\caption{Histogram of ${\rm{Tr}}\left(\rho_A^2\right)$ in the random singlet regime. Exact diagonalization results obtained for random XX chains with $\sim 10^5$ independent samples, with $L=16,\ldots,512$ lattice sites at moderate disorder (full box distribution). 
Inset: Best fits are $\langle S_1\rangle =0.236(4)\ln x+0.60(3)-1.4(4)/x$, $\langle S_2\rangle=0.229(2)\ln x +0.37(1)-0.15(10)/x$ and ${\widetilde{\langle S_2\rangle}}=0.180(5)\ln x +0.47(3)-1.6(5)/x$, which compare very well to $\frac{\ln 2}{3}\simeq 0.231$ and $f_2\frac{\ln 2}{3}\simeq 0.177$. }
\label{fig:S2}
\end{center}
\end{figure}

The distribution of ${\rm{Tr}}\left(\rho_A^2\right)$, shown in the main panel of Fig.~\ref{fig:S2} for the random singlet state of random XX chains (exact diagonalization results) for increasing system lengths $L=16,\ldots,512$, displays broadening at small values with increasing size,  suggesting non-self-averaging entanglement\footnote{One may also try to interpret ${\rm{Tr}}\left(\rho_A^2\right)$ as the inverse participation ratio in the Schmidt basis. Using this analogy, ${\widetilde{\langle S_q\rangle}}$ would correspond to the average while $\langle{S}_q\rangle$ would be the typical value.}.  In the inset of 
Fig.~\ref{fig:S2}, the log scalings are plotted for $q=1,~2$, showing a very good agreement with the analytical prediction Eq~\eqref{eq:Sq} and \eqref{eq:Sqtilde}.\\

\paragraph{Random spin chains with higher spin---} Infinite randomness fixed points also occur for $s>1/2$ chains~\cite{hyman_impurity_1997,monthus_phases_1998,refael_spin_2002,damle_permutation-symmetric_2002}, where Refael and Moore have shown~\cite{refael_entanglement_2007} that 
\be
\langle S_{\rm vN}\rangle = \frac{\ln(2s+1)}{3}\ln L +{\rm constant}.
\ee
Non-abelian random-singlet states are also expected for disordered chains of Majorana or Fibonacci anyons~\cite{bonesteel_infinite-randomness_2007,fidkowski_textitc_2008,fidkowski_permutation-symmetric_2009}, with a logarithmic von-Neumann entropy whose "effective central charge" pre-factor is given by $\ln d$, where $d$ is the quantum dimension, {\it{e.g.}} $d=\sqrt{2}$ for a Majorana chain, and $d=(1+\sqrt{5})/2$ for Fibonacci.
\subsubsection{Entanglement, disorder, and RG flows}
\begin{figure}[hb]
\begin{center}
\includegraphics[width=0.6\columnwidth,clip]{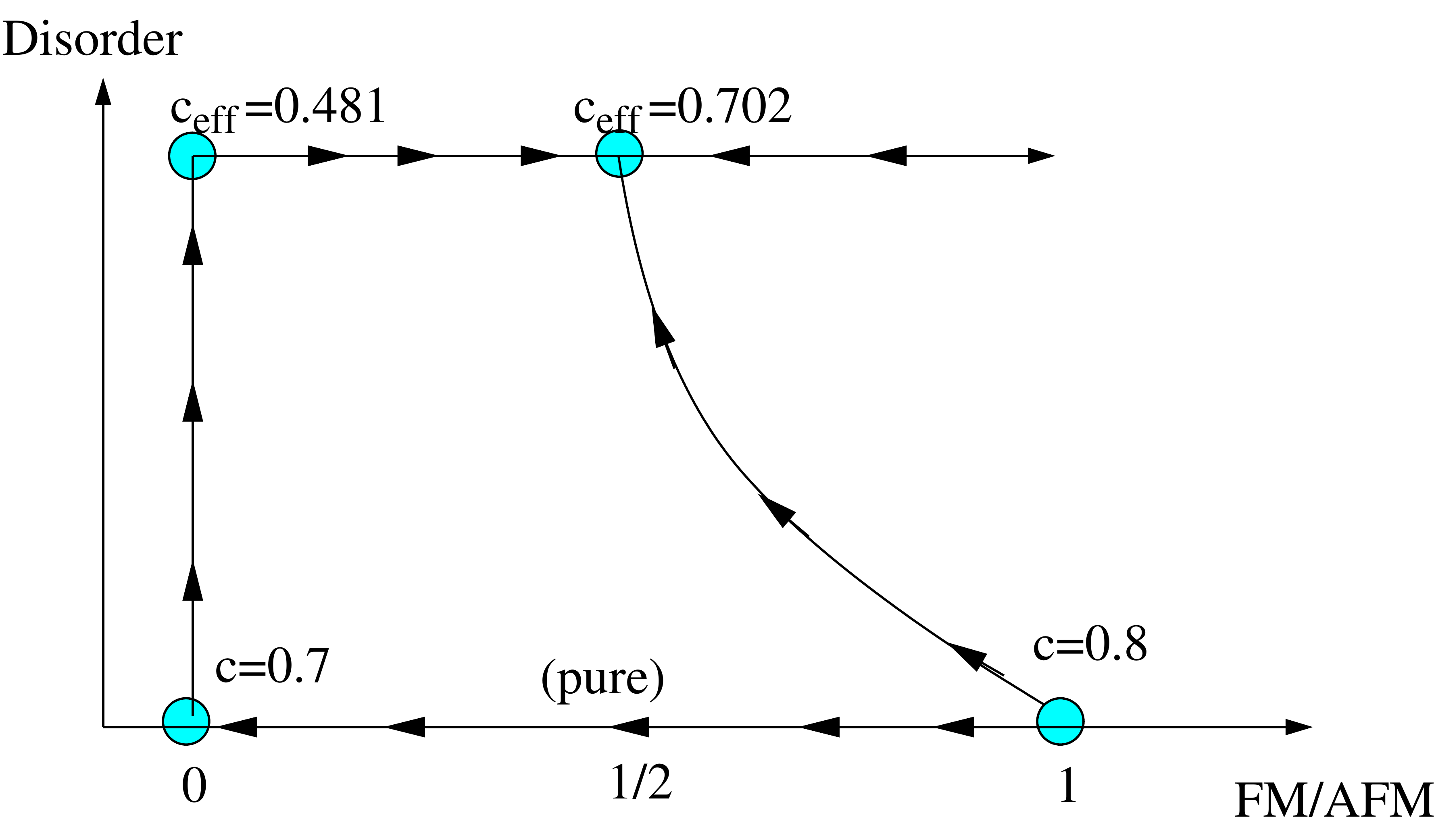}
\caption{RG flow diagram of the pure and random golden chain with ferromagnetic and antiferromagnetic couplings. Reprinted from~\cite{fidkowski_textitc_2008}.}
\label{fig:golden}
\end{center}
\end{figure}

\paragraph{Absence of $c$-theorem---}
Following Zomolodchikov $c$-theorem~\cite{zomolodchikov_``irreversibility_1986} which yields for clean (disorder-free) fixed points a decreasing of entanglement entropy along RG flows, the question whether this also holds in the presence of disorder was addressed after the discovery of decreasing entropies along infinite randomness flows~\cite{refael_entanglement_2004,laflorencie_scaling_2005,refael_entanglement_2007}. Two counter examples have shown that this cannot be true for disordered fixed points, as first discussed by Santachiara~\cite{santachiara_increasing_2006} for generalized quantum Ising chains including the $N$-states random Potts chain, and later by Fidkowski and co-workers~\cite{fidkowski_textitc_2008} for disordered chains of Fibonacci anyons. The phase diagram of this disordered golden chain model is sketched in Fig.~\ref{fig:golden}, as also discussed by Refael and Moore in Ref.~\cite{refael_criticality_2009}.\\

\paragraph{Engineered disorder---}
Interestingly,  a volume-law scaling of the entanglement entropy can be achieved for the ground-state of a class of random spin chains where disorder is not random, but engineered, building on the decimation rules such that the probability of finding a singlet at any distance $x$ is uniform. Designed with very fast decaying couplings, the so-called concentric singlet phase can be constructed~\cite{vitagliano_volume-law_2010,ramirez_conformal_2014,ramirez_entanglement_2015}, with a very strong entanglement entropy proportional to the number of sites inside the subsystem.\footnote{See also Refs.~\cite{irani_ground_2010,movassagh_power_2014} for power-law violation of the area-law in translationally invariant models.}

\begin{figure}[b]
\begin{center}
\includegraphics[width=0.6\columnwidth,clip]{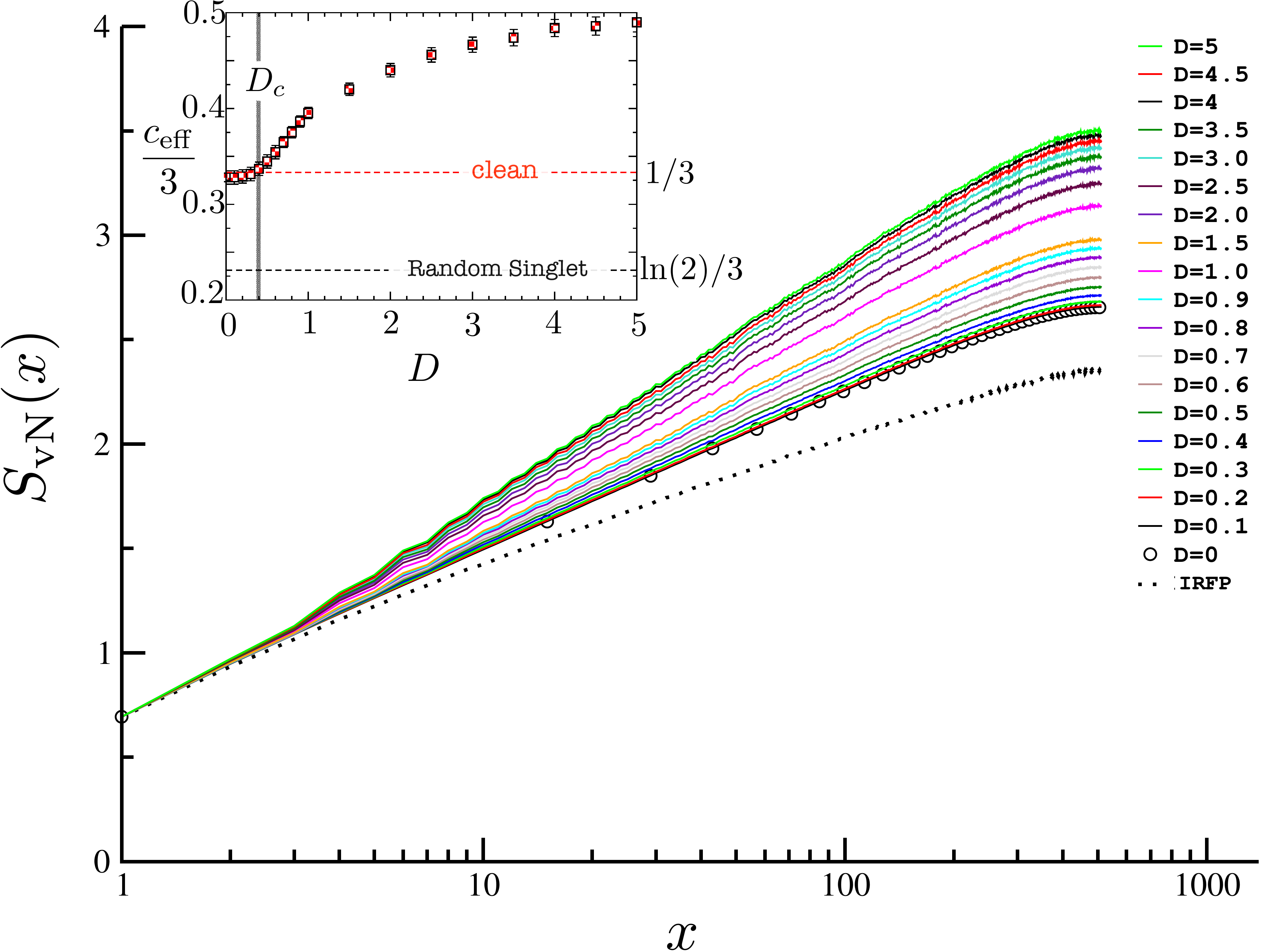}
\caption{Disorder averaged entanglement (von-Neumann) entropy plotted against subsystem size $x$ for the critical ground-state of Eq.~\eqref{eq:Hcorr} with correlated disorder for various disorder strengths $D$ ($P(J)=J^{-1+1/D}/D$ with chains of 1024 sites averaged over 5 000 disorder realizations). Inset: coefficient of the logarithmic increase of the entanglement entropy plotted against the disorder strength $D$. Reprinted from Ref.~\cite{hoyos_protecting_2011}.}
\label{fig:correlated}
\end{center}
\end{figure}

Another model where disorder is partly controlled was proposed by Binosi and co-workers~\cite{binosi_increasing_2007}, through the following disordered quantum Ising chain model
\be
{\cal{H}}=-\sum_i J_i\left(S_i^z S_{i+1}^z +S_i^x\right),
\label{eq:Hcorr}
\ee
where one sees that the independent random couplings $J_i$ act on both a site and its adjacent link, such that a perfect (but purely local) correlation is achieved. Using field theory, strong disorder RG, and large scale numerical diagonalization techniques, we have investigated this interesting model in Ref.~\cite{hoyos_protecting_2011}. For perfect correlation weak disorder is irrelevant, whereas any small breaking of the perfect correlation between  field and coupling in Eq.~\eqref{eq:Hcorr} brings the system back to infinite-randomness physics. For larger disorder (and perfect local correlation), there is a line of critical points with unusual properties such as an increase of the entanglement entropy with the disorder strength, as shown in Fig.~\ref{fig:correlated}.

This model brings an interesting example where by construction the disordered system is always strictly critical at the local level. This apparent suppression of local randomness protects the clean physics against small disorder, but at strong enough disorder a new physics appears where entanglement increases with the strength of disorder. The XXZ extension of this model was recently studied in Ref.~\cite{getelina_entanglement_2016}, reaching similar conclusions as compared to free fermions.

Finally, let us note the interesting case of aperiodic quantum spin chains~\cite{hida_new_2004,vieira_low-energy_2005,vieira_aperiodic_2005} where the entanglement entropy was studied using strong disorder RG~\cite{igloi_entanglement_2007,juhasz_entanglement_2007}. A logarithmic growth with subsystem size was found, with a non-universal prefactor depending on the aperiodic sequence, and possibly larger than the corresponding clean case.

\subsubsection{$d>1$ Infinite randomness}
Infinite randomness physics is not specific to one dimension, as shown for the $d\ge 2$ disordered quantum Ising model~\cite{motrunich_infinite-randomness_2000,kovacs_renormalization_2010,kovacs_infinite-disorder_2011,monthus_random_2012}, disordered contact process~\cite{vojta_infinite-randomness_2009}, or dissipative systems~\cite{vojta_infinite-randomness_2009-1}. However, note that random singlet physics does not describe $d>1$ random exchange antiferromagnets~\cite{lin_low-energy_2003,laflorencie_random-exchange_2006}, where long-range order is surprisingly robust to disorder~\cite{laflorencie_random-exchange_2006}. 

\begin{figure}[b]
\begin{center}
\includegraphics[width=0.5\columnwidth,clip]{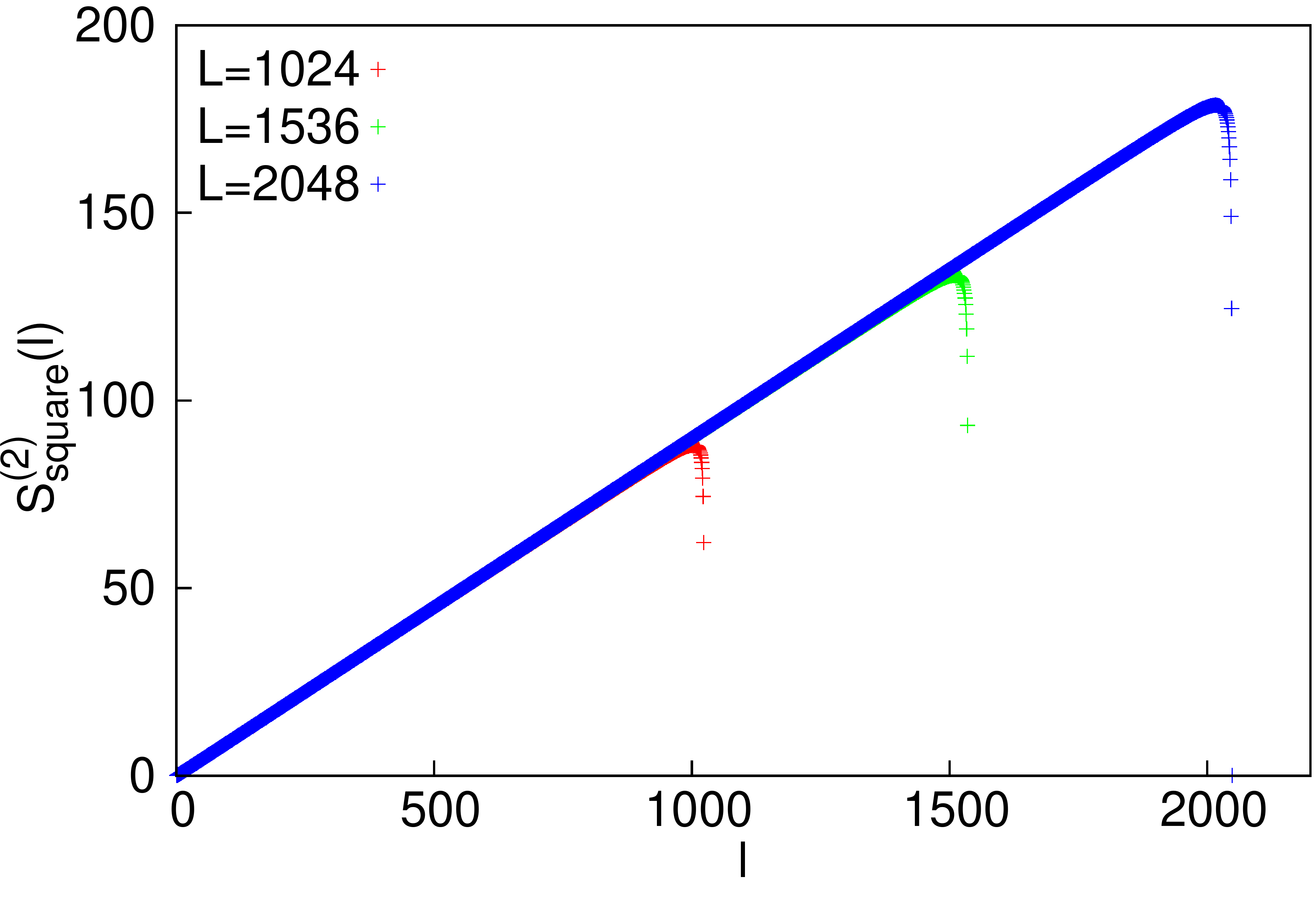}
\caption{Area law scaling of the von-Neumann entropy of the disordered transverse field Ising model on the square lattice, computed at criticality (infinite disorder fixed point) using strong disorder RG. Reprinted from Ref.~\cite{kovacs_infinitely_2012}.}
\label{fig:kovacs1}
\end{center}
\end{figure}

Entanglement entropy at the IRFP of the $d=2$ random transverse field Ising model on the square lattice has been first studied numerically in 
\cite{lin_entanglement_2007} where a surprising  double logarithmic enhancement of the area law was found $S\sim L\ln (\ln L)$, based on strong disorder RG arguments\footnote{Nevertheless, another type of IRFP in higher dimensions occurs in the bond-diluted quantum Ising ferromagnet~\cite{senthil_higher_1996} for which the same authors~\cite{lin_entanglement_2007} found (using much larger system sizes) a pure area law contribution at the percolation threshold.}. Later, using larger systems, Yu and co-workers~\cite{yu_entanglement_2008} concluded for a pure area law with additive (negative) logarithmic corrections.

\begin{figure}[b]
\begin{center}
\includegraphics[width=.54\columnwidth,clip]{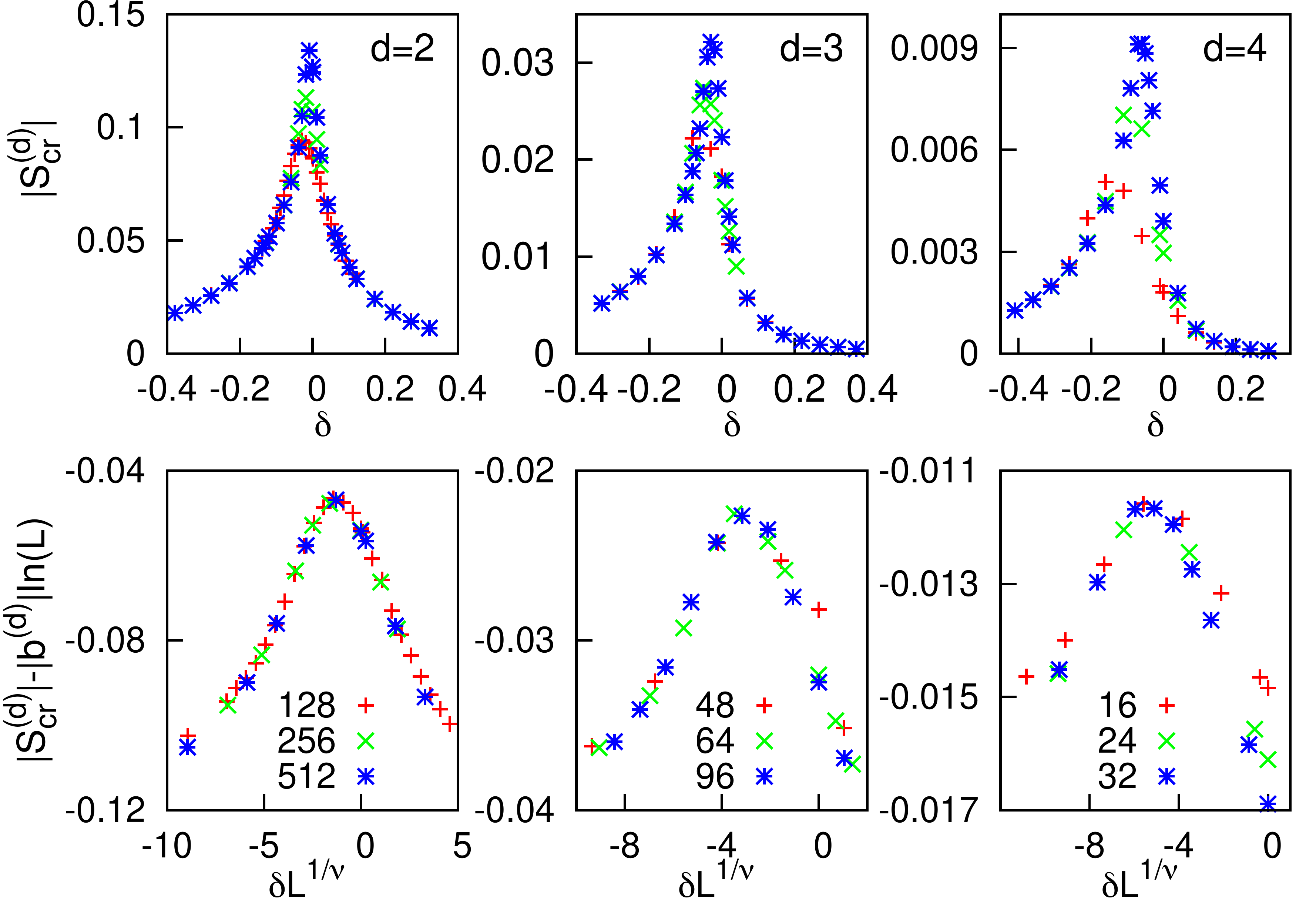}
\caption{Upper panels: entropy contribution of the
corners $|S_{\rm cr}^{(d)}|$ in dimension  $d= 2, 3$ and 4 for different system sizes as
a function of distance to criticality $\delta$. Bottom panels: Universal scaling function using $b^{(2)}=-0.029$, $b^{(3)}=0.012$, and $b^{(4)}=-0.006(2)$. Reprinted from Ref.~\cite{kovacs_universal_2012}.}
\label{fig:kovacs2}
\end{center}
\end{figure}

Using a greatly improved strong disorder RG algorithm where the ${\cal O}(N^3)$ running time of the na{\"{\i}}ve algorithm was brought down to ${\cal O}(N \ln N)$ for arbitrary dimension, Kov{\'a}cs and Igl{\'o}i~\cite{kovacs_renormalization_2010,kovacs_infinite-disorder_2011} have studied entanglement of $d=2,3,4$ disordered quantum Ising models up to $N\sim 10^6$ spins~\cite{kovacs_universal_2012}. This allowed to get a very good control of finite size scaling, as shown in Fig.~\ref{fig:kovacs1} where the square lattice entropy, plotted as a function of the subsystem size, clearly displays a pure area law as a leading term. Interestingly, corrections to this area law scaling can be precisely computed: in two dimensions, for square (or slab in higher dimension) subsystems, additive logarithmic corrections are found~\cite{kovacs_universal_2012}, coming from the 4 corners of a square subsystem, such that
\be
S_{\rm vN}=aL+4\l_{1}(\pi/2)\ln L +{\rm constant}
\ee
with $\l_{1}(\pi/2)=-0.029(1)$ obtained for very large clusters, up to $2048 \times 2048$ (to be compared with $-0.019(5)$ obtained for smaller systems $160 \times 160$ by Yu {\it{et al.}}~\cite{yu_entanglement_2008}). This log correction, arising from $\pi/2$ corners at a strong disorder fixed point, can be compared to clean critical points contributions studied in Section~\ref{sec:corners}, and reported in Table~\ref{tab:corner}. More generally, Kov{\'a}cs and Igl{\'o}i found that additive logarithmic contributions  induced by the sharp subsystems boundaries only occur at criticality, as shown in the bottom part of Fig.~\ref{fig:kovacs2} where the corner part appears to be a universal singular function, describing the log divergence at the critical point with a prefactor whose magnitude and sign depend on the dimension: $l_1(3d)=0.012(2)$, and $l_1(4d)=-0.006(2)$.

Such universal features of the additive logarithmic corrections to the area law appear to be a promising way to further characterize quantum criticality~\cite{bueno_universality_2015,bueno_universal_2015,bueno_corner_2015}.
\subsection{Entanglement and Kondo physics}
\label{sec:kondo}
\subsubsection{Generalities}
The resistivity minimum observed in dilute magnetic alloys received a beautiful theoretical explanation by Jun Kondo~\cite{kondo_resistance_1964} based on a perturbative calculation which breaks down at low-energy. This is a remarkable condensed matter example of asymptotic freedom~\cite{gross_twenty_1999}, with the effective interaction between magnetic impurities and conduction electrons growing when the temperature decreases. Non-perturbative calculations, based on a scaling approach~\cite{anderson_poor_1970}, the renormalization group (RG)~\cite{wilson_renormalization_1975}, Fermi liquid theory~\cite{nozieres_fermi-liquid_1974}, Bethe Ansatz~\cite{andrei_diagonalization_1980,wiegmann_exact_1981} provide a very powerful framework to study this non-trivial many-body problem (for a thorough review, see~\cite{hewson_kondo_1993}). In particular, the anomalous scattering from magnetic impurities leads to an enhancement of both specific heat and magnetic susceptibility at low temperature, below the Kondo temperature $T_K$. The thermal entropy $S^{\rm Th}(T)$ is a non-trivial function of $T$, reflecting the RG fixed points of the Kondo problem. In the limit of a weak bare Kondo 
coupling, the impurity contribution  
$S^{\rm{Th}}_{\rm{imp}}(T)\approx \ln 2$ at $T\gg T_K$ and
$S^{\rm{Th}}_{\rm{imp}}(T)\to 0$ at $T\ll T_K$. This reflects the fact that the 
bare coupling of the magnetic impurity to the conduction electrons is very weak 
so that we obtain essentially the full entropy of a free spin-1/2, $\ln 2$ at 
$T\gg T_K$. However, as the temperature is lowered the spin becomes "screened" {\it{i.e.}}
it goes into a singlet state and the impurity entropy is accordingly lost. The 
asymptotic values of $S^{\rm{Th}}_{\rm{imp}}(T)$ at high and low temperatures are characteristic 
of the RG fixed points of the Kondo Hamiltonian. 

However much less is known about real space spatial properties of the Kondo effect, in particular the so-called Kondo screening length $\xi_K\sim v_F/T_K$ (where $v_F$ is the Fermi velocity) which is expected to govern spatial correlations at $T\ll T_K$~\cite{affleck_kondo_2009}. Such a Kondo screening cloud\footnote{Probably better described as a "faint fog"~\cite{bergmann_quantitative_2008}}, which can be seen as the region over which a magnetic impurity form a singlet with a conduction electron, is very hard to experimentally observe in realistic systems because it occurs over quite large distances $\sim \mu$m, and decays as $1/r^d$~\cite{affleck_kondo_2009}, thus making reduced dimensionality better candidates to observe it~\cite{affleck_detecting_2001,sorensen_kondo_2005,pereira_kondo_2008}. 

\subsubsection{Impurity entanglement in the one-channel Kondo problem}
Recently, several developments have been made to study how the Kondo length $\xi_K$ emerges in the entanglement properties of quantum impurity problems~\cite{sorensen_impurity_2007,sorensen_quantum_2007,affleck_entanglement_2009,sodano_kondo_2010,eriksson_impurity_2011,deschner_impurity_2011,bayat_entanglement_2012,saleur_entanglement_2013,vasseur_universal_2014,lee_macroscopic_2015,pixley_entanglement_2015,erdmenger_entanglement_2016}.
Assuming a direct analogy with the thermal impurity entropy which is a scaling function $f(T/T_K)$, one migh expect the zero temperature impurity entanglement entropy to be also a scaling function $g(r/\xi_K)$, where $r$ is the spatial extension of the subsystem over which entanglement is computed.

\begin{figure}[!ht]
\begin{center}
\includegraphics[width=.7\columnwidth,clip]{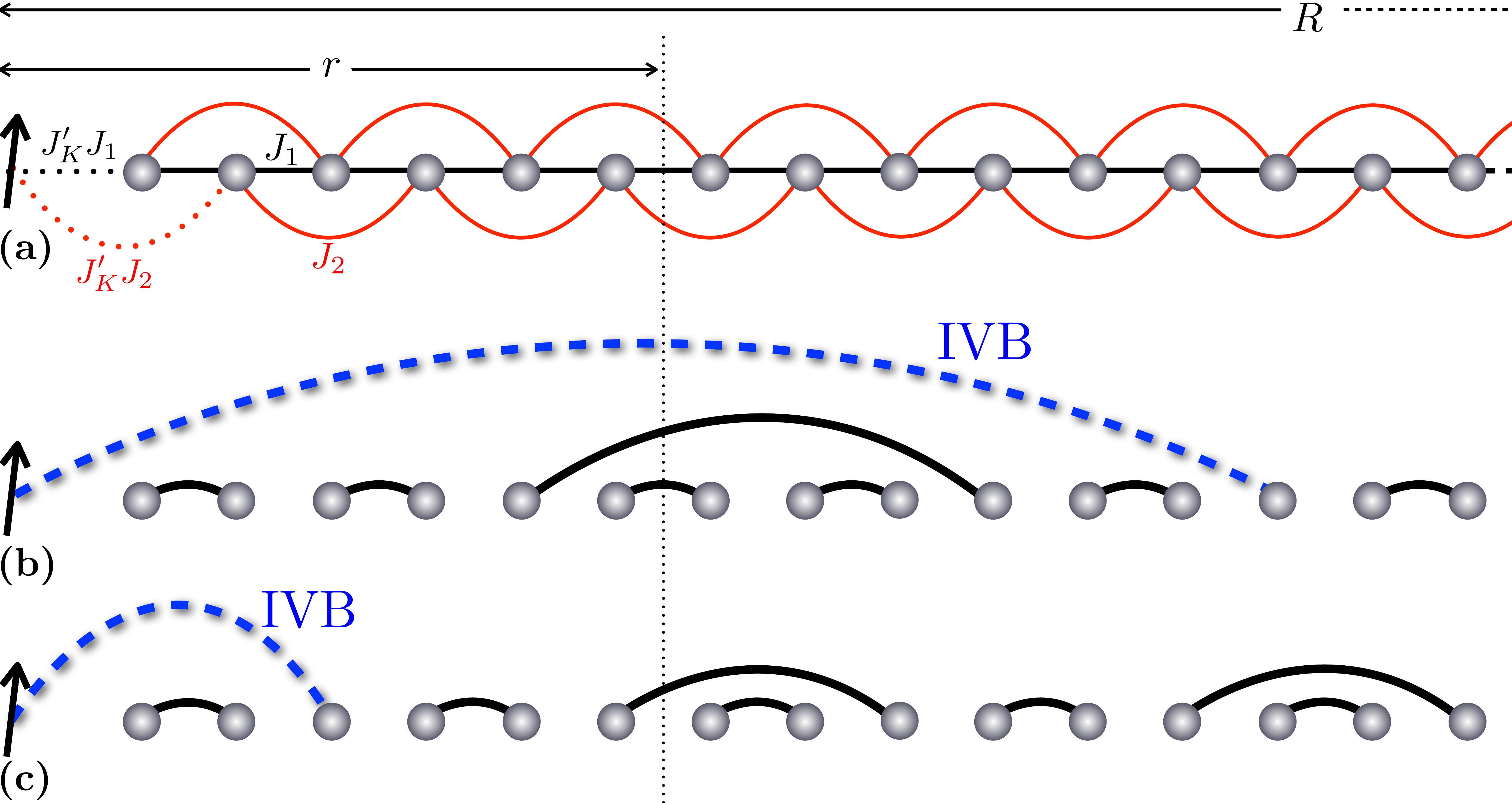}
\end{center}
\caption{(a) Schematic picture for the open frustrated Heisenberg chain model Eq.~\eqref{eq:sckm} coupled to a spin impurity (arrow).
(b) and (c) Example of valence bond configurations with the Kondo singlet materialized by the impurity valence bond (IVB) shown (blue dashed bond) for weak (b) and strong (c) coupling regimes. The total chain length is $R$ and the subsystem has $r$ sites, including the impurity spin.}
\label{fig:IVB}
\end{figure}

The usual Kondo Hamiltonian~\cite{andrei_solution_1983} contains a Heisenberg interaction between a 
$s_{\rm imp}=1/2$ impurity spin
and  otherwise non-interacting electrons:
\begin{equation}
{\cal H}=\int d^{3}r\left[\psi ^{\dagger }(-\nabla ^{2}/2m)\psi +J _{K}\delta^3(\vec
    r)\psi ^{\dagger }(\vec{\sigma}/2)\psi \cdot \vec{s}_{\rm imp}\right].  \label{eq:H3DKondo}
\end{equation}
At zero temperature, the impurity spin is screened by the conduction electrons through the formation
of a Kondo singlet, expected to take place over a length scale:
\begin{equation}
\xi_K=v_F/T_K\propto e^{1/(\nu J_K)},
\end{equation}
where $\nu$ is the density of states per spin band, $T_K$ is the Kondo temperature and $v_F$ the velocity of the fermions. Due to the $\delta-$function form of the interaction
Eq.~(\ref{eq:H3DKondo}) can be reduced to a one-dimensional model on a semi-infinite line with the impurity spin at the origin~\cite{sorensen_quantum_2007}. A further simplification is possible using the spin chain Kondo model~\cite{eggert_magnetic_1992,laflorencie_kondo_2008}, depicted in Fig.~\ref{fig:IVB} (a), and governed by the following spin-$\frac{1}{2}$ Heisenberg chain Hamiltonian
\be
{\cal H} =
J_{1}\sum_{i=1}^{L-1}\vec{S}_{1}\cdot \vec{S}_{i+1}+J_{2}^c\sum_{i=2}^{L-2}\vec{%
S}_{i}\cdot \vec{S}_{i+2} + J_{K}'{\vec{s}}_{\rm imp}\cdot \left(J_1\vec{S}_{1}+J_{2}\vec{S}_{2}\right),
\label{eq:sckm}
\ee
where the frustrated second neighbor antiferromagnetic coupling is tuned at the critical point $J_2^c\simeq 0.2412$~\cite{eggert_numerical_1996,rachel_detecting_2012} where logarithmic corrections vanish\footnote{For $J_2=0$ Eq.~\eqref{eq:sckm} was shown to be integrable with Bethe Ansatz by Frahm and Zvyagin~\cite{frahm_open_1997}, with also a strong analogy with the Kondo problem but with a modified Kondo temperature $T_K\sim \exp(-b/\sqrt{J'_K})$. For $J_2>J_2^c$, the spin chain enters a dimerized
phase~\cite{haldane_spontaneous_1982} with a gap and the relation between
Eq.~(\ref{eq:sckm}) and Kondo physics no longer holds.}.

Numerical simulations on this simplified Kondo model Eq.~\eqref{eq:sckm} have been carried out~\cite{laflorencie_boundary_2006,sorensen_impurity_2007,sorensen_quantum_2007,sodano_kondo_2010}, showing that (i) open boundary conditions lead to oscillations of the entanglement entropy, slowly decaying away from the boundaries with an exponent governed by the Luttinger parameter~\cite{laflorencie_boundary_2006}; and (ii) the impurity contribution to the entanglement entropy appears to be a universal scaling function $S^{\rm{E}}_{\rm{imp}}(R,r,J'_K)=g(r/\xi_K,R/\xi_K)$, where $r$ ($R$) is the subsystem (full system) size, see Fig.~\ref{fig:IVB}, and $\xi_K(J'_K)$ is the Kondo length.

\begin{figure}[t]
\begin{center}
\includegraphics[width=.825\columnwidth,clip]{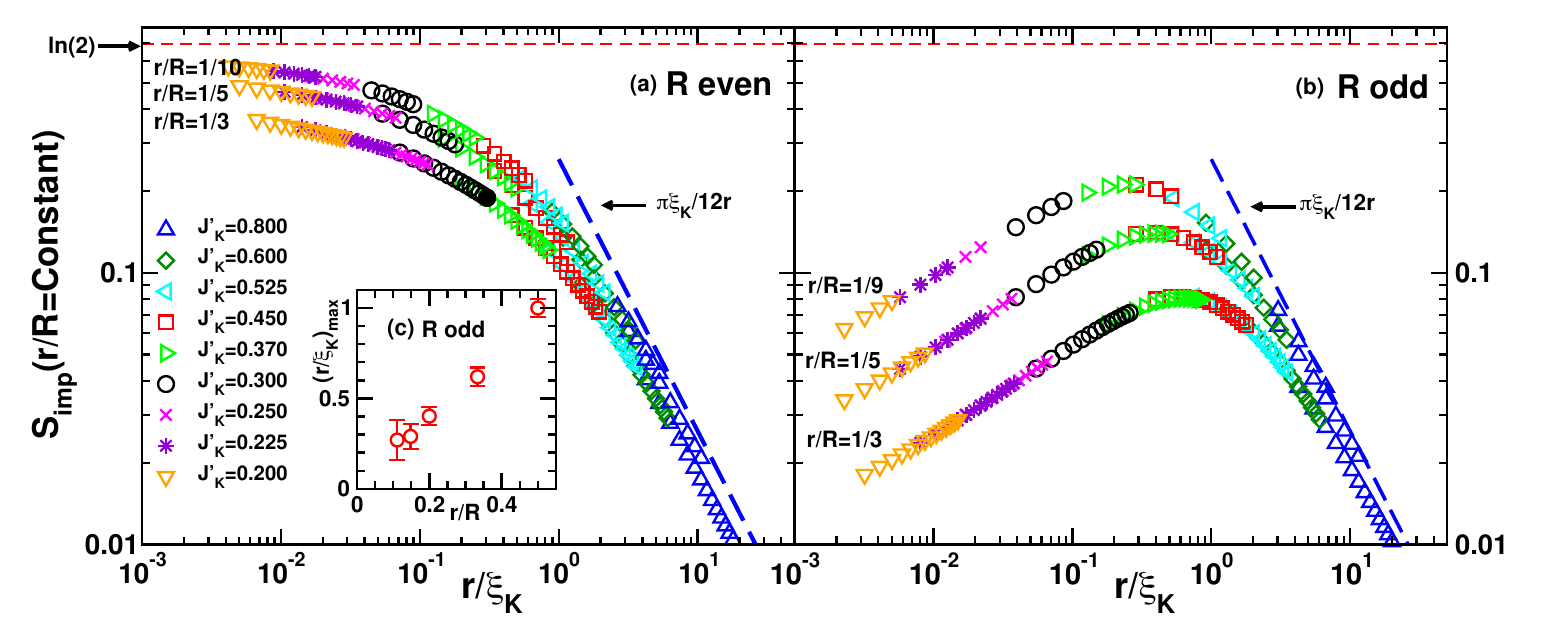}
\caption{Universal scaling plot of $S_{\rm{imp}}$ at fixed $r/R$,
(a) for $R$ even, (b) $R$ odd.
DMRG results for the $J_{1}-J_{2}$ spin chain Kondo model Eq.~\eqref{eq:sckm} 
for various Kondo couplings $J_{K}^{\prime }$. 
Dashed line $\pi \xi_K/(12r)$ is the Fermi liquid theory prediction. Inset (c):  location of the maximum, $(r/\xi_K)_{\rm max}$,
of $S_{\rm imp}$ for odd ${R}$, plotted versus $r/{R}$. From Ref.~\cite{sorensen_impurity_2007}.}
\label{fig:simp}
\end{center}
\end{figure}

The simplest quantity to study is the single site impurity entanglement entropy (defined for $r=1$) which is directly proportional to the impurity magnetization $\langle s^z_{\rm imp}\rangle$. This local observable, studied for the spin-boson model~\cite{leggett_dynamics_1987,kopp_universal_2007,le_hur_entanglement_2007} and for the usual Kondo problem~\cite{sorensen_scaling_1996,barzykin_kondo_1996} was shown to violate scaling in Ref.~\cite{sorensen_quantum_2007}. Indeed, the Kondo screening mechanism being a non-local many-body quantum effect, scaling properties are expected to occur for many-body correlators, as for instance in the entanglement of a finite size $r>1$ region. The impurity contribution to the entanglement entropy was therefore studied~\cite{sorensen_impurity_2007,sorensen_quantum_2007} in the limit $1\ll r\ll R$, defined by
\be
S_{\rm imp}(R,r,J'_K)=S_{\rm u}(R,r,J'_K)-S_{\rm u}(R-1,r-1,1),
\ee
where $S_{\rm u}(R,r,J'_K)$ is the uniform part of the entropy for a total system of length $R$, a subsystem of length $r$ and a boundary coupling $J'_K$. DMRG simulation results are displayed in Fig.~\ref{fig:simp} for $R$ even and odd. In the strong coupling regime $r/\xi_K\gg 1$, the local Fermi liquid theory of Nozi\`eres~\cite{nozieres_fermi-liquid_1974} predicts
$S_{\rm imp}=\pi\xi_K/(12 r)$, perfectly captured by the numerical results in Fig.~\ref{fig:simp}. In the opposite limit, when the Kondo cloud is much larger than the subsystem $r/\xi_K\ll 1$, DMRG data from Fig.~\ref{fig:simp} suggest  $\lim_{r/\xi_K\to 0}\lim_{r/R\to 0}S_{\rm imp}(r/\xi_K,r/{R})=\ln 2$, 
for either parity of $R$. 
%

A heuristic picture giving the general features of $S_{\rm imp}(r/\xi_K,r/{R})$,
for both even and odd ${R}$, sketched in Ref.~\cite{sorensen_impurity_2007,sorensen_quantum_2007},
can be derived from a resonating valence bond view of the ground state, as depicted if Fig.~\ref{fig:IVB} (b,c).
We loosely identify 
$S_{\rm imp}$ with $\ln 2 ~\times$ the probability of an ``impurity valence bond'' (IVB) stretching 
from the impurity site into the region $>r$ (Fig.~\ref{fig:IVB}).
The typical 
length of the IVB is expected to be $\sim \xi_K$ (for $\xi_K<{R}$), leading to the monotonic decrease of $S_{\rm imp}$ 
with increasing $r/\xi_K$, for $R$ even, and its 
vanishing as $r/\xi_K\to \infty$. For ${R}$ odd, the 
ground state contains one unpaired spin.  When $\xi_K\to\infty$ this unpaired spin 
is the impurity itself, implying no IVB and hence $S_{\rm imp}=0$.  As $\xi_K$ decreases 
the probability of having an IVB increases, since the unpaired spin 
becomes more likely to be at another site, due to Kondo screening, 
but the average length of the IVB, when it is present, decreases. 
These two effects trade off to give $S_{\rm imp}$ a maximum when $\xi_K\propto R$, in good agreement 
with the inset of Fig.~\ref{fig:simp}, at which 
point the probability of having an IVB becomes $O(1)$ but the decreasing average size of the IVB 
starts to significantly reduce the probability of it stretching into the region $>r$.

\subsubsection{Boundary effects}
\label{sec:OBC}
Conformally invariant boundary fixed points are expected to present a term
in the entanglement entropy which corresponds
to the zero temperature Affleck-Ludwig impurity entropy $\ln g$~\cite{affleck_universal_1991}, as pointed out in Ref. \cite{calabrese_entanglement_2004}. Therefore, in a semi-infinite chain with the subsystem staring at the boundary, Eq.~\eqref{eq:S1d} becomes
\be
S^{\rm OBC}_q(r)=\frac{c}{12}\left(1+\frac{1}{q}\right)\ln r +s_q/2 +\ln g +\ldots
\label{eq:S1dobc}
\ee  
For the single 
channel Kondo model $\ln g = \ln 2$ at the weak 
coupling fixed point where the impurity is unscreened, and 
$\ln g = 0$ at the strong coupling fixed point where 
it is screened. The thermodynamic
impurity entropy decreases monotonically from $\ln 2$ to $0$ with decreasing $T/T_K$, in the same way as the impurity entanglement entropy with increasing $r/\xi_K$. 

The validity of Eq.~\eqref{eq:S1dobc} has been checked numerically for various examples of conformally invariant boundary conditions, for quantum Ising chains~\cite{zhou_entanglement_2006,sorensen_quantum_2007,taddia_entanglement_2013,taddia_entanglement_2013-1}, a case where there is no boundary induced oscillations, as well as for the two-channel Kondo problem~\cite{alkurtass_entanglement_2016}. Conversely, for critical open chain models having a continuous symmetry, slowly decaying oscillations have been reported~\cite{laflorencie_boundary_2006,legeza_entropic_2007,szirmai_spatially_2008,lauchli_spreading_2008,roux_spin_2008} and interpreted as $2k_F$-Friedel oscillations~\cite{friedel_metallic_2007}, whose decay is governed by the Luttinger parameter~\cite{laflorencie_boundary_2006,fagotti_universal_2011}. An exact computation for the XX (free-fermions) case was done by Fagotti and Calabrese~\cite{fagotti_universal_2011} who derive rigorously the asymptotic behavior for large block sizes on the basis of a recent mathematical theorem for the determinant of Toeplitz plus Hankel matrices. Interestingly, unusual 
oscillating corrections have been reported for higher R\'enyi indices $q>1$ for open systems, and also for periodic chains, {\it{i.e.}} in the absence of boundary~\cite{calabrese_parity_2010,cardy_unusual_2010,calabrese_universal_2010}.
%

\begin{figure}[ht!]
\begin{center}
\includegraphics[width=0.5\columnwidth,clip]{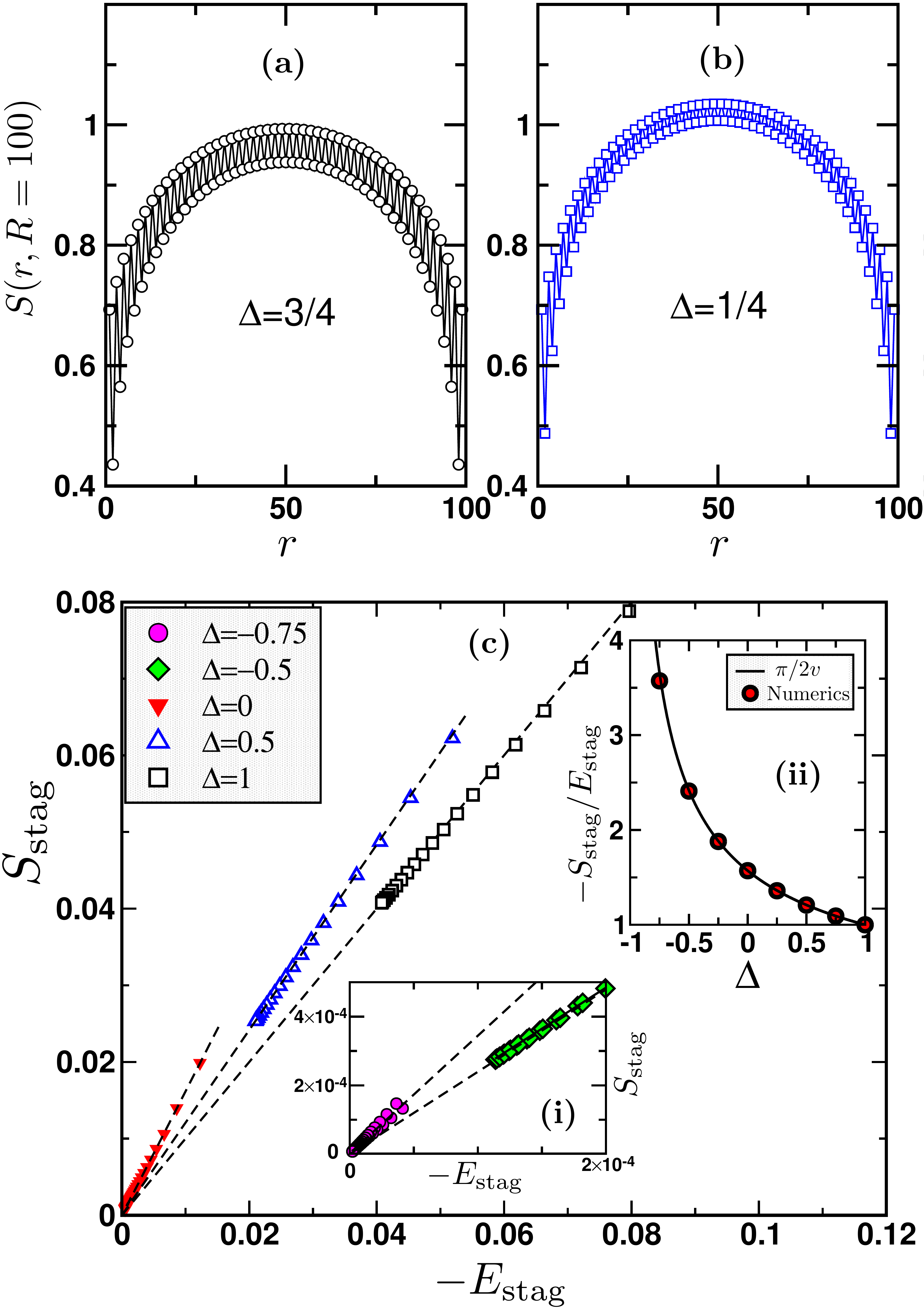}
\caption{(a) and (b) Von Neuman entropy of XXZ chains of lengths $R=100$ with
  OBC computed with DMRG for various anisotropies $\Delta$. One sees the oscillating feature, whose amplitude is reduced when $\Delta$ decreases. Linear behavior of the staggered part of the
  entanglement entropy, ${{S}}_{\rm stag}$ as a function of the staggered energy
    density, $-E_{\rm stag}$, both computed using DMRG  on critical open XXZ chains of
    size $200\le R\le 1000$ for various anisotropies $\Delta$. Data
    from free-fermions diagonalization at $\Delta=0$ are also shown for $R=2000$.  Dashed lines are linear
    fits of the form  Eq.~\eqref{eq:SE}.  Inset (i) is a
    zoom close to 0, showing data for $\Delta=-3/4$ and $-1/2$.  Inset (ii)
    shows the proportionality factor against $\Delta$ extracted from the numerical data
    (circles), for a larger set of values of $\Delta$,  which is compared with
    $\pi /2v$.Reprinted from \cite{affleck_entanglement_2009} (a,b) and from \cite{laflorencie_boundary_2006} (c).}
\label{fig:stag}
\end{center}
\end{figure}
Despite the good analytical description for free fermions~\cite{fagotti_universal_2011}, boundary induced oscillations in the von-Neumann entropy still lack a complete analytical understanding for interacting critical systems. A phenomenological grasp was conjectured, based on DMRG data analysis and a valence bond picture in Ref.~\cite{laflorencie_boundary_2006} where the alternating part in the entropy for critical XXZ chains was found to be directly proportional to the
alternating term in the energy density.  Indeed, open end breaks
translational invariance and a slowly decaying alternating term (at $2k_F = \pi$ in the absence of external magnetic field)
or ''dimerization'' in the energy density $E_{\rm u}(r,R)+(-1)^{r}E_{\rm stag}(r,R)$ appears at distance $r$ for chains of length $R$, where for critical chains $E_{\rm stag}(r,R)\propto [\frac{2{R}}{\pi }\sin (\frac{\pi
    r}{{R}})]^{-K}$,
$K$ being the Luttinger liquid parameter. The oscillating term in the entropy was then found to be intimately related to such energy density oscillations, such that 
\be {S}_{\rm stag}=-(\pi a^2/2v)E_{\rm stag}, \label{eq:SE} \ee
where $v$ is the velocity of excitations, and $a$ the lattice spacing, introduced to respect the dimensionless character of the entanglement
entropy. Fig.~\ref{fig:stag} summarizes these results, obtained in Refs.~\cite{laflorencie_boundary_2006,affleck_entanglement_2009}.

It is important to note that boundary induced oscillations are relevant for the DMRG technique which better perform with OBC, and also in the experimental context of cold atoms where such oscillations have been predicted for trapped bosons~\cite{campostrini_quantum_2010}. Likewise, open ends lead to oscillations in gapped models, such as the $s=1$ Heisenberg chain \cite{fan_boundary_2007}, but with an alternating part which decays exponentially fast with the distance from the boundary, confirming the fact that this alternating component is controlled by the spin-spin correlations. Let us finally notice that boundary critical phenomena have also been studied using entanglement renormalization schemes~\cite{evenbly_boundary_2010,silvi_entanglement_2010}.

\subsubsection{Other examples of quantum impurity problems probed by entanglement}

The two-impurity Kondo problem was investigated using two-site entanglement witnesses in Ref.~\cite{cho_quantum_2006}, or the more complex many-body Schmidt gap, expected to play the role of an order parameter close to quantum criticality~\cite{bayat_order_2014}, likewise observed for the two-channel Kondo problem~\cite{alkurtass_entanglement_2016}. 
Another example of quantum criticality was studied for the spin-boson model~\cite{leggett_dynamics_1987} where quantum entanglement of the impurity was studied in great detail~\cite{costi_entanglement_2003,kopp_universal_2007,le_hur_entanglement_2007,hur_entanglement_2008}.
Entanglement entropy was also studied near Kondo-destruction quantum critical points~\cite{pixley_entanglement_2015}.

Local defects in quantum wires, reminiscent of the so-called Kane-Fisher problem~\cite{kane_transport_1992,eggert_magnetic_1992}, have been studied quite intensively using entanglement estimates~\cite{peschel_entanglement_2005,zhao_critical_2006,apollaro_entanglement_2006,calabrese_entanglement_2012-2,freton_infrared_2013,vasseur_universal_2014,petrescu_fluctuations_2014,pouranvari_effect_2015}. 
Dynamical properties have also been investigated for quantum impurity problems. For instance in quantum Ising chains with local or extended defects, the time evolution of the entanglement entropy, studied by Igl{\'o}i and co-workers~\cite{igloi_entanglement_2009}, displays a logarithmic growth with a non-universal prefactor which depends on the  details of the defects.

The Kondo cloud dynamics has been probed after a quantum quench in the resonant level model. At long enough time inside the light cone, the Kondo screening cloud relaxes exponentially to the final equilibrium structure, with a relaxation rate given by the emergent energy scale of impurity screening~\cite{ghosh_dynamics_2015}. Finally,  universal oscillations in the entanglement levels have also been reported by Bayat and co-workers~\cite{bayat_universal_2015} for the two-impurity Kondo spin chain model after a local quench of the RKKY interaction from RKKY to Kondo regime.
\subsection{Many-body localization}
\label{sec:mbl}
\subsubsection{General properties}
The so-called many-body localization phenomenon has attracted a huge interest in
recent years, following precursor
works~\cite{fleishman_interactions_1980,altshuler_quasiparticle_1997,jacquod_emergence_1997,georgeot_integrability_1998,
gornyi_interacting_2005,basko_metalinsulator_2006} which discussed whether Anderson localization~\cite{anderson_absence_1958} can survive
interactions. A new paradigm has then emerged for many-body eigenstates of strongly disordered interacting systems which may not obey the "eigenstate thermalization hypothesis" (ETH)~\cite{deutsch_quantum_1991,srednicki_chaos_1994,rigol_thermalization_2008} and would fail to thermalize~\cite{nandkishore_many-body_2015}. Such a many-body localization (MBL) clearly challenges the very foundations of quantum statistical physics~\cite{nandkishore_many-body_2015,altman_universal_2015} as thermalization in such closed systems cannot occur without external bath (for a recent discussion, see also Ref. \cite{nandkishore_spectral_2014}). Another interesting property is that MBL may lead to long-range (possibly topological) order, otherwise absent for equilibrated systems~\cite{huse_localization-protected_2013,bahri_localization_2015,chandran_many-body_2014,bauer_area_2013,vosk_dynamical_2014}.
Furthermore, the MBL
 shows some similarities with integrable systems, with an extensive number of quasi-local integrals of
motion~\cite{vosk_many-body_2013,serbyn_local_2013,ros_integrals_2015,chandran_constructing_2015}. 

Several numerical studies have recently focused on the MBL phase, and associated phase transitions. Most of the studied models are one-dimensional, mostly for practical reasons: {\it{e.g.}} the random field Heisenberg or XXZ chain~\cite{znidaric_many-body_2008,pal_many-body_2010,bardarson_unbounded_2012,luca_ergodicity_2013,luitz_many-body_2015}, $d=1$ interacting fermions~\cite{oganesyan_localization_2007,bauer_area_2013,mondaini_many-body_2015} (relevant to the recent experiments~\cite{schreiber_observation_2015} on quasi-periodic chains~\cite{Iyer_many-body_2013}), quantum Ising chains~\cite{huse_localization-protected_2013,vosk_dynamical_2014,kjall_many-body_2014}, spin-glasses~\cite{laumann_many-body_2014}, and also translationally-invariant models~\cite{grover_quantum_2014,de_roeck_scenario_2014,schiulaz_dynamics_2015,yao_quasi_2014}, or driven systems~\cite{ponte_many-body_2015}. While exact diagonalization are typically limited to $\sim 20$ sites~\cite{luitz_many-body_2015}, larger systems can also be studied using approximate methods such as strong disorder RG techniques, the so-called RSRG-X method~\cite{pekker_hilbert-glass_2014,vosk_dynamical_2014,huang_excited-state_2014,vasseur_quantum_2015-1}, or Matrix Product States approaches exploiting the low entanglement property of the MBL regime~\cite{friesdorf_many-body_2015,chandran_spectral_2015,pekker_encoding_2014,khemani_obtaining_2015,yu_finding_2015}. However, these techniques are practically limited to strong disorder on the MBL side of the phase diagram where they are well controlled. At weaker disorder, in the ergodic regime, only exact numerical approaches are expected to give quantitatively reliable results.

\begin{figure}[h]
\begin{center}
\includegraphics[width=.725\columnwidth,clip]{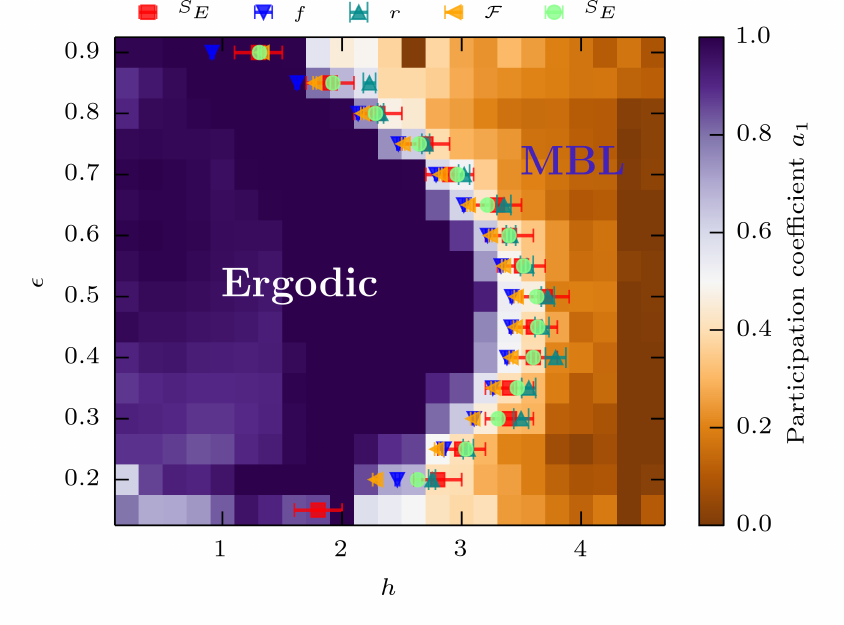}
\caption{Phase diagram Disorder ($h$) {\it{vs.}} Energy density ($\epsilon$) of the
            random-field Heisenberg chain Eq.~\eqref{eq:mbl}. The ergodic phase (dark region with a
            participation entropy coefficient $a_1\simeq 1$) is separated from the
            MBL regime (bright region with $a_1 \ll 1$). Various symbols show the
        energy-resolved MBL transition points extracted from finite size scaling
        performed over system sizes $L\in\{14,15,16,17,18,19,20,22\}$ for different quantities. 
Reprinted from \cite{luitz_many-body_2015}.}
\label{fig:mbl}
\end{center}
\end{figure}

A canonical example of interacting model which exhibits a MBL transition is the random field Heisenberg $s=1/2$ chain\footnote{Equivalent via a  Jordan-Wigner (Matsubara-Matsuda) transformation to interaction spinless fermions (hard-core bosons) in a random potential.}, governed by the  Hamiltonian
\be
{\cal H}=\sum_{i=1}^{L} \left(J{\vec{S}}_i\cdot {\vec{S}}_{i+1}-h_i S_i^z\right),
\label{eq:mbl}
\ee
where the random field is drawn from a uniform distribution in $[-h,h]$.  Using a spectral transformation, this model has been studied by exact diagonalization at varying energy densities across the entire spectrum up to $L =22$ spins~\cite{luitz_many-body_2015} in order to extract the many-body mobility edge shown in Fig.~\ref{fig:mbl}.

Because the MBL phenomenon leads to the absence of thermalization, we do not expect any sort of thermodynamic signatures of a localization/delocalization transition in finite temperature observables. Instead, the transition has to be studied either at infinite temperature, {\it{i.e.}} with all eigenstates equally weighted~\cite{oganesyan_localization_2007,pal_many-body_2010}, or by targeting eigenstates at finite energy density above the ground-state~\cite{kjall_many-body_2014,luitz_many-body_2015}, which is better justified for systems having a many-body mobility edge at finite energy density $\epsilon$ (Fig.~\ref{fig:mbl}). Another possibility to study such a dynamical transition is to make a global quench from a high-energy unentangled product state~\cite{bardarson_unbounded_2012,vosk_many-body_2013}. 

Numerical studies of the many-body eigenstates provide a very precise and quantitive way to characterize the localized/delocalized nature of the system. For example, a popular way to differentiate extended and localized phases relies on the spectral statistics from random
matrix theory~\cite{rmt}. As exploited in several works~\cite{jacquod_emergence_1997,georgeot_integrability_1998,oganesyan_localization_2007,pal_many-body_2010,cuevas_level_2012,laumann_many-body_2014,luitz_many-body_2015}, the ergodic regime harbors a statistical distribution of level spacings
which follows Wigner's surmise of the Gaussian orthogonal ensemble (GOE), while a Poisson distribution is
expected for localized states. At the transition, it is not clear whether there is a continuous family of critical theories~\cite{laumann_many-body_2014} or semi-Poisson statistics~\cite{serbyn_spectral_2016,monthus_many-body-localization_2016,monthus_level_2016}. 
\begin{table}[hb]
\centering
\begin{tabular}{l||c|c|c}
&Delocalized&Transition&MBL\\
\hline
Spectral statistics&GOE&?&Poisson\\
Entanglement entropy $S^{\rm E}(L)$&volume-law&volume-law\footnote{See discussions in Refs.~\cite{grover_certain_2014,chandran_finite_2015,monthus_many-body-localization_2016}.}&area law\\
Entanglement variance $\sigma_{\rm E}^2(L)$&vanishes&diverges&finite\\
Entanglement dynamics $S^{\rm E}(t)$& $t^{1/z}$&$\ln t$&$\ln t$\\
\hline
\hline
\end{tabular}
\caption{\label{tab:ent} Various properties of ergodic and MBL phases, as well as at the transition.}
\end{table}

The level statistics properties are summarized in Table~\ref{tab:ent}, together with entanglement features across the different regimes. Below we discuss in detail how entangled are the eigenstates in both delocalized and MBL regimes, as well as at the transition between the two regimes. We first focus on physics at equilibrium, looking at eigenstates for which a clear transition from a volume to an area law scaling of the entropy is observed. Then, the non-equilibrium situation is examined through quantum quenches, starting from high-energy untangled states.
\subsubsection{Area {\it{vs.}} volume law for highly excited states}
Thermalization in isolated quantum systems implies that the system itself acts as its own heat bath~\cite{gogolin_equilibration_2016}. This is the case for the so-called ergodic regime (adjacent of the MBL phase, see Fig.~\ref{fig:mbl}) where ETH~\cite{deutsch_quantum_1991,srednicki_chaos_1994} is expected to hold.  In this delocalized phase, the RDM of a high energy eigenstate can be viewed as a thermal density matrix at high temperature. Therefore, the entanglement entropy of such a highly excited eigenstate is very close to the thermodynamic entropy of the subsystem at high temperature, thus exhibiting a volume-law scaling. Such delocalized eigenstates can therefore be described as pure random states~\cite{page_average_1993}. Volume-law entanglement at high temperature has been verified for disorder-free quantum spin chains~\cite{sorensen_quantum_2007,alba_entanglement_2009,sato_computation_2011,alba_eigenstate_2015,keating_spectra_2015}, as well as for the  ergodic regime of weakly disordered chains~\cite{bauer_area_2013,kjall_many-body_2014,luitz_many-body_2015}.

\begin{figure}[b]
\begin{center}
\includegraphics[width=.7\columnwidth,clip]{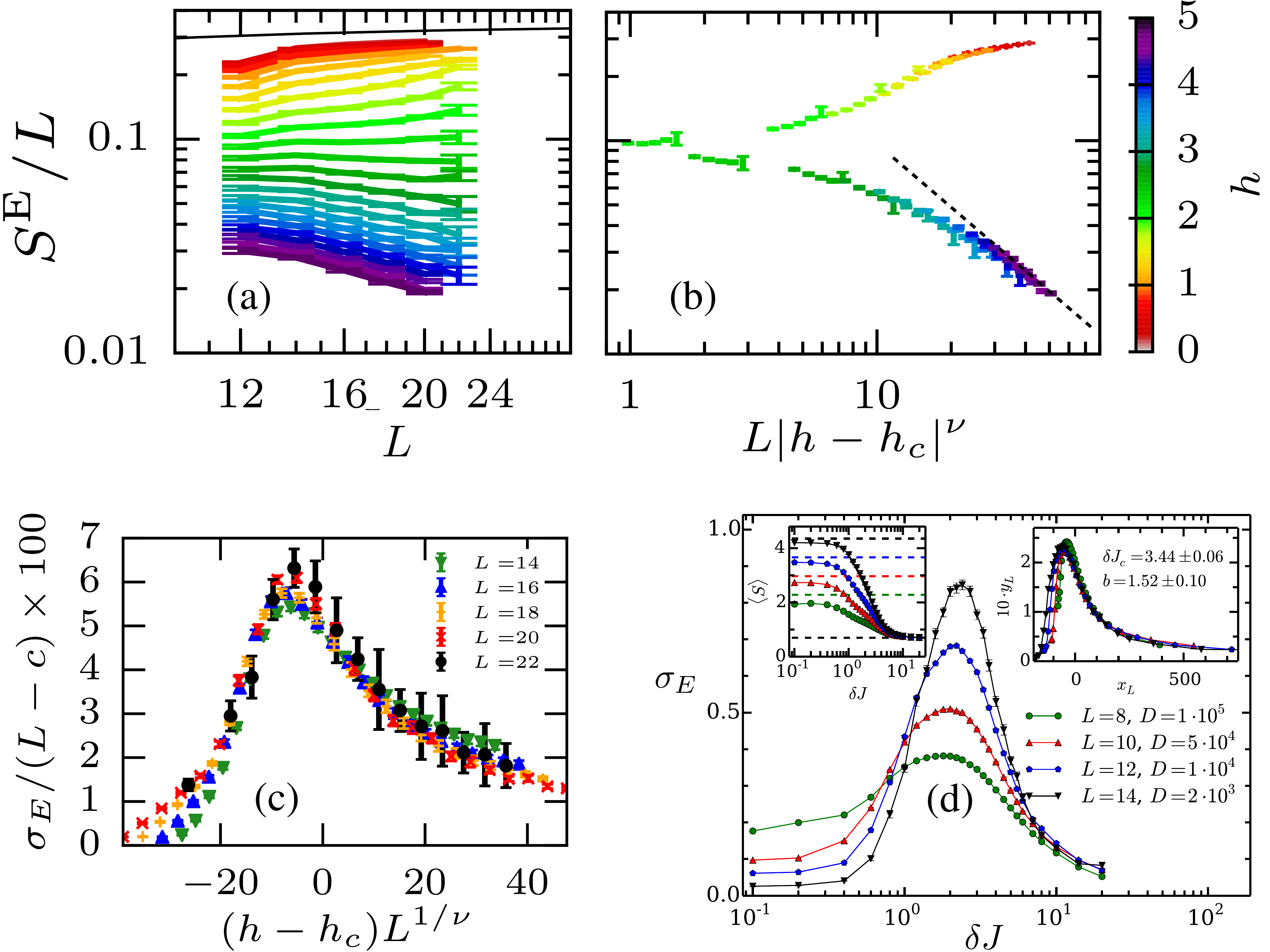}
\caption{Exact diagonalization results for the disorder-average entanglement (von-Neumann) entropy at half-chain for (a-c) the random field Heisenberg chain model Eq.~\eqref{eq:mbl} at a fixed energy density $\epsilon=0.8$ with $L=12,\ldots, 22$~\cite{luitz_many-body_2015}, and (d) on a disordered quantum Ising chain computed at a fixed energy density in the middle of the many-body spectrum,with $L=8,\ldots,14$~\cite{kjall_many-body_2014}. (a) There is a clear transition from a volume-law at small disorder $h$ to a sub-volume law at $h>h_c\sim 2.3$, compatible from the scaling plot (b) with a strict area law (dased line). (c-d) Singular behavior of the  standard deviation $\sigma_{\rm E}$ of the entropy at the transition.
Reprinted from \cite{luitz_many-body_2015} (a-c) and from \cite{kjall_many-body_2014} (d).}
\label{fig:volumearea}
\end{center}
\end{figure}

\paragraph{Many-body localization---}
On the other hand, 
contrary to the thermal phase, the MBL regime does not obey ETH~\cite{pal_many-body_2010} and the eigenstates sustain a much smaller (area law) entanglement, qualitatively closer to the entanglement entropy of a ground-state~\cite{eisert_colloquium:_2010}. 
Such qualitatively distinct properties have been clearly shown numerically in various exact diagonalization studies~\cite{bauer_area_2013,kjall_many-body_2014,luitz_many-body_2015}. In Fig.~\ref{fig:volumearea} (a-b) exact diagonalization results of model Eq.~\eqref{eq:mbl} are shown for the disorder-average von-Neumann  entropy of half-chains for eigenstates lying in the middle of the many-body spectrum ($\epsilon=0.5$ in Fig.~\ref{fig:mbl}) as a function of size $L$ and disorder strength $h$. The transition from volume to sub-volume scaling is clearly visible, as provided by the scaling plot in panel (b). Numerical data are compatible with a volume-law entanglement entropy at criticality~\cite{grover_certain_2014}, and with a strict area law scaling for the MBL regime (dashed line in panel (b) of Fig.~\ref{fig:volumearea})\footnote{In the MBL regime, Bauer and Nayak~\cite{bauer_area_2013} reported a weak logarithmic violation of the area law for the maximum entropy, obtained from the (sample-dependent) optimal bipartition. A logarithmic violation was also found using matrix product states~\cite{kennes_entanglement_2015}.}. Breakdown of  the area law  at the delocalization transition was recently exploited in a numerical linked cluster expansion study~\cite{devakul_early_2015} to locate the MBL transition\footnote{This study concluded for a critical boundary $h_c(\epsilon)$ larger than the one of Fig.~\ref{fig:mbl} from~\cite{luitz_many-body_2015}.}.

Another very interesting point concerns the entropy histograms, as discussed in  Refs.~\cite{bauer_area_2013,kjall_many-body_2014,luitz_many-body_2015,lim_nature_2015,luitz_long_2016}, for which a qualitative change is observed across the different regimes. The evolution of the standard deviation $\sigma_{\rm E}$ with increasing system length $L$ provides a quantitative tool, plotted in Fig.~\ref{fig:volumearea} (c-d). While in the delocalized regime $\sigma_{\rm E}(L)\to 0$ (in agreement with Ref.~\cite{nadal_phase_2010}), it remains constant in the localized phase, and diverges with $L$ at the transition. This is likely a signature of an absence of self-averaging for the entropy, and a possible signature of infinite randomness physics at criticality~\cite{vosk_theory_2015}. \\

A quite surprising feature concerns the structure of the ES, as explored by Yang and co-workers~\cite{yang_two-component_2015} for high-energy levels in both ETH and MBL regimes. In the delocalized phase, they found a "two-component" structure of the entanglement levels, with a universal part corresponding to genuine random states~\cite{page_average_1993}, and a non-universal (model-dependent) part. Interestingly, the universal part lies in the "high-energy" sector of the ES, in contrast with usual expectations (see Section~\ref{sec:ES}).  The universal fraction decreases when the MBL transitions is approached and vanishes in the MBL regime. This observed effect for high-energy eigenstates clearly contrasts with ground-states of either pure~\cite{li_entanglement_2008,calabrese_entanglement_2008}  or disordered systems~\cite{leiman_correspondence_2015} for which the universal part (if it exists~\cite{chandran_how_2014}) is rather expected in the "low-energy" part. However, it is not excluded that such a two-component structure will vanish in the thermodynamic limit. Spectral statistics of the entanglement levels, studied in Ref.~\cite{geraedts_many-body_2016}, shows GOE statistics in the delocalized regime while an intriguing semi-Poisson behavior was found in the MBL regime. Interestingly, Serbyn and co-workers have found a power-law decay of the ES in the MBL phase, with a decay exponent related to the localization length~\cite{serbyn_universal_2016}.

\begin{figure}[h]
\begin{center}
\includegraphics[width=0.6\columnwidth,clip]{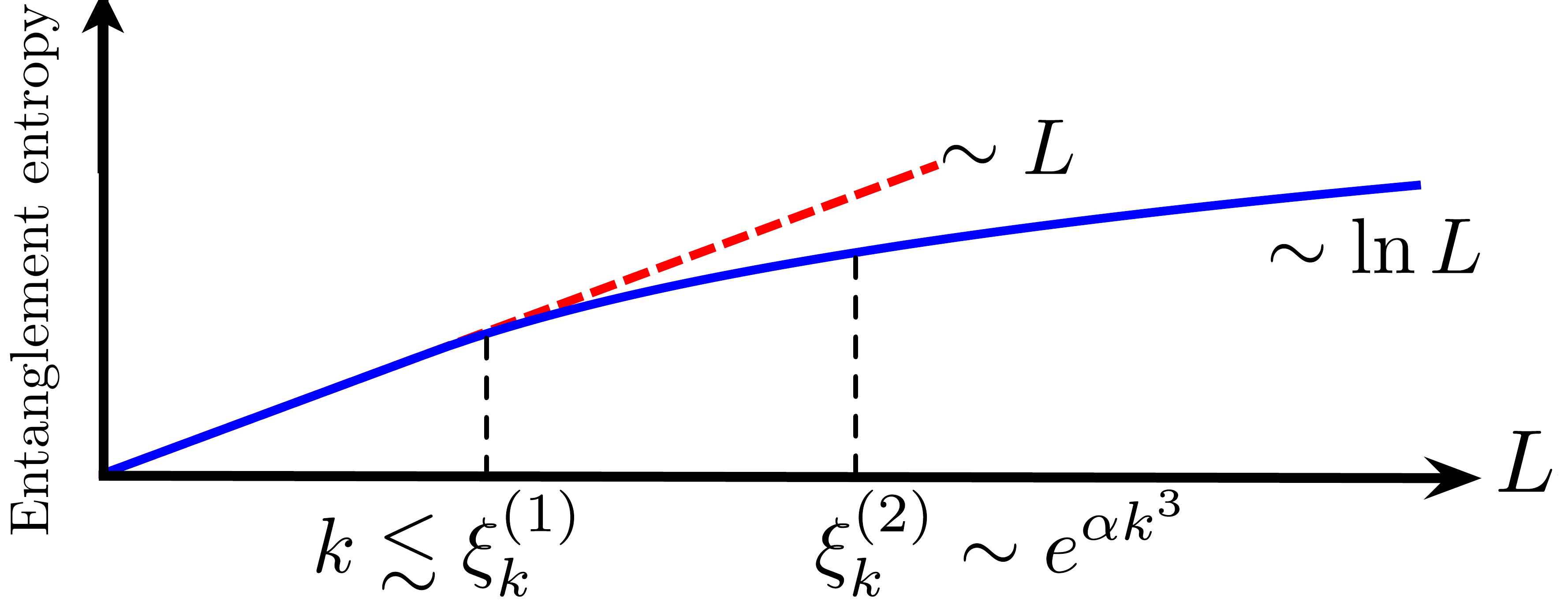}
\caption{Entanglement entropy at high energy for random bonds SU(2)$_k$ chains. Reprinted from \cite{vasseur_quantum_2015-1}.}
\label{fig:su2k}
\end{center}
\end{figure}

\paragraph{Random bonds---} SU(2)$_k$ anyonic chains~\cite{feiguin_interacting_2007} with random bonds~\cite{bonesteel_infinite-randomness_2007,fidkowski_textitc_2008,fidkowski_permutation-symmetric_2009} have also been studied in their high-energy regime~\cite{vasseur_quantum_2015-1}, using RSRG-X techniques~\cite{pekker_hilbert-glass_2014}. The well-known Ising ($k=2$), Potts ($k=4$) , and Heisenberg ($k=\infty$) cases are included in this family of models. In contrast with the random field case Eq.~\eqref{eq:mbl}, Vasseur and co-workers~\cite{vasseur_quantum_2015-1} have found for  random SU(2)$_k$ chains a set of infinite randomness fixed points that are the infinite-temperature analogs of Damle-Huse fixed points~\cite{damle_permutation-symmetric_2002}. The entanglement entropy of high-energy eigenstates, sketched in Fig.~\ref{fig:su2k}, displays a crossover from a volume-law scaling $\sim L$  at small lengths $L\ll k$ to the asymptotic regime $\sim \ln L$ for $L\gg \xi_{k}^{(2)}\sim \exp(k^3/4\pi^2)$. This logarithmic asymptotic behavior is also confirmed for the quantum Ising case $k=2$ by Huang and Moore in Ref.~\cite{huang_excited-state_2014}. Consequences for the Heisenberg case are interesting since we expect volume-law entropy and self-thermalization only in this $k\to\infty$ limit. Note also that for random-bonds XXZ chains, the strong disorder MBL regime is of spin-glass type with a strict area law entropy, as found in~\cite{vasseur_particle-hole_2016}.

%

\subsubsection{Entanglement growth after global quantum quenches}
The spreading in time of entanglement after a quantum quench is a topic which has attracted a huge amount of work, in the wake of nonequilibrium dynamics in isolated interacting quantum systems~\cite{polkovnikov_textitcolloquium_2011}. For integrable systems, it was originally shown~\cite{calabrese_evolution_2005,chiara_entanglement_2006} that the block entropy grows linearly with time $S(t)\sim t$ after a sudden global quench\footnote{There is a qualitative difference between global and local quenches~\cite{alba_entanglement_2014}. While the former displays ballistic spreading of entanglement, the later exhibits a much slower logarithmic growth~\cite{eisler_evolution_2007}. }, using both analytical (CFT) and numerical (exact diagonalization and time-dependent DMRG). This was also confirmed for non-integrable systems~\cite{kim_ballistic_2013} where the ballistic growth of $S(t)$ contrasts with energy transport which is diffusive.

Intuitively, due to the finite velocity of excitations we expect a finite velocity of propagation of information in quantum systems with local interactions, resulting in a 
Lieb-Robinson bound for entanglement spreading~\cite{bravyi_lieb-robinson_2006,eisert_general_2006}: $S(t)\le v_{\rm LR}t$. Quite remarkably, Burrell and Osborne have shown that for disordered XY spin chains~\cite{burrell_bounds_2007}, the effective light cone grows much slower $\sim \ln t$, thus yielding an entanglement spreading at most growing logarithmically with time, at least for non-interacting disordered fermionic chains.

We now discuss in details these properties of entanglement spreading in disordered interacting quantum systems after a global quench, starting at $t=0$ from a high-energy untangled product state $|\Psi(0)\rangle$, and letting the system evolve under the unitary evolution $\exp(-iHt)|\Psi(0)\rangle$. 

\paragraph{Logarithmic scaling in quantum glass phases and in the MBL regime---} 
One of the first numerical observation of a slow logarithmic growth of entanglement entropy after a global quench in disordered systems was achieved by De Chiara and co-workers~\cite{chiara_entanglement_2006} for XX spin chains with random bonds,\footnote{And no random field, thus preserving the particle-hole symmetry of the Hamiltonian for which we do not expect Anderson localization.}
following a quench from the simple (high-energy) N\'eel state $|\hskip -0.11cm \uparrow\downarrow\uparrow\downarrow\ldots\rangle$. A very slow spreading of entanglement $S(t)\sim \ln(\ln t)$ was also observed for the disordered transverse field Ising chain~\cite{igloi_entanglement_2012}, with a saturation value at long time $\sim \ln L$. However, when interactions are turned on, Vosk and Altman have shown using a dynamical RG approach~\cite{vosk_many-body_2013} for random bonds XXZ (in the easy-plane regime) that starting from the N\'eel state, the entanglement entropy would increase much faster with time: $S(t)\sim (\ln t)^{2/\phi}$ where $\phi=(1+\sqrt 5)/2$ is the golden mean. 
Moreover, Vasseur and co-workers~\cite{vasseur_quantum_2015-1} have also shown that for a large class of random bonds systems, a modified logarithmic growth $S(t)\sim (\ln t)^{1/\psi_k}$ is expected~\cite{vosk_dynamical_2014,vasseur_quantum_2015-1}, with an exponent $\psi_k<1$ which depends on the nature of the quantum critical glass phase. For the general class of SU(2)$_k$ quantum critical anyonic glasses~\cite{bonesteel_infinite-randomness_2007,fidkowski_textitc_2008,fidkowski_permutation-symmetric_2009,vasseur_quantum_2015-1}, the non-trivial tunnelling exponent for excited states $\psi_k$ has also been computed by Vasseur and co-workers~\cite{vasseur_quantum_2015-1}. For the Heisenberg SU(2) point, $\psi_{k\to \infty}\to 0$, yielding a faster growth for $S(t)$. 

It is interesting to remark that interactions lead to an entanglement growth which exceeds the upper bound $\propto \ln t$ proved for disordered free-fermions~\cite{burrell_bounds_2007}. Moreover, the large time asymptotic value is a volume-law $S_{t\to \infty}\sim L$, albeit with a small (non-thermal) prefactor, and which may also depend on the initial state~\cite{nanduri_entanglement_2014}, as well as on disorder strength. Nevertheless, the SU(2) Heisenberg case is special since RG breaks down~\cite{vosk_many-body_2013,vasseur_quantum_2015-1,agarwal_$1/f^ensuremathalpha$_2015} and ETH may still hold.

\begin{figure}[h]
\begin{center}
\includegraphics[width=.45\columnwidth,clip]{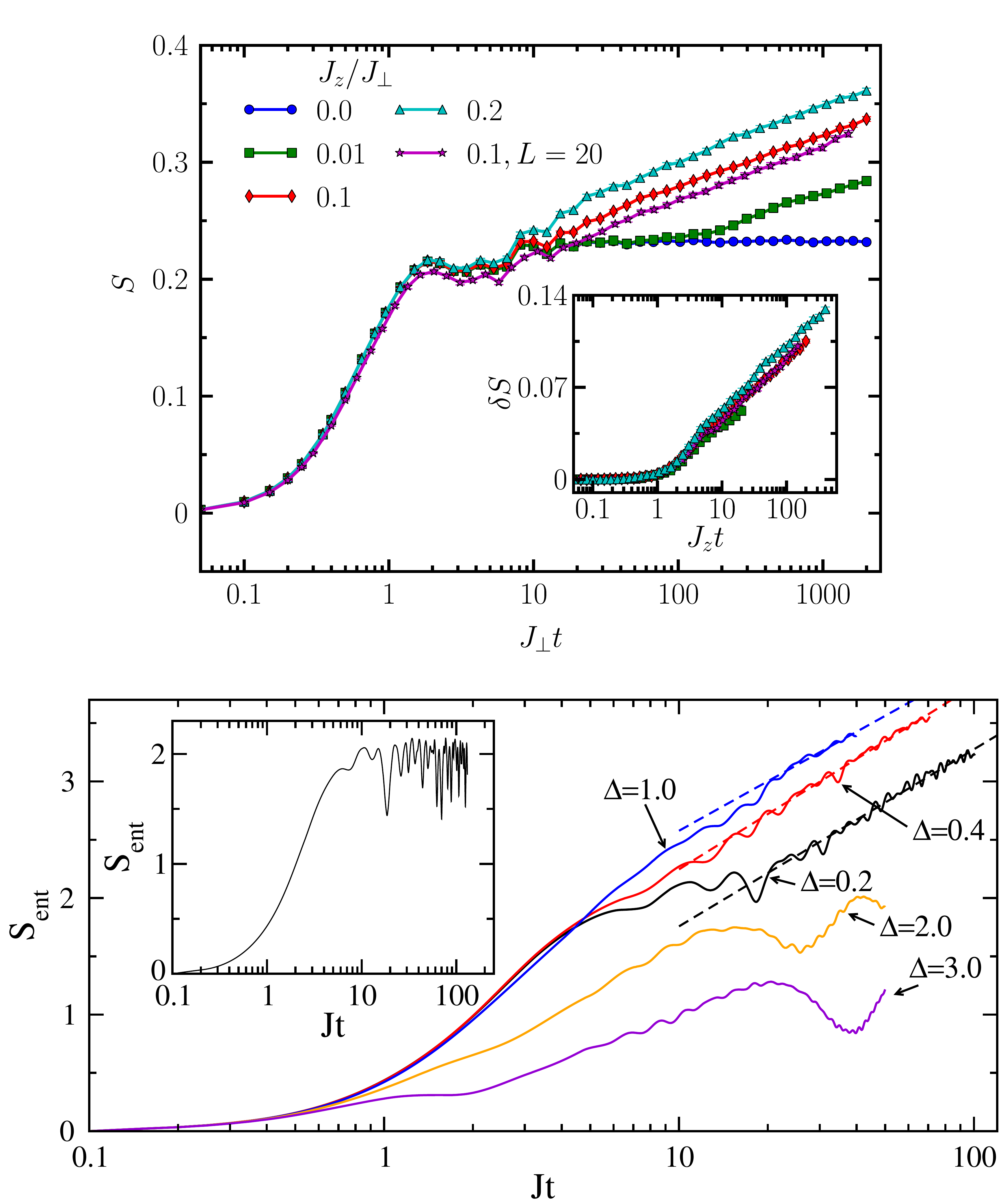}
\caption{Entanglement entropy growth from an unentangled product state for different interaction strengths $J_z$ or $\Delta$ of the XXZ $s=1/2$ chain for very strongly disordered random fields, in the MBL regime. Reprinted from \cite{bardarson_unbounded_2012} (Top) and from \cite{andraschko_purification_2014} (Bottom).}
\label{fig:time}
\end{center}
\end{figure}

The situation also changes drastically (as compared to non-interacting) for random fields or random potentials~\cite{znidaric_many-body_2008,bardarson_unbounded_2012,serbyn_universal_2013,vosk_many-body_2013,andraschko_purification_2014,zhao_entanglement_2016}. This is illustrated in Fig.~\ref{fig:time} where we see that only infinitesimal interactions yield an unbounded growth of entanglement entropy~\cite{bardarson_unbounded_2012} in the random field XXZ chain, in great contrast with free fermions for which the entropy quickly saturate to a strict area law. Interactions induce dephasing~\cite{serbyn_universal_2013,serbyn_local_2013} and give rise to a classical picture based on the so-called "l-bits" representation for the MBL regime~\cite{nandkishore_many-body_2015}. Distant spins are weakly coupled $V(\ell)\sim \exp(-\ell/\xi)$ and therefore get slowly entangled, after a long time $\ln t\sim \ell$, leading to a very slow entanglement growth
\be
S_{\rm MBL}(t)\propto \ln t,
\ee
up to saturation to a {\it{volume-law}}\footnote{This does not mean an equilibrated thermal state though.}. The above scaling has a prefactor which is controlled by both the initial state before the quench and the localization length in the MBL phase~\cite{nanduri_entanglement_2014}.
This dephasing is triggered by the interactions, as confirmed for long-range interacting models~\cite{pino_entanglement_2014} where the entanglement growth is faster, algebraic in time with an exponent controlled by the power-law interaction\footnote{This is relevant for dipolar cold gases, good candidates for observing MBL~\cite{yao_many-body_2014,serbyn_interferometric_2014,jurcevic_quasiparticle_2014}.}.
Below, we summarize in Table~\ref{tab:growth} the different behaviors of entanglement entropy growth after a global quench from a non-entangled initial product state in the strongly disordered regime of various models. At the MBL-ETH transition, while the l-bits picture  breaks down,  logarithmic spreading is still expected as a consequence of the infinite randomness nature of this critical point with a tunneling exponent $\psi=1$~\cite{vosk_theory_2015}.

\begin{table}[h!]
\centering
\begin{tabular}{c||c|c|c}
Interaction&Model&EE growth&$t=\infty$\\
\hline
&Random field/potential (AL)&bounded&area law\\
No&Transverse field Ising&&\\
& or&$\ln(\ln t)$&$\propto\ln L$\\
&Random bonds/hopping&&\\
\hline
&Random field/potential (MBL)&$\ln t$&$\propto L$\\
Yes&Random bonds XXZ&$\ln^{2/\phi}t$&$\propto L$\\
&Random bond SU(2)$_k$&$\ln^{1/\psi_k}t$&$\propto L$\\
\hline
\hline
\end{tabular}
\caption{\label{tab:growth}Entanglement entropy growth after a quench from a high-energy unentangled state for different one-dimensional quantum spin/particles models, either interacting or not, in their strong disorder regime.  The size $L$ scaling of the infinite time saturation limit is also given.}
\end{table}

\paragraph{Sub-ballistic entanglement growth in the delocalized regime---} Upon decreasing the disorder strength, the (dynamical) transition from the MBL phase to the ETH regime displays sharp signatures in the entanglement growth properties, changing from a slow logarithmic spreading to a ballistic growth expected in the absence of disorder~\cite{calabrese_evolution_2005,chiara_entanglement_2006,kim_ballistic_2013}. Asking whether ballistic spreading holds in the entire delocalized ETH regime, numerical~\cite{agarwal_anomalous_2015,bar_lev_absence_2015,torres-herrera_dynamics_2015,luitz_extended_2016,varma_energy_2015,khait_transport_2016} and analytical~\cite{vosk_theory_2015,potter_universal_2015,gopalakrishnan_griffiths_2016} works have revealed the existence of an intermediate delocalized regime where the dynamical response is anomalously slow. Sometimes dubbed "sub-diffusive", this regime is expected close to criticality in one dimension where the entanglement spreading may be blocked by bottlenecks made of rare critical segments of length $\lambda$. In such "Griffiths"~\cite{griffiths_nonanalytic_1969} regions, the entanglement spreading is drastically slowed, and occurs on time scales $\ln t \sim \lambda$. For a system of length $L$, the longest of such rare regions typically scales $\lambda^*\sim \xi\ln(L/\xi)$~\cite{vosk_theory_2015,potter_universal_2015}, $\xi$ being a characteristic length diverging at the transition. Therefore the waiting time for entanglement spreading is dominated by $\tau^*\sim L^{\xi/a}$, yielding 
\be
S(t)\sim t^{1/z},
\ee
with a dynamical exponent $z\sim \xi$ diverging when the ETH-MBL transition is approached, in agreement with  infinite randomness  ($z=\infty$) and a logarithmic growth at the transition.

\begin{figure}[b]
\begin{center}
\includegraphics[width=.85\columnwidth,clip]{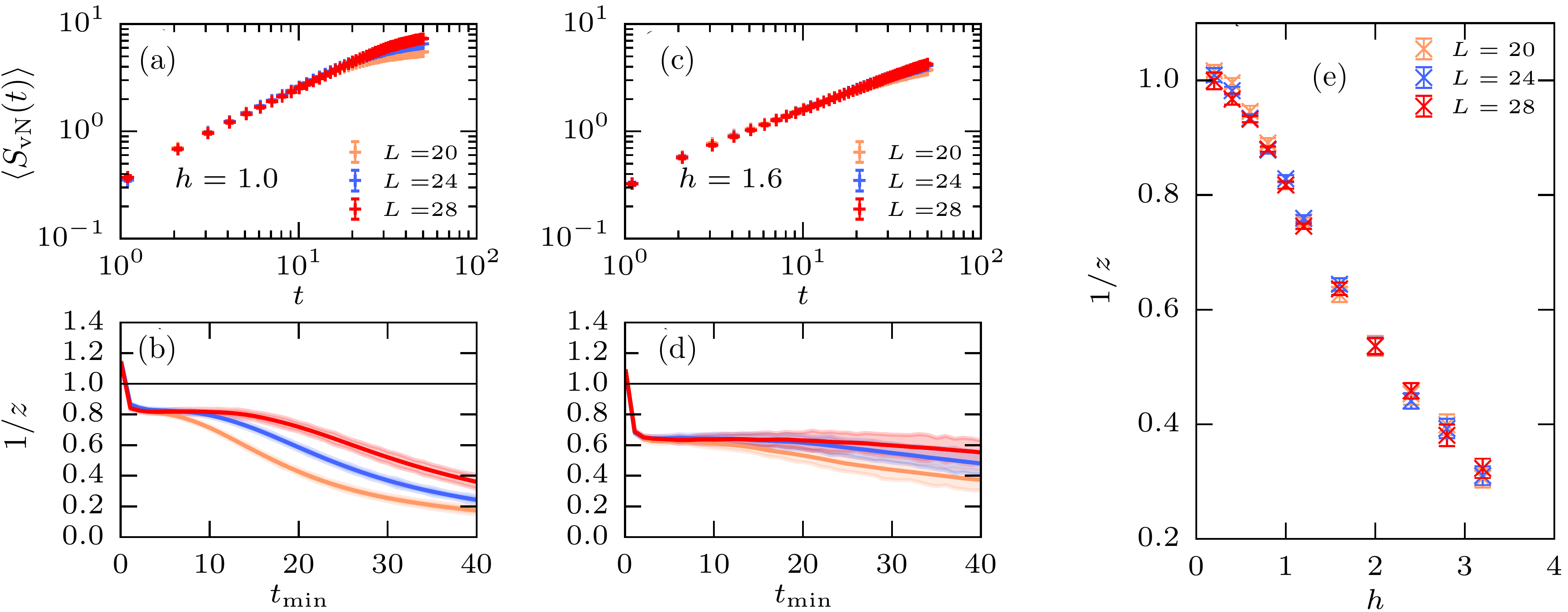}
\caption{(a) and (c): Disorder averaged time evolution of the entanglement entropy $\langle S_{\rm vN}(t)\rangle$ in open random-field Heisenberg chains
for different system sizes and two values of disorder. (b) and (d): Logarithmic derivative of
the disorder averaged time evolution of $\langle S_{\rm vN}(t)\rangle$, obtained by power-law fits over 8 points in time,
starting from $t_\text{min}$. The formation of plateaus corresponds to the power-law regime, with
growing extent with $L$. The plateaus determine the range of the power-law regime, over
which the exponent $1/z$ is extracted, displayed as a function of disorder in panel (e). Reprinted from \cite{luitz_extended_2016}).}
\label{fig:time2}
\end{center}
\end{figure}

This sub-ballistic entanglement spreading was verified numerically~\cite{luitz_extended_2016} using a Krylov space method allowing to preform exact diagonalization calculations for random-field Heisenberg chains Eq.~\eqref{eq:mbl} up to $L=28$ sites. Results~\cite{luitz_extended_2016} are displayed in Fig.~\ref{fig:time2} where the slow algebraic growth of the von-Neuman entropy after a sudden quench from an initial non-entangled basis state is clearly visible\footnote{In order to respect the energy-resolved character of this ETH-MBL transition discussed above (see Fig.~\ref{fig:mbl} for the mobility edge extracted in~\cite{luitz_many-body_2015}), the quenched states are chosen to be in the middle of the spectrum.}. The dynamical exponent $z$ is extracted as a function of the disorder strength $h$ of the random longitudinal field, and plotted in Fig.~\ref{fig:time2} (e).

The anomalous sub-ballistic
 regime for the entanglement entropy inside the ETH phase is found to persist in an extended parameter region, seemingly down to very small $h$. At first sight, this seems hard to reconcile with the fact that a sub-diffusive regime is
attached to rare Griffiths
regions~\cite{vosk_theory_2015,potter_universal_2015,agarwal_anomalous_2015}, thus only expected close to the MBL transition. One should remember however that in the
 quench protocol, inhomogeneity is also present in the initial product state where energy density  fluctuate locally leading to anomalously "hot" and "cold"
regions. The presence of a mobility edge~\cite{luitz_many-body_2015} in the random-field Heisenberg model
Eq.~\eqref{eq:mbl} may be responsible for such a large extension for the anomalous dynamical regime observed in Ref.~\cite{luitz_extended_2016} using
such a global quench protocol. 

\subsubsection{Open questions}
An important open issue concerns experimentally relevant realizations of MBL.  While very new, it appears to become a flourishing field with a growing number of striking proposals, coming either from traditional condensed matter systems~\cite{ovadia_evidence_2015} or from quantum gas labs~\cite{schreiber_observation_2015,smith_many-body_2015,bordia_coupling_2016,choi_exploring_2016}. The measurement of entanglement growth in cold atom setups is in principle possible (see next section) using newly developed techniques~\cite{islam_measuring_2015}. For systems of trapped ionic chains with controlled disorder, the growth of quantum Fisher information~\cite{d._petz_introduction_2011} (giving a lower bound on the entanglement) has been measured~\cite{smith_many-body_2015}.

Logarithmic entanglement spreading is expected to be the smoking gun of MBL, but one should ask whether the sub-ballistic growth found in most of the delocalized regime~\cite{luitz_extended_2016} is a generic feature in the presence of disorder, or is it model-dependent? A natural question follows regarding a possible transition between a ballistic and a sub-ballistic regime for entanglement spreading  inside the delocalized regime.

Another important open issue touches the question of the nature of the localized - delocalized transition and the possible ergodicity breaking at the transition, as well as in its vicinity~\cite{grover_certain_2014,pino_nonergodic_2016,monthus_many-body-localization_2016}.

Finally, let us also mention the interesting and strongly debated question of MBL in disorder-free systems~\cite{grover_quantum_2014,roeck_asymptotic_2014,schiulaz_ideal_2014,de_roeck_scenario_2014,yao_quasi_2014,schiulaz_dynamics_2015,van_horssen_dynamics_2015,papic_many-body_2015,hickey_signatures_2016}.

\section{Towards entanglement measurement}
\label{sec:exp}
Experimental detection of entanglement is a vast topic~\cite{guhne_entanglement_2009} for which many different setups have been explored~\cite{makhlin_josephson-junction_1999,berkley_entangled_2003,ghosh_entangled_2003,haffner_scalable_2005,childress_coherent_2006,bertaina_rare-earth_2007,stamp_spin-based_2009,simmons_entanglement_2011,bernien_heralded_2013,dolde_room-temperature_2013,pla_high-fidelity_2013,pfaff_unconditional_2014,pirandola_advances_2015}. Among them, bulk solid-state systems, where correlations and quantum entanglement usually go hand in hand~\cite{verstraete_entanglement_2004}, are potentially interesting systems to measure and exploit entanglement. 
Strongly correlated phases and quantum dynamics can be interpreted as a signature of quantum entanglement, as for instance discussed for the dipolar spin glass LiHo$_x$Y$_{1-x}$F$_4$~\cite{ghosh_entangled_2003}, spin-$1/2$ antiferromagnets 
CN[Cu(NO$_3$)$_2 \cdot 2.5$D$_2$O]~\cite{brukner_crucial_2006,das_experimental_2013,singh_experimental_2013} or [Cu(DCOO)$_2 \cdot 4$D$_2$O]~\cite{christensen_quantum_2007}, supramolecular antiferromagnetic rings (Purple-Cr$_7$Ni)~\cite{candini_entanglement_2010}, Kondo screening in atomic chains~\cite{choi_entanglement-induced_2015}, and more generally in neutron scattering experiments~\cite{cowley_quantum_2003,cramer_measuring_2011}. Similar signatures are also expected for optical lattice experiments~\cite{cramer_spatial_2013}.

\begin{figure}[h]
\begin{center}
\includegraphics[width=\columnwidth,clip]{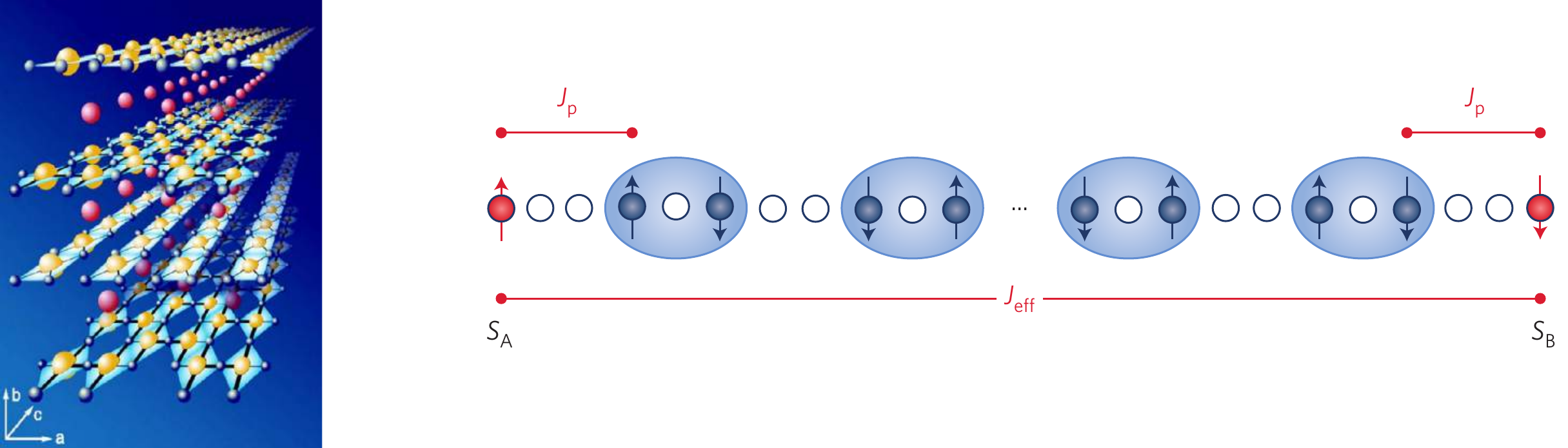}
\caption{Left: three-dimensional representation of the spin-ladder material ${\mathrm{Sr}}_{14}{\mathrm{Cu}}_{24}{\mathrm{O}}_{41}$. Right: one-dimesnioanl structure made of spin-dimer chains with isolated spins (red) coupled over long distance $\ell\simeq 220-250~\AA$ through the effective coupling $J_{\rm eff}$. Reprinted from \cite{sahling_experimental_2015}}
\label{fig:spinchain}
\end{center}
\end{figure}

Long-distance entanglement between solid-state qubits is also a major issue for quantum communication schemes~\cite{bennett_teleporting_1993,bennett_quantum_2000}. In this context, spin networks have been proposed to be promising setups to realize a quantum data bus~\cite{bose_quantum_2003,christandl_perfect_2004,osborne_propagation_2004,campos_venuti_qubit_2007,trifunovic_long-distance_2013}. Interestingly, the spin-ladder compound ${\mathrm{Sr}}_{14}{\mathrm{Cu}}_{24}{\mathrm{O}}_{41}$ (see Fig.~\ref{fig:spinchain})~\cite{lorenzo_macroscopic_2010} appears to be a good candidate for long-range entanglement between unpaired spins which gets entangled at low temperature, coupled by an effective coupling $J_{\rm eff}$ across $\ell=220-250~\AA$~\cite{sahling_experimental_2015}, as sketched in Fig.~\ref{fig:spinchain}.

Nevertheless, in most of these works the low-temperature entanglement is seen through few-body correlations that are measured using standard solid-state physics probe, such as heat capacity, susceptibility, neutron scattering. It is however much more difficult to access many-body entanglement estimates, like the entanglement entropy for instance. In the rest of this section we discuss the few proposals that have been recently discussed, either using noise and fluctuations in solid-state and mesoscopic systems, or in cold atom experiments.

\subsection{Quantum noise and bipartite fluctuations}
Fluctuations of globally conserved quantities, such as the number of particles for U(1) symmetic itinerant systems, or the total magnetization for O(2) magnets, are key mesurable quantities which can reflect the amount of entanglement present in a subsystem. First discussed in the context of mesoscopic systems~\cite{loss_probing_2000,burkard_noise_2000,chtchelkatchev_bell_2002,beenakker_quantum_2003,crepieux_electron_2003,jordan_entanglement_2004}, the link between entanglement entropy and bipartite fluctuations has been explored later by several authors~\cite{klich_measuring_2006,klich_quantum_2009,
hsu_quantum_2009,song_general_2010,song_entanglement_2011,song_entanglement_2011-1,song_bipartite_2012,rachel_detecting_2012,calabrese_exact_2012,susstrunk_free_2012,vicari_entanglement_2012,eisler_universality_2013,petrescu_fluctuations_2014,thomas_entanglement_2015,
dasenbrook_dynamical_2015}.

Below we first describe free fermions where there is an exact correspondence between cumulants and entanglement entropies~\cite{klich_quantum_2009,song_entanglement_2011,song_bipartite_2012}. Then we discuss interacting systems in one dimension, and beyond. While there is a quantitative link for Luttinger liquids between bipartite fluctuations and entanglement entropy~\cite{song_general_2010,song_bipartite_2012,laflorencie_spin-resolved_2014}, this is no longer true for interacting two-dimensonal systems~\cite{song_entanglement_2011-1}, although quantum critical points can be detected using fluctuations~\cite{rachel_detecting_2012}. We finally address the key issue of experimental detection for nano- or meso-electronic devices such as quantum point contacts and also for more traditional solid-state bulk systems such as quantum magnets.

\subsubsection{Bipartite fluctuation as an entanglement meter}
\label{sec:BF}
\paragraph{Free fermions---} For systems that can be mapped to non-interacting fermions, the entanglement entropy can be accessed experimentally through the full counting statistics of a conserved U(1) charge, such as the particle number ${N}$ for instance when studying the charge transfer across mesoscopic conductors~\cite{levitov_charge_1993,blanter_shot_2000,esposito_nonequilibrium_2009,kambly_factorial_2011}.
The idea of using quantum noise as an entanglement meter, first proposed by Klich and Levitov~\cite{klich_quantum_2009} following a local quench, was further explored by Song and co-workers for both out-of-equilibrium and at equilibrium~\cite{song_entanglement_2011,song_bipartite_2012} who derived  exact expressions Eq.~\eqref{eq:S_C} for the von-Neuman and the R\'enyi entropies in terms of cumulants $C_n$.
The second cumulant $C_2$, also called the fluctuations is defined by
\be
	{\cal{F}} = \langle{\left({N}_A -\langle{{N}_A}\rangle\right)^2}\rangle \label{eq:def_F}
\ee where $\langle N_A\rangle$ is the ground-state expectation value of the particle number in a subsystem $A$. Higher order cumulants, defined by
\be
	C_n = (-i\partial_\lambda)^n\ln \chi(\lambda)|_{\lambda=0},\label{eq:def:cumulants}
\ee where the generating function $\chi(\lambda)=\mean{\exp(i\lambda{N}_A)}$, enter in the following expansions~\cite{song_entanglement_2011,song_bipartite_2012}, only involving even cumulant $C_{2p}$:
\be \label{eq:S_C}S_{\rm vN}=\lim_{R\to \infty}\sum_{p=1}^{R/2} \alpha_{2p}(R) C_{2p},
\ee
and
\be
S_{q>1}=\lim_{R\to \infty}\sum_{p=1}^{qR/2} \beta_{2p}(q,R) C_{2p},\ee 
with
$\beta_{2p}(q,R) =\frac{1}{1-q}\sum_{r=1}^{R}\sum_{m=0}^{r}\sum_{s=2p}^{qr}(-1)^{r+s+qr+qm}\times\frac{1}{r}
\binom{R}{r}\binom{r}{m}\binom{qm}{qr-s}\frac{{\cal{S}}_1(s,2p)}{(s-1)!}$, and 
$\alpha_{2p}(R)=2\sum_{k=2p-1}^{R} \frac{{\cal{S}}_1(k,2p-1)}{k!k}$, and 
where ${\cal{S}}_1(s,2p)$ are the unsigned Stirling numbers of the first kind. In the case of gaussian fluctuations, only the variance $C_2$ is non-zero, which considerably simplifies the expressions, yielding
\be
S_q=\frac{\pi^2}{6}\left(1+\frac{1}{q}\right)C_2.
\ee

\paragraph{Interacting systems: Luttinger liquids---} 
For one-dimensional critical (zero-temperature) Luttinger liquids, taking a subsystem of length $\ell$, its RDM can be described as the thermal density matrix of an open Luttinger liquid with the same velocity $u$ and Luttinger parameter $K$~\cite{lauchli_operator_2013,laflorencie_spin-resolved_2014} at an effective low temperature $T_{\rm ent}={u\ln(\ell/\ell_0)}/({\pi\ell})$, where $\ell_0$ is an order one length scale. Ignoring irrelevant corrections, a Luttinger liquid is equivalent to a free boson model whose partition function at inverse temperature $\beta$ is exactly known~\cite{eggert_magnetic_1992}:
\be
Z(\ell,\beta)= {\zeta}(\ell,\beta)\sum_{m=-{\ell}/{2}}^{{\ell}/{2}}\exp\left(-\beta\frac{\pi u}{2K \ell}m^2\right),
\label{eq:Z}
\ee
where ${\zeta}(\ell,\beta)=\prod_{n=1}^{\infty}\left[2\sinh\left(\frac{u\pi}{4\ell T}n\right)\right]^{-1}$, and $m=-\ell/2,\ldots,\ell/2$ are the conserved charges (here the magnetization of a spin-1/2 chain, see \cite{eggert_magnetic_1992}).
From this expression, we immediately see that the normalized weights $p_m$ of the $S^z=\pm m$ sectors have a Gaussian distribution
with a variance $C_2(\ell,T)=(K\ell T) /(u\pi)$.
Therefore using the quantum/thermal correspondence~\cite{laflorencie_spin-resolved_2014}, the leading scaling of zero-temperature bipartite fluctuations in a subsystem of length $\ell$ is given by
\be
C_2(\ell)=\frac{K}{\pi^2}\ln(\ell/\ell_0),
\label{eq:C2}
\ee
while the R\'enyi entropy $S_q^{\rm th}(\beta)=\frac{q\beta}{1-q} \left[ F(\beta) - F(q \beta) \right]=\frac{\pi}{6u}\left(1+\frac{1}{q}\right)\ell/\beta$ gives for the general relation between entanglement entropies and bipartite fluctuations:
\be
S_q(\ell)=\frac{\pi^2}{6K}\left(1+\frac{1}{q}\right)C_2(\ell)+O(1),
\ee
in agreement with previous works~\cite{song_general_2010}.

\begin{figure}[t!]
\begin{center}
\includegraphics[clip,width=.85\columnwidth]{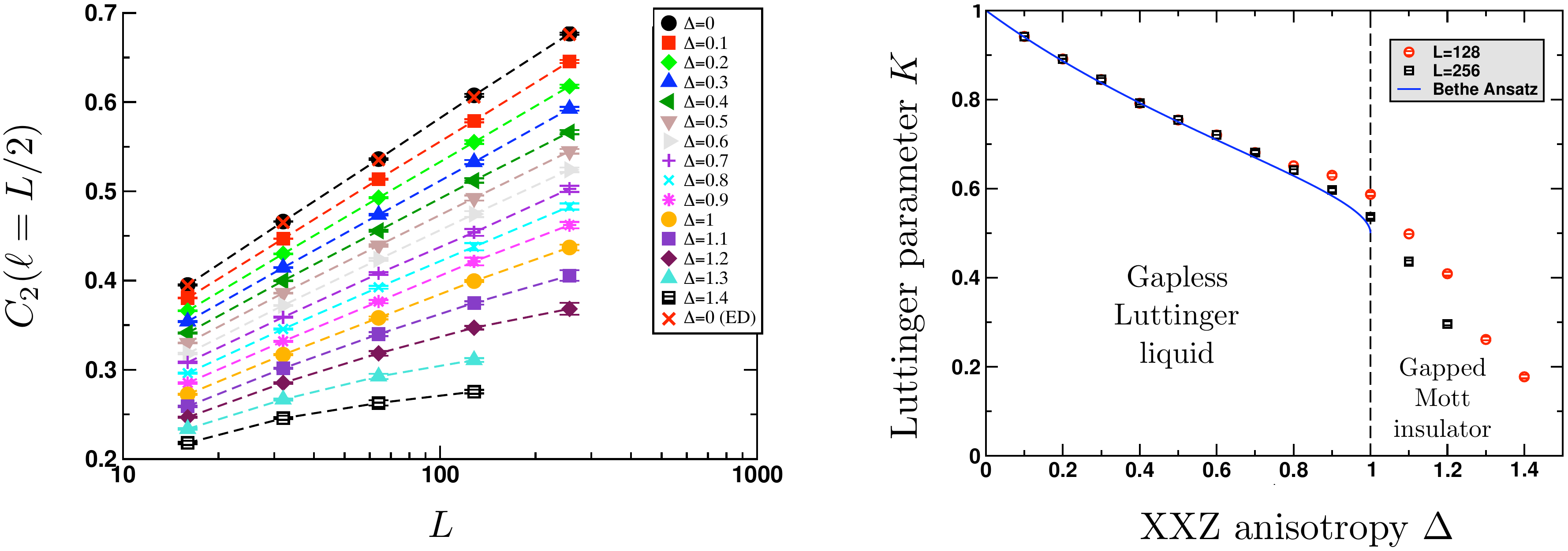}
\caption{Left: QMC results for the fluctuations $C_2(\ell=L/2)$ plotted against the total system size $L$ for periodic chains of lengths $L=16,~32,~64,~128,~256$. Different  symbols are for different values of the XXZ anisotropy For the XX point $\Delta=0$, QMC data (black circles) are also compared to exact diagonalization (free-fermions) results (red crosses). Logarithmic scaling is observed in the critical regime $\Delta\le 1$ whereas for $\Delta>1$ $C_2$ tends to saturate at large $L$. Inset: schematic picture for the periodic ring where subsystem A is taken with $x=L/2$ sites. Right: Prefactor $K_{\rm eff}$ of the logarithmic scaling Eq.~(\ref{eq:C2}) extracted by fitting QMC data for $L\in[2^{p-1},2^{p}]$ with $L=128$ ($p=7$) and $L=256$ ($p=8$). The blue line shows the exact Bethe Ansatz result for $K(\Delta)$ Eq.~\eqref{eq:K}.}
\label{fig:F1DQMC}
\end{center}
\end{figure}

The logarithmic scaling Eq.~\eqref{eq:C2} of the bipartite fluctuations $C_2(\ell)$ can be checked for using DMRG or QMC simulations~\cite{song_general_2010,song_bipartite_2012}. 
Here we show QMC results in the ground-state of spin-$1/2$ XXZ chains in the antiferromagnetic critical regime for $C_2$ of half-chains $\ell=L/2$ in Fig.~\ref{fig:F1DQMC} for various anisotropies $\Delta$ and sizes $L=16,\ 32,\ 64,\ 128,\ 256$. In the entire antiferromagnetic critical regime $\Delta\in [0,1]$ the scaling of $C_2$ is logarithmic, while it tends to saturate once $\Delta> 1$, as clearly visible in the left panel of Fig.~\ref{fig:F1DQMC}. 
A logarithmic fit of the QMC data to the form $C_2=K_{\rm eff}/\pi^2\ln L + {\rm constant}$ yields estimates for the Luttinger parameters plotted in the right part of Fig.~\ref{fig:F1DQMC}, in very good agreement with the exact expression known from Bethe ansatz Eq.~\eqref{eq:K}. 

It is fair to say that the second cumulant is a quite simple observable to measure within QMC or DMRG simulations. It therefore provides a very interesting quantity to access the Luttinger liquid parameter $K$  when it is unknown. Conversely, if the value of $K$ at a quantum critical point $g_c$ is known but the precise value of $g_c$ is not, we will see below that measuring the second cumulant $C_2$ provides a very efficient way to estimate the critical coupling.

\subsubsection{Bipartite fluctuations to detect quantum criticality}

\paragraph{One dimension: Luttinger liquids---} Following the previous discussion, we show how efficient the use of Eq.~\eqref{eq:C2} is to locate 1D quantum critical points with a high accuracy.

\begin{figure}[t]
\begin{center}
\includegraphics[width=0.45\columnwidth,clip]{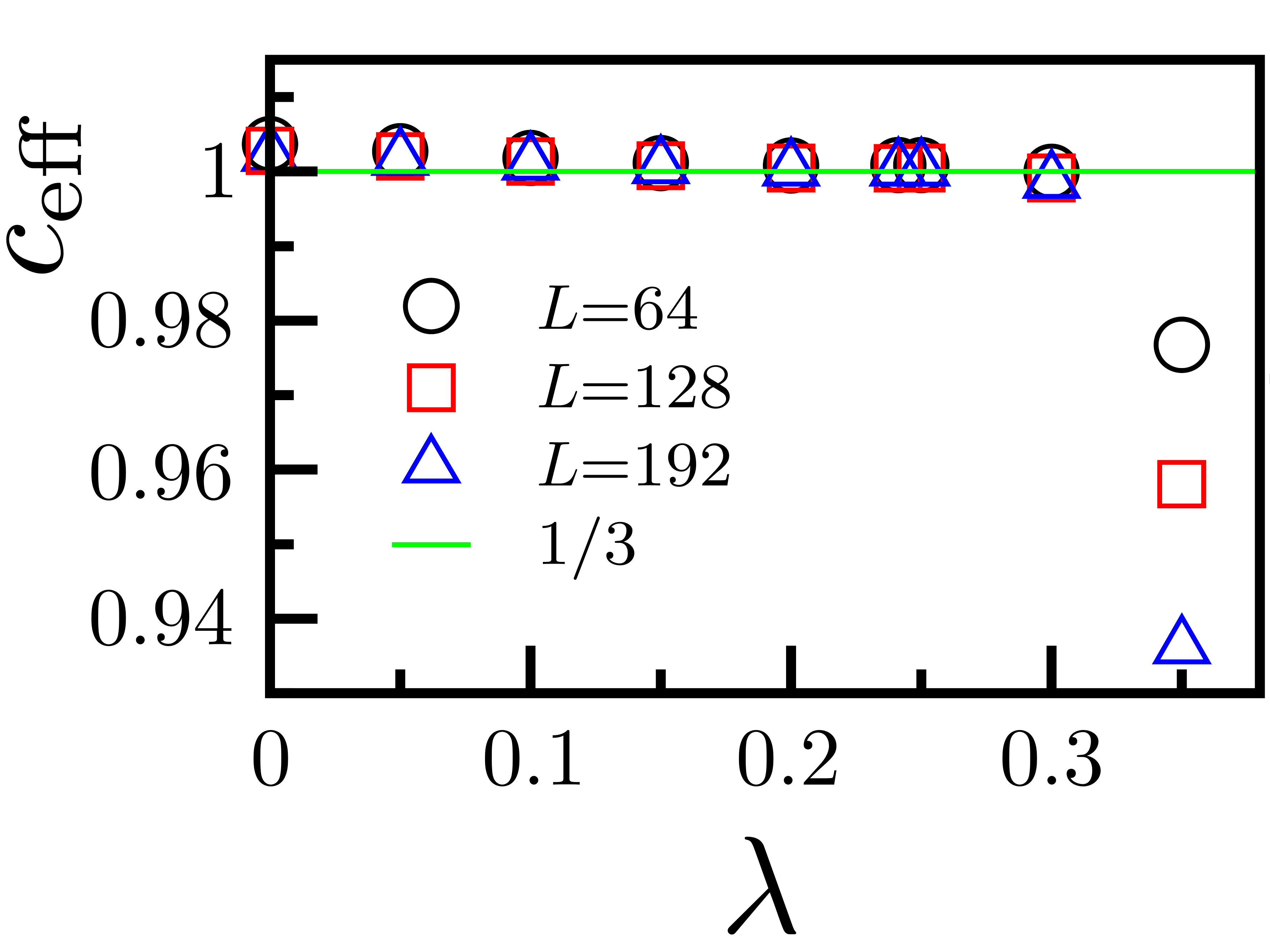}
\caption{Coefficient of a logarithmic fit for the von Neumann entanglement entropy as a function of the second neighbor coupling $J_{2}$. DMRG results for $L=64,128,192$ sites. Reprinted from \cite{alet_valence-bond_2010}.}
\label{fig:j1j2}
\end{center}
\end{figure}

{\it{(i) Frustrated spin chain:}}
The first model we study
is the frustrated spin-$1/2$ $J_1 - J_2$ chain, governed by the Hamiltonian 
\begin{equation}\label{eq:j1j2}
\mathcal{H}_{J_1 - J_2}=\sum_i \left(\,J_1{\bf{S}}_i\cdot{\bf{S}}_{i+1} +J_2 ~{\bf{S}}_i\cdot{\bf{S}}_{i+2}\,\right)\ ,
\end{equation}
where we define $\lambda=J_2/J_1\ge 0$. For $\lambda\le \lambda_c$, this model has power-law critical correlations and a logarithmic scaling of entanglement entropy Eq.~\eqref{eq:S1d}, while at $\lambda_c\simeq 0.2412$ a Berezinsky-Kosterlitz-Thouless (BKT) transition into a dimerized phase occurs~\cite{haldane_spontaneous_1982,okamoto_fluid-dimer_1992,eggert_numerical_1996} where the entropy obeys a strict area law (see Fig.~\ref{fig:xydim}).  A precise estimate of $\lambda_c$ using entanglement entropy turns out to be very difficult, as explored by Alet and co-workers in~\cite{alet_valence-bond_2010}. This is illustrated in Fig.~\ref{fig:j1j2} where the effective (size-dependent) central charge, estimated from DMRG data for various system sizes, is plotted against the frustrated coupling $\lambda$. The crossing of the data to the value $c_{\rm eff}=1$, expected to occur at $\lambda_c$ where irrelevant logarithmic corrections vanish~\cite{eggert_magnetic_1992} is very difficult to detect, and in practice less precise as compared to spin-spin correlation function~\cite{eggert_numerical_1996}.

\begin{figure}[t]
\begin{center}
\includegraphics[width=.85\columnwidth,clip]{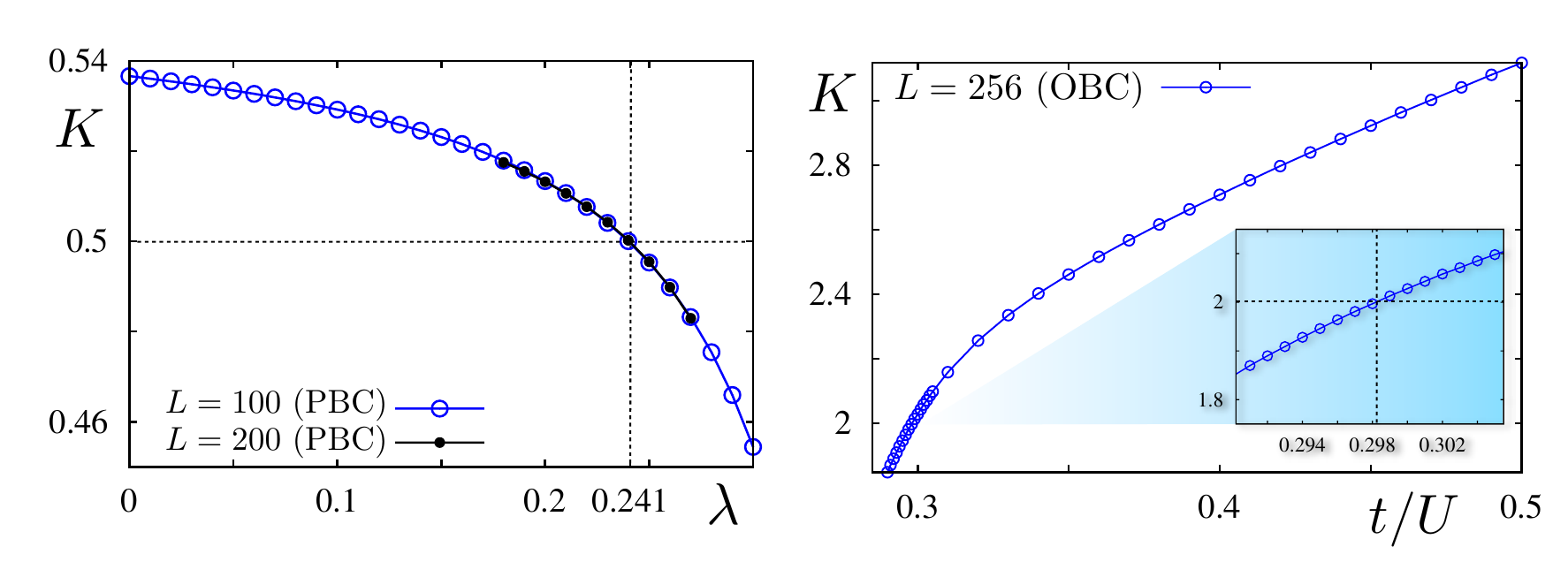}
\caption{Left: Effective Luttinger parameter $K$ of the $J_1-J_2$ chain extracted via \eqref{eq:C2} and plotted against the frustration parameter $\lambda$. DMRG data for $L=100$ (open circles) and $L=200$ (black dots) for periodic chains. Right: Same for the Bose-Hubbard chain at filling unity plotted against $t/U$ for $L=256$ and open boundary conditions.  Inset: zoom close to the transition. Reprinted from \cite{rachel_detecting_2012}.}
\label{fig:KJ1J2}
\end{center}
\end{figure}

The effect of marginal corrections is much more pronounced in the bipartite fluctuations of the magnetization which involve the uniform part of the spin correlations inside subsystem $A$
\be
C_2=\sum_{i,j\in A} [\langle{{S}^z_i{S}^z_j}\rangle - \langle{{S}^z_i}\rangle\langle{{S}^z_j}\rangle ].\label{eq:FA}\ee
where additive logarithmic corrections are well-known~\cite{eggert_susceptibility_1994}. Fitting DMRG data to the form Eq.~\eqref{eq:C2}, one can extract an effective (size-dependent) Luttinger parameter as a function of $\lambda$. This is plotted in the left panel of Fig.~\ref{fig:KJ1J2} where one clearly sees the vanishing of the marginal corrections leading to $K_c=0.5$ at $\lambda_c=0.2412(3)$, in perfect agreement with the best estimate $\lambda_c=0.241167(5)$~\cite{eggert_numerical_1996}.

\begin{figure}
\begin{center}
\includegraphics[width=.65\columnwidth,clip]{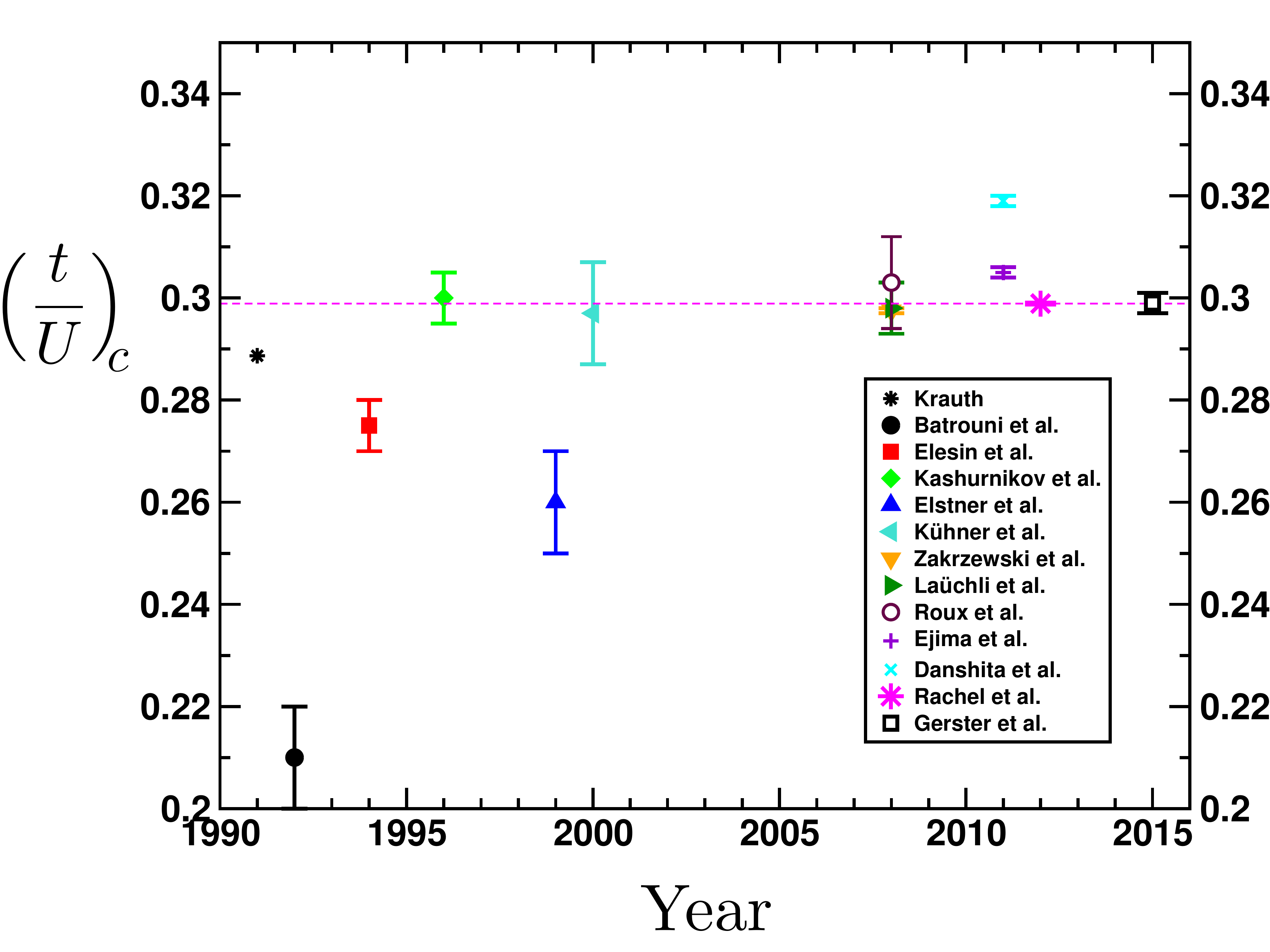}
\caption{Various critical point estimates of the Superfluid - Mott Insulator transition in the one-dimensional Bose-Hubbard model at unit filling. The plotted values in function of time refers to those of Table~\ref{tab:K}.}\label{fig:K}
\end{center}
\end{figure}

{\it{(ii) Bose-Hubbard chain at unit filling:}}
Another very interesting example is provided by the one-dimensional Bose-Hubbard model (for reviews, see~\cite{cazalilla_one_2011,krutitsky_ultracold_2016}) with one particle per site, governed by the following Hamiltonian
\begin{equation}\label{eq:bhm}
\mathcal{H}=-t\sum_{\langle ij \rangle} b_i^\dag b_{j}^{\phantom{\dag}} +
\frac{U}{2}\sum_i n_i \big(n_i - 1\big) -\sum_i \mu\, n_i\ ,
\end{equation}
where $t$ is the hopping amplitude, $U$ the on-site repulsion, and $\mu$ the chemical potential chosen such that $\langle n_i\rangle=1$ in the ground-state. 
For unit filling, the quantum phase transition from a superfluid to a Mott insulator is of BKT type~\cite{haldane_effective_1981}.
The exponent governing the algebraic decay of the Green's function
$\langle b_r^\dag b_0^{\phantom{\dag}}\rangle\propto r^{-1/2K}$ in the superfluid phase $K\ge 2$ jumps at the transition from $K_c=2$ to 0 in the Mott insulator~\cite{giamarchi_quantum_2003}.  The precise determination of the critical ratio $(t/U)_c$ turns out to be very difficult using finite size numerics, as sumarized in Table~\ref{tab:K} and Fig.~\ref{fig:K}. As discussed in details in \cite{rachel_detecting_2012}, bipartite fluctuations of the particle number provide a very simple and efficient tool to detect the transition with a very high accuracy, as shown in panel right of Fig.~\ref{fig:KJ1J2}.  The Luttinger parameter $K$ is extracted from fitting $C_2$ to Eq.~\eqref{eq:C2} for various sizes $L=64$, $128$, and $256$ (the latter is shown in Fig.~\ref{fig:KJ1J2}). By performing finite size scaling we obtain a very precise estimate $\lambda_c = 0.2989(2)$, in good agreement with the most reliable estimates (see Table~\ref{tab:K} and Fig.~\ref{fig:K}). Note that the bipartite fluctuations allow to estimate the critical point with a very high accuracy, the numerical cost being quite low.\\
\\

\begin{table}
\begin{tabular}{l|l|l|l|l}
\hline
Year&Reference&Technique&Observable&Estimate\\
\hline\hline
1991&Krauth \cite{krauth_bethe_1991} & Bethe Ansatz &&$0.2887$\\
\hline
1992&Batrouni {\it et al.} \cite{batrouni_world-line_1992} & QMC & Stiffness &$0.21(1)$\\
\hline
1994&Elesin {\it et al.} \cite{elesin_mott-insulator-superfluid-liquid_1994} & ED & Gap &$0.275(5)$\\
\hline
1996&Kashurnikov {\it et al.} \cite{kashurnikov_mott-insulator-superfluid-liquid_1996} & QMC & Gap & $0.300(5)$\\
\hline
1999&Elstner {\it et al.} \cite{elstner_dynamics_1999}& Strong coupling & Gap & $0.26(1)$\\
\hline
2000& K\"uhner {\it et al.} \cite{kuhner_one-dimensional_2000} & DMRG & Correlations & $0.297(10)$\\
\hline
2008&Zakrzewski {\it et al.} \cite{zakrzewski_accurate_2008} & TEBD & Correlations& $0.2975(5)$\\
\hline
2008 & La\"uchli {\it et al.} \cite{lauchli_spreading_2008}& DMRG &Entanglement &$0.298(5)$\\
\hline
2008 & Roux {\it et al.} \cite{roux_quasiperiodic_2008}& DMRG & Gap &$0.303(9)$\\
\hline
2011& Ejima {\it et al.} \cite{ejima_dynamic_2011} & DMRG & Correlations &$0.305(1)$\\
\hline
2011 & Danshita {\it et al.} \cite{danshita_superfluid--mott-insulator_2011} & TEBD & Excitations&$0.319(1)$\\
\hline
{{2012}}&{Rachel {\it{et al.}}} \cite{rachel_detecting_2012}& {DMRG}&{Fluctuations}&{${{0.2989(2)}}$}\\
\hline
{{2016}}&{Gerster {\it{et al.}}} \cite{gerster_superfluid_2016}& {Tensor network}&{Stiffness}&{$0.299(2)$}\\
\hline
\hline
\end{tabular}
\caption{\label{tab:K} Various estimates for the critical coupling of the one-dimensional Bose-Hubbard model Eq.~\eqref{eq:bhm} at unit filling. These estimates are also plotted {\it{vs.}} year in Fig.~\ref{fig:K}. One should emphasize on the most accurate result (second last line) obtained using bipartite fluctuations~\cite{rachel_detecting_2012}. See also \cite{krutitsky_ultracold_2016} for an exhaustive list of estimates.}
\end{table}
%

\paragraph{Beyond one dimension---} One can also study quantum critical points in higher dimension, as done in \cite{rachel_detecting_2012} for two-dimensional coupled $s=1/2$ Heisenberg ladders (see Fig.~\ref{fig:F2D}), described by the following Hamiltonian
\be
{\cal H} = \sum_{\rm ladd.}\,\,{\bf{S}}_{i}\cdot{\bf{S}}_j \,\,\,+\sum_{\rm inter-ladd.} \lambda\,{\bf{S}}_{i}\cdot{\bf{S}}_j.
\label{eq:model}
\ee
This model displays two gapped rung-singlet phases if the inter-ladder coupling $\lambda_c^1>\lambda>\lambda_{c}^2$ with $\lambda_{c}^1=0.31407(5)$~\cite{matsumoto_ground-state_2001} and $\lambda_c^2=1.9096(2)$~\cite{wenzel_comprehensive_2009}, and a gapless N\'eel ordered phase in between, as shown in Fig.~\ref{fig:F2D} (a).
$T=0$ fluctuations of the total magnetization in a sub-system $\cal A$ of size $L/2 \times L$, embedded in a periodic square lattice $L\times L$ are shown in Fig. ~\ref{fig:F2D} (b-c), with square lattices size up to $L\times L = 10^4$, for the isotropic square lattice $\lambda=1$ (N\'eel) and for weakly coupled ladders with $\lambda=0.1$ (gapped rung singlet I). In contrast with the entanglement (or R\'enyi) entropy which displays a strict area law both phases, the second cumulant follows different scalings \cite{song_entanglement_2011}:
\begin{figure}[b]
\begin{center}
\includegraphics[width=.9\columnwidth,clip]{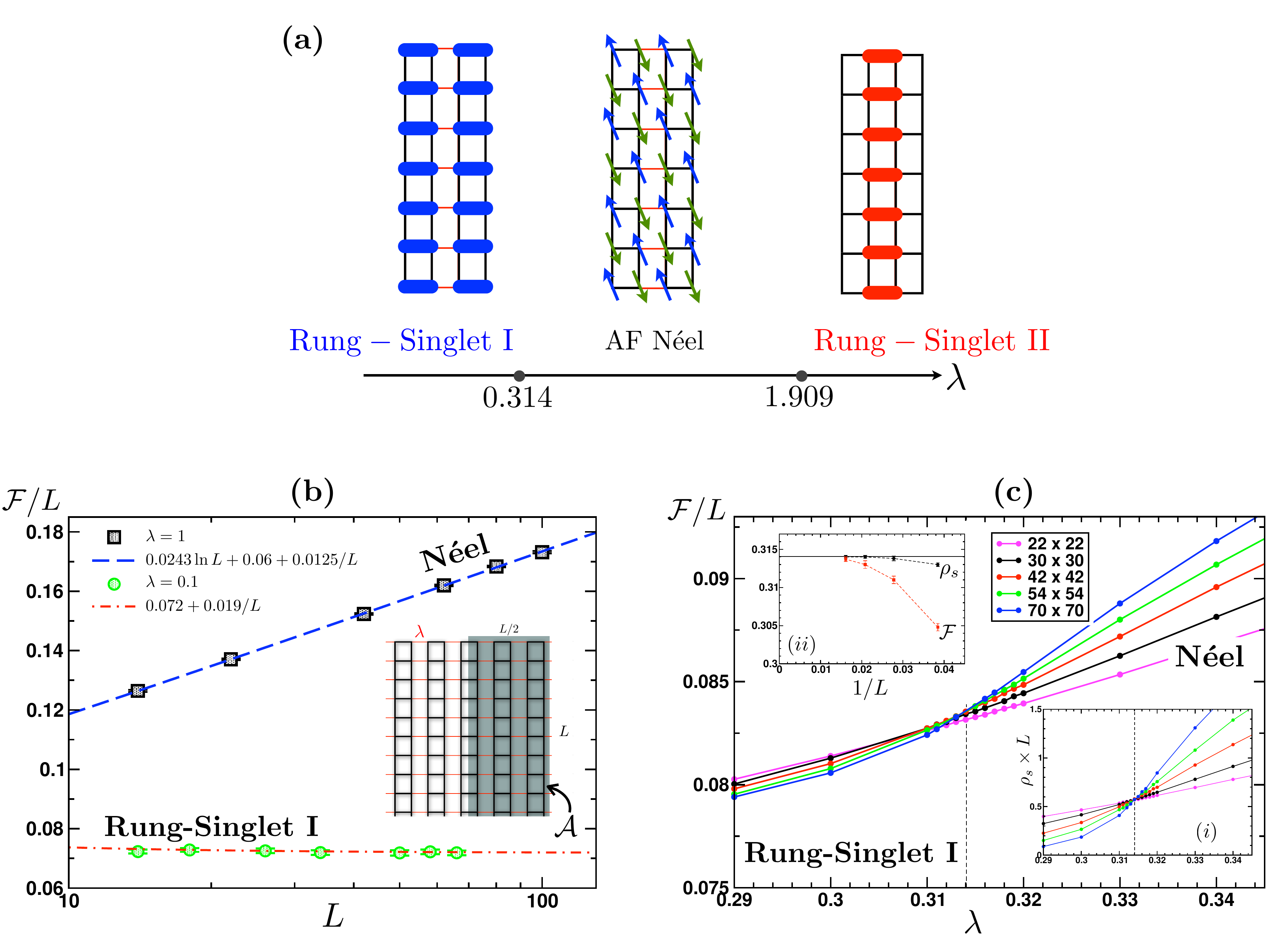}
\caption{QMC results for $T=0$ fluctuations $\cal F$ of the total magnetization in a region $\cal A$ for 2D coupled spin-$\frac{1}{2}$ ladders [Eq.~\eqref{eq:model}], depicted in the inset of (a). Panel (b): ${\cal F}/L$ increases logarithmically with $L$ in the N\'eel regime (black squares $\lambda=1$) whereas it saturates to a constant in the gapped state (green circles $\lambda=0.1$). Panel (c): ${\cal F}/L$, plotted {\it{vs.}} $\lambda$ for various system sizes, displays a crossing point at $\lambda_c$. Insets: $(i)$ crossing of the stiffness $\rho_s\times L$ at $\lambda_c$ for the same sizes; $(ii)$ $1/L$ convergence of the crossing point for $\cal F$ (red squares) and $\rho_s$ (black circles) to the critical value (horizontal black line) $\lambda_c=0.31407$~\cite{matsumoto_ground-state_2001}. Reprinted from \cite{rachel_detecting_2012}.}
\label{fig:F2D}
\end{center}
\end{figure}
%
\be
\label{eq:F2D}
{\cal F}(\ell)\sim\left\{
\begin{array}{lr}
\alpha \ell\ln \ell +\beta \ell +\gamma &{\rm{~~{{(Gapless~N\acute{e}el)}}}}\\
\beta' \ell +\gamma'&{\rm{~(Gapped~Rung~Singlet).}}
\end{array}
\right.
\ee
Therefore, $\cal F/\ell$ plotted for different sizes displays a crossing point at $\lambda_c$, as observed in panel (c) of Fig.~\ref{fig:F2D}.
The spin stiffness $\rho_s$, also known to be a useful quantity to locate quantum criticality, is shown in the right inset (ii) of Fig.~\ref{fig:F2D} (c) where a similar crossing is observed for $\rho_s\times L^{d+z-2}$, with $z=1$ and $d=2$. As usual for such a technique, a drift of the crossing point is observed with $L$, as visible in the left inset (i) of Fig.~\ref{fig:F2D} (b). Already known for a few other models~\cite{wang_high-precision_2006,wenzel_comprehensive_2009}, the crossing points obtained from the stiffness converge very rapidly with $1/L$ to the bulk value $\lambda_c$, whereas we found a slower convergence for the estimates obtained from ${\cal F}/L$. Despite such effect (which may not be generic but model dependent), this simple example clearly shows that $\cal F$ is a very useful quantity to locate a quantum critical point between ordered and disordered phases for $d>1$.

\subsubsection{Experimental proposal}
\begin{figure}[h]
\begin{center}
\includegraphics[width=.45\columnwidth,clip]{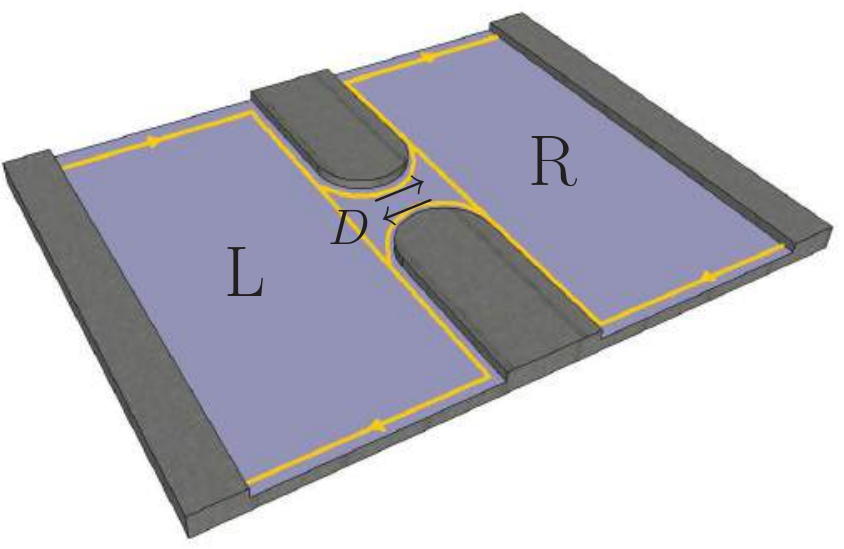}
\caption{Quantum point contact with chiral edge states (yellow lines) along the edges of a two-dimensional electron gas in a perpendicular magnetic field. The sample is divided into a left (L) and a right (R) region by a split gate acting as a quantum point contact. Electrons in the two in-coming edge states are transmitted with probability $D$ or reflected with probability $1-D$. Reprinted from \cite{song_bipartite_2012}.}
\label{fig:qpc_}
\end{center}
\end{figure}

\paragraph{Quantum point contact---}
A quantum point contact is a beam splitter with tunable transmission and reflection that serves as a "door" between electron reservoirs \cite{klich_quantum_2009,song_entanglement_2011-1}, as depicted in Fig.~\ref{fig:qpc_}. Time fluctuations of the current measured at this quantum point contact are exactly the bipartite fluctuations when replacing the temporal window $t$ by the spatial extent $x$. Due to space-time duality, such a replacment is always allowed for conformally invariant sytems \cite{song_entanglement_2011}, even when they are interacting. Quantum point contacts with free-electron reservoirs offer the opportunity to directly measure entanglement entropy. The first few cumulants, in particular $C_2$, have already been successfully measured \cite{gershon_detection_2008,le_masne_asymmetric_2009,ubbelohde_measurement_2012}.

\begin{figure}[h]
\begin{center}
\includegraphics[width=.55\columnwidth,clip]{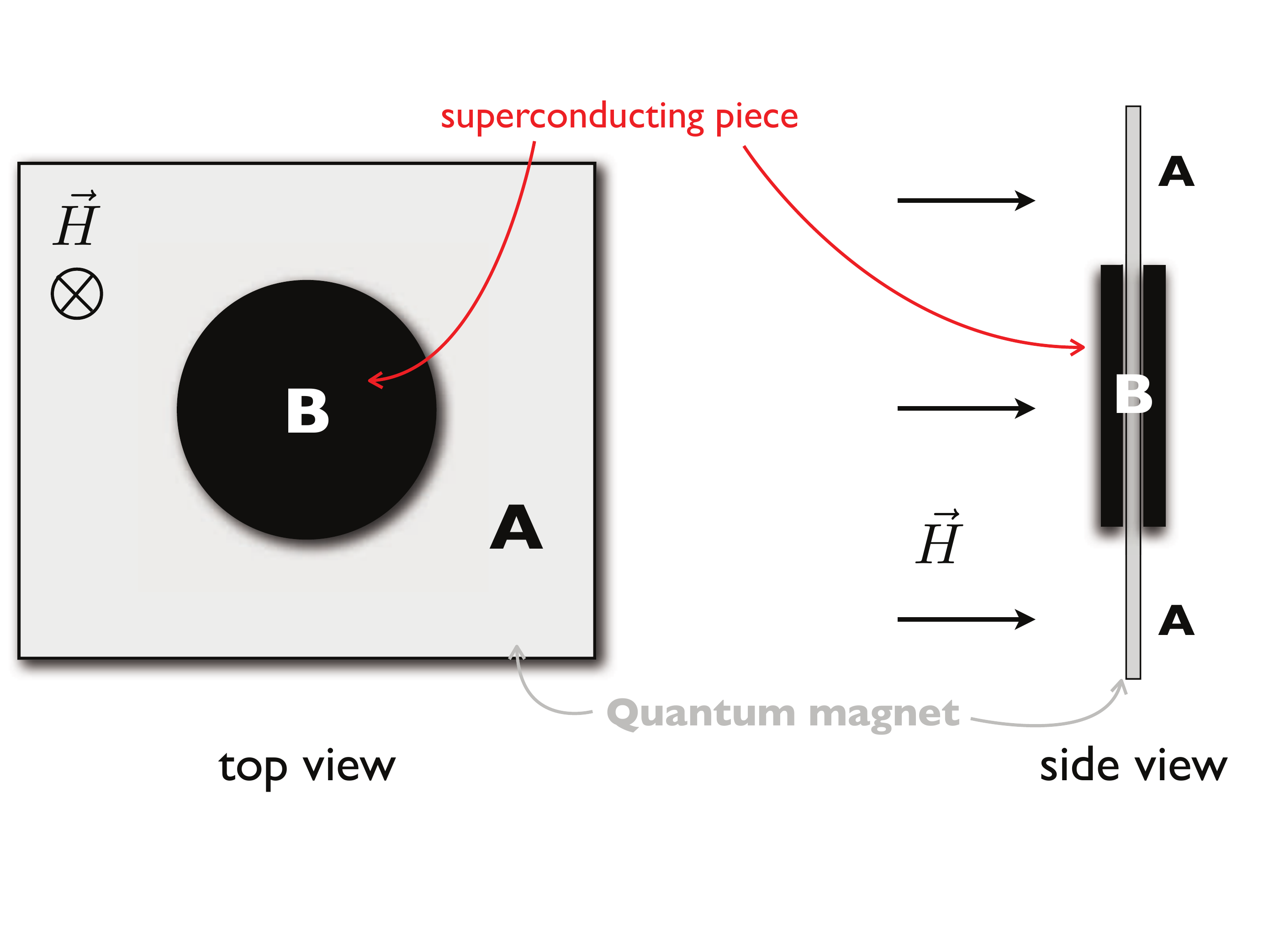}
\caption{Experimental setup proposed to extract the second cumulant (bipartite fluctuation) ${\cal{F}}_A$ of the magnetization within a subsystem $A$. The superconducting device (black disk) is placed on top of the quantum antiferromagnet (on region $B$) so that the Meissner effect will cancel the external field $h_B=0$ in the antiferromagnet whereas $h_A=H$. This Meissner screen is also placed on the other side such that only the $A$ contribution of the field-induced magnetization is measured. Reprinted from \cite{song_bipartite_2012}}
\label{fig:setup}
\end{center}
\end{figure}

\paragraph{Quantum antiferromagnets---}
Bipartite fluctuations can also be measured in O(2) symmetric quantum magnets, where the $z$-component of spin ${S}^z$ being the conserved charge, playing the role of particle number in mesoscopic and cold atom systems. 
The second cumulant of a subsystem $A$
\be
	{\cal{F}}_A = \sum_{i,j\in A} [\langle{{S}^z_i{S}^z_j}\rangle - \langle{{S}^z_i}\rangle\langle{{S}^z_j}\rangle ].\label{eq:FA}
\ee
can be understood as the Curie constant of the partial susceptibility, defined by
${\cal{F}}_A= T \times \left(\frac{d\langle S^z_A \rangle}{d h_A}\right)_{h_A\to 0}$ where $h_A$ is a small uniform external magnetic field applied to region $A$  {\it{only}}. As proposed in~\cite{song_entanglement_2011,song_bipartite_2012}, such a setup can be realized by applying the magnetic field over the entire sample while region $B$ (the rest of the system) is protected by superconducting Meissner screens. Sketched in Fig.~\ref{fig:setup}, such screens would eliminate the external field as well as the magnetic response outside region $A$. By varying the size of $A$ and extrapolating to very low temperature, the scaling of ${\cal{F}}_A$ could be measured.

\subsection{Entanglement detection and measure in cold atom experiments}

Quantum information processing using cold atoms trapped in optical lattices has been intensively discussed by many authors~\cite{garcia-ripoll_spin_2003,mandel_controlled_2003,zoller_implementing_2004,treutlein_quantum_2006,bloch_many-body_2008,bloch_quantum_2008}. Recently, theoretical proposals to directly extract entanglement witnesses, such as the R\'enyi entropies, have been debated in the context of artificial matter loaded in optical lattices. Below, we discuss the few theoretical proposal that have been suggested in the context of ultra-cold atoms, and present the very first experimental achievements.

\subsubsection{Theoretical proposals using SWAP-based protocols}
\begin{figure}[h]
\begin{center}
\includegraphics[width=.8\columnwidth,clip]{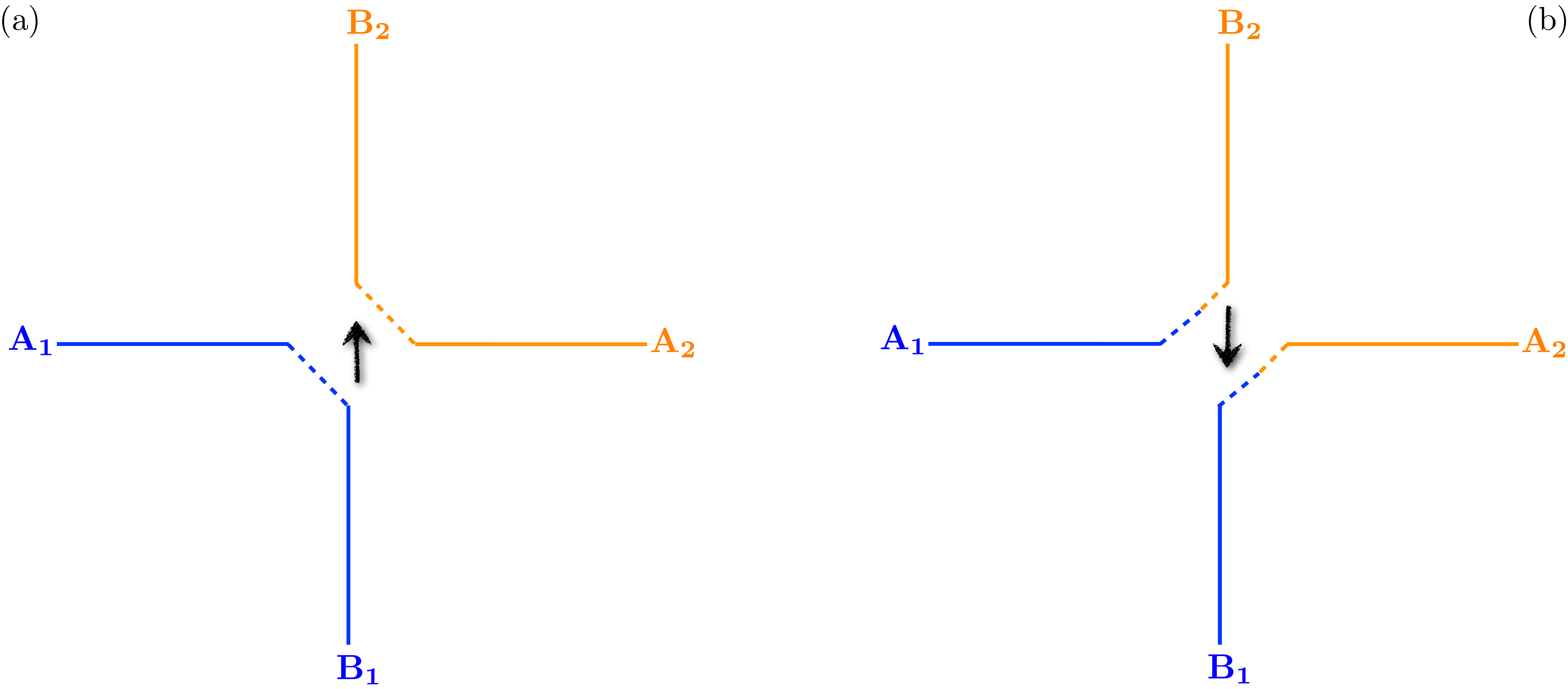}
\caption{Proposal for measuring the $q=2$ R\'enyi entropy, inspired from Abanin and Demler~\cite{abanin_measuring_2012}. Two identical copies are split in pairs of half-chains arranged in a cross geometry. A quantum switch (two-level system $\uparrow$ or $\downarrow$ at the center) controls the way in which the half-chains are connected by selectively allowing tunneling to one of the neighbors. The ground-states overlap Eq.~\eqref{eq:ol} of the two configurations is directly proportional to the $q=2$ R\'enyi entropy, and can be measured by studying Rabi oscillations of the quantum switch.}
\label{fig:swap}
\end{center}
\end{figure}

Using previous ideas based on copying a quantum system $q$ times~\cite{horodecki_method_2002}, and applying a SWAP operation~\cite{zanardi_entangling_2000,hastings_measuring_2010,cardy_measuring_2011}, Abanin and Demler~\cite{abanin_measuring_2012} proposed a general method to measure $q$-R\'enyi entropies for integer $q\ge 2$. The idea is based on the SWAP protocol, illustrated in Fig.~\ref{fig:swap} for $q=2$ copies of a one-dimensional system partitioned in two parts A$_{1,2}$ and B$_{1,2}$, controlled by a two-level system which plays the role of a quantum switch. Writing the ground-states $|{\rm{GS}}\rangle=\Bigl(\sum_i\lambda_i|\phi_i^{\rm A_1}\rangle\otimes|\phi_i^{\rm B_1}\rangle\Bigr)\otimes\Bigl(\sum_j\lambda_j|\phi_j^{\rm A_2}\rangle\otimes|\phi_j^{\rm B_2}\rangle\Bigr)$ and $|{\rm{GS'}}\rangle=\Bigl(\sum_i\lambda_i|\phi_i^{\rm A_1}\rangle\otimes|\phi_i^{\rm B_2}\rangle\Bigr)\otimes\Bigl(\sum_j\lambda_j|\phi_j^{\rm A_2}\rangle\otimes|\phi_j^{\rm B_1}\rangle\Bigr)$ of the two configurations (a) and (b) in Fig.~\ref{fig:swap}, the second R\'enyi entropy is given by the overlap
\be
\langle {\rm{GS}}|{\rm{GS'}}\rangle=\sum_i\lambda_i^4=\exp(-S_2).
\label{eq:ol}
\ee
In order to evaluate this overlap, a weak tunneling is introduced between the two states of the quantum switch, leading to an hybridization of the two ground-states: $|{\rm{GS}}\rangle\otimes |\hskip-0.1cm \uparrow\rangle$ and $|{\rm{GS'}}\rangle\otimes |\hskip-0.1cm \downarrow\rangle$. Studying Rabi oscillations of the quantum switch  between the two ground-states gives a direct access to the above overlap Eq.~\eqref{eq:ol} which is directly proportional to the Rabi frequency.

In the same time, Daley and co-workers~\cite{daley_measuring_2012} proposed a slightly different but related theoretical setup where entanglement growth is tracked during a quench dynamics of two identical copies of a boson chain whose mutual coupling is reduced. After tunneling has occurred, a measurement of the SWAP operator is achieved by measuring the parity number,

\subsubsection{First measurements}
\begin{figure}[t]
\begin{center}
\includegraphics[width=.65\columnwidth,clip]{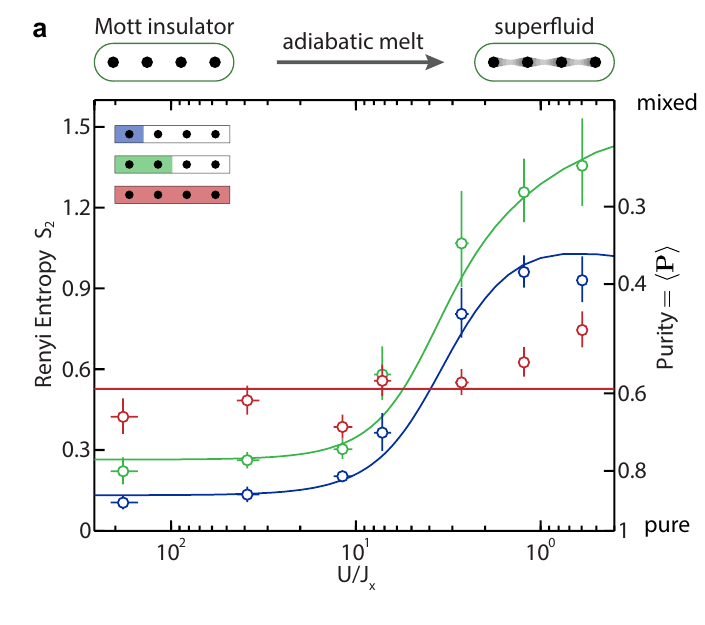}
\caption{Ground-state entanglement entropy and purity measured on a 4-site Bose-Hubbard chain across the Mott insulator to superfluid transition for different sub-systems (different colors). Circles are data and solid lines are exact diagonalization results. Reprinted from \cite{islam_measuring_2015}.}
\label{fig:islam}
\end{center}
\end{figure}

Based on some earlier theoretical proposals~\cite{ekert_direct_2002,moura_alves_multipartite_2004}, Islam {\it{et al.}}~\cite{islam_measuring_2015} realized the first measurement of the second R\'enyi entropy by making interference between two copies of a 4-site Bose-Hubbard chain in an optical lattice. Building on the quantum gas microscope technique~\cite{bakr_quantum_2009,sherson_single-atom-resolved_2010}, the second moment of the RDM (the purity) is accessed without resorting to quantum tomography~\cite{james_measurement_2001}. Using a $50\%$ - $50\%$ beam splitter the purity is directly estimated by measuring the average particle number parity. The measured second R\'enyi entropy is shown across the superfluid - insulator transition in Fig.~\ref{fig:islam}.

Using a different protocol, essentially based on particle number fluctuations~\cite{mazza_detecting_2015}, Fukuhara and co-workers~\cite{fukuhara_spatially_2015} have measured two-site entanglement between the spins of ultra-cold atoms loaded in an optical lattice. Starting with a localized spin-flip excitation, the time evolution of the concurrence between distant spins has been recorded through the measure of transverse spin correlations.

Note also that the quantum Fisher information~\cite{d._petz_introduction_2011}, already addressed in the context of many-body localization of trapped ions~\cite{smith_many-body_2015}, has been recently discussed by Hauke and co-workers~\cite{hauke_measuring_2016}. There, they made a stricking theoretical proposal for accessing quantum Fisher information via a direct measurement of dynamical susceptibilities.
\newpage

\section{Conclusion}
\label{sec:con}
\subsection{Summary}

In this review we have tried to give a general, while non-exhaustive, survey of the flourishing recent activity in the field of quantum entanglement in condensed matter physics, focusing on bipartite entanglement for clean and disordered systems.

Let us briefly summarize the main results we have reviewed. The ground-state of most correlated quantum systems exhibits an area law entanglement entropy, at most logarithmically enhanced for Luttinger, Fermi or Bose liquids. While the area law prefactor is expected to display a cusp singularity at quantum critical points, universality is encoded in sub-dominant terms, such as additive constants or corner contributions. Symmetry breakings or topological order can be clearly identified through such sub-leading corrections beyond area law in generalized R\'enyi entanglement entropies.
Regarding entanglement spectroscopy, we have seen that entanglement levels can contain more information than the entropies, and the "low-energy" part carries some universality for 1+1 conformal field theory, gapped states, broken symmetry phases, topological order. Quantum disordered systems such as random spin models, quantum (Kondo) impurity problems, or many-body localized systems all display fascinating entanglement properties. The most recent example being the many-body localization problem where highly excited states exhibit a dynamical transition between ergodic and non-ergodic regimes with qualitatively different entanglement properties, as well as anomalous entanglement spreading following a global quantum quench. On the experimental side, bipartite fluctuations in mesoscopic and solid state systems represent a promising tool to measure entanglement, as well as recent realizations in ultra-cold atom experiments.

\subsection{Open questions}
Let us give some directions towards a few open questions. The recently debated universality of entanglement spectra remains a largely open issue, leading to the question of the conditions under which an entanglement spectrum is universal.  A particularly interesting aspect is the spatial variation of 
the entanglement temperature, decaying away from the (real-space) cut for area law ground states.

Another key point concerns the computational aspects of entanglement in strongly correlated systems, in particular for highly entangled states for which going beyond brute force exact diagonalization (restricted to small systems) is a central issue, in particular to understand ergodicity breaking in interacting disordered systems at high energy. Systems with long-range couplings, relevant to cold atom experiments, are also very challenging for numerics since ground-state entanglement is no longer bounded to an area law.

Several aspects remain to be understood for $d\ge 2$ systems, in particular for non-Fermi liquid states where the logarithmic enhancement of the area law appears to be related but distinct from the free-fermion result. Shape dependence is also a very interesting topic where for instance in $d=2$ corner contributions carry a universal additive logarithmic correction in the entropy for conformal field theories. However, is universality in such corner terms also present at disordered quantum critical points?

Finally, the idea of a quantum revolution promoted by X.-G. Wen~\cite{wen_topological_2013} where there is unification between matter and quantum information is a very attractive one, but whether or not the standard model does emerge from long-ranged entangled qubits remains certainly an open issue.

\section*{Acknowledgments}
It is a great pleasure to warmly thank all my collaborators in works devoted to entanglement in condensed matter systems during the past decade: 
Ian Affleck, Fabien Alet, Sylvain Capponi, Ming-Shyang Chang, Christian Flindt, Jos\'e Hoyos, Israel Klich, Karyn Le Hur, David Luitz, Matthieu Mambrini, Eduardo Miranda, Alexandru Petrescu, Xavier Plat, Stephan Rachel, Nicolas Regnault, Zoran Ristivojevitch, Francis Song, Erik S\o rensen, Andr\'e Vieira, Thomas Vojta. 
I also want to thank Markus Holzmann, Andreas L\"auchli,  Gr\'egoire Misguich, and William Witczak-Krempa for their critical reading of the manuscript.

This work was supported by the Agence Nationale de la Recherche under programs Quapris (ANR-11-IS04-005-0) and BOLODISS (ANR-14-CE32-0018).
\end{document}